\documentclass[twocolumn,twocolappendix]{aastex631}
 \usepackage{mathptmx,courier}
  \usepackage[scaled=.92]{helvet}
  \usepackage{CJK}
  \DeclareMathAlphabet{\mathcal}{OMS}{cmsy}{m}{n}
\usepackage{natbib}
\usepackage{graphicx}
\usepackage[figuresright]{rotating}
\usepackage {amsmath,amssymb,eqnarray,tabularx}
\usepackage{enumerate}
\usepackage{enumitem}
\usepackage{mathtools}
\hypersetup{bookmarksnumbered=true,bookmarksopen=true}
\usepackage[all]{hypcap}

\newcommand{\hi}{{\rm H}{\textsc i}}

\newcommand{\degree}{\ensuremath{\text{\textdegree}}}

\def\jyb{\rm{Jy~beam^{-1} }}
\def\jykms{\rm{Jy~km~s^{-1} }}

\def\mjyb{\rm{mJy~beam^{-1} }}

\def\kms{\rm{km~s^{-1} }}
\def\cmsq{\rm{ cm^{-2} } }

\def\Msun{\rm{M_{\odot}}}

\defcitealias{Wang23}{W23}
\defcitealias{Walter08}{W08}

\begin{document}

\begin{CJK*}{UTF8}{gbsn}

\title{ FEASTS Combined with Interferometry (I): Overall Properties of Diffuse $\hi$ and Implications for Gas Accretion in Nearby Galaxies}
\correspondingauthor{Jing Wang}
\email{jwang\_astro@pku.edu.cn}

\author[0000-0002-6593-8820]{Jing Wang (王菁)}
\affiliation{ Kavli Institute for Astronomy and Astrophysics, Peking University, Beijing 100871, China}

\author{Xuchen Lin (林旭辰)}
\affiliation{ Kavli Institute for Astronomy and Astrophysics, Peking University, Beijing 100871, China}

\author{Dong Yang (杨冬)}
\affiliation{ Kavli Institute for Astronomy and Astrophysics, Peking University, Beijing 100871, China}

\author{Lister Staveley-Smith}
\affiliation{International Centre for Radio Astronomy Research, University of Western Australia, 35 Stirling Highway, Crawley, WA 6009, Australia}
\affiliation{ARC Centre of Excellence for All-Sky Astrophysics in 3 Dimensions (ASTRO 3D), Australia}

\author{Fabian Walter}
\affiliation{Max-Planck-Institut fu{\" r} Astronomie, K{\" o}nigstuhl 17, D-69117, Heidelberg, Germany}

\author{Q. Daniel Wang}
\affiliation{Department of Astronomy, University of Massachusetts, Amherst, MA 01003, USA}

\author{Ran Wang (王然)}
\affiliation{ Kavli Institute for Astronomy and Astrophysics, Peking University, Beijing 100871, China}

\author{A. J. Battisti}
\affiliation{Research School of Astronomy and Astrophysics, Australian National University, Cotter Road, Weston Creek, ACT 2611, Australia}
\affiliation{ARC Centre of Excellence for All-Sky Astrophysics in 3 Dimensions (ASTRO 3D), Australia}

\author{Barbara Catinella}
\affiliation{International Centre for Radio Astronomy Research, University of Western Australia, 35 Stirling Highway, Crawley, WA 6009, Australia}
\affiliation{ARC Centre of Excellence for All-Sky Astrophysics in 3 Dimensions (ASTRO 3D), Australia}

\author{Hsiao-Wen Chen}
\affiliation{Department of Astronomy and Astrophysics, The University of Chicago, 5640 S. Ellis Avenue, Chicago, IL 60637, USA}

\author{Luca Cortese}
\affiliation{International Centre for Radio Astronomy Research, University of Western Australia, 35 Stirling Highway, Crawley, WA 6009, Australia}
\affiliation{ARC Centre of Excellence for All-Sky Astrophysics in 3 Dimensions (ASTRO 3D), Australia}

\author{D. B. Fisher}
\affiliation{Centre for Astrophysics and Supercomputing, Swinburne University of Technology, P.O. Box 218, Hawthorn, VIC 3122, Australia}
\affiliation{ARC Centre of Excellence for All Sky Astrophysics in 3 Dimensions (ASTRO 3D), Australia}

\author{Luis C. Ho (何子山)}
\affiliation{ Kavli Institute for Astronomy and Astrophysics, Peking University, Beijing 100871, China}

\author{Suoqing Ji(季索清)}
\affiliation{Astrophysics Division \& Key Laboratory for Research in Galaxies and Cosmology, Shanghai Astronomical Observatory, Chinese Academy of Sciences, Shanghai 200030, China}

\author{Peng Jiang(姜鹏)}
\affiliation{National Astronomical Observatories, Chinese Academy of Sciences, 20A Datun Road, Chaoyang District, Beijing, China}

\author{Guinevere Kauffmann}
\affiliation{Max Planck Institut f\"ur Astrophysik, Karl-Schwarzschild-Strasse 1, 85748, Garching, Germany}

\author{Xu Kong(孔旭)}
\affiliation{Deep Space Exploration Laboratory / Department of Astronomy, University of Science and Technology of China, Hefei 230026, China }
\affiliation{School of Astronomy and Space Science, University of Science and Technology of China, Hefei 230026, China}

\author{Ziming Liu(刘孜铭)}
\affiliation{National Astronomical Observatories, Chinese Academy of Sciences, 20A Datun Road, Chaoyang District, Beijing, China}

\author{Li Shao(邵立)}
\affiliation{National Astronomical Observatories, Chinese Academy of Sciences, 20A Datun Road, Chaoyang District, Beijing, China}

\author{Jie Wang(王杰)}
\affiliation{National Astronomical Observatories, Chinese Academy of Sciences, 20A Datun Road, Chaoyang District, Beijing, China}

\author{Lile Wang(王力乐)}
\affiliation{ Kavli Institute for Astronomy and Astrophysics, Peking University, Beijing 100871, China}

\author{Shun Wang(王舜)}
\affiliation{ Kavli Institute for Astronomy and Astrophysics, Peking University, Beijing 100871, China}

\begin{abstract}

We present a statistical study of the properties of diffuse $\hi$ in ten nearby galaxies, comparing the $\hi$ detected by the single-dish telescope FAST (FEASTS program) and the interferometer VLA (THINGS program), respectively. 
The THINGS' observation missed $\hi$ with a median of 23\% due to the short-spacing problem of interferometry and limited sensitivity. 
We extract the diffuse $\hi$ by subtracting the dense $\hi$, which is obtained from the THINGS data with a uniform flux-density threshold, from the total $\hi$ detected by FAST. 
Among the sample, the median diffuse-$\hi$ fraction is 34\%, and more diffuse $\hi$ is found in galaxies exhibiting more prominent tidal-interaction signatures. 
The diffuse $\hi$ we detected seems to be distributed in disk-like layers within a typical thickness of $1\,\text{kpc}$, different from the more halo-like diffuse $\hi$ detected around NGC 4631 in a previous study. 
Most of the diffuse $\hi$ is cospatial with the dense $\hi$ and has a typical column density of $10^{17.7}$--$10^{20.1}\,\text{cm}^{-2}$. 
The diffuse and dense $\hi$ exhibits a similar rotational motion, but the former lags by a median of 25\% in at least the inner disks, and its velocity dispersions are typically twice as high. 
Based on a simplified estimation of circum-galactic medium properties and assuming pressure equilibrium, the volume density of diffuse $\hi$ appears to be constant within each individual galaxy, implying its role as a cooling interface. 
Comparing with existing models, these results are consistent with a possible link between tidal interactions, the formation of diffuse $\hi$, and gas accretion.

\end{abstract}


\section{Introduction} 
\label{sec:introduction}

In modern astrophysics, galaxies and their environments are considered an ecosystem of dark matter, stars and their remnants, and gaseous components \citep{Naab17}. Gas flows link the different parts of this system, cause matter/momentum/energy to exchange, and thus drive the growth and evolution of galaxies. The details of how these processes work remains a major question in galaxy astrophysics \citep{Somerville15}. From large to small scales, the key processes include gas-removing environmental effects, gas accretion into circum-galactic medium (CGM) from the inter-galactic medium (IGM), gas condensation into star-forming disks from the CGM, the flow of gas within the disk, the fueling of gas to star formation and black hole growth, and the feedback afterwards \citep{Crain23}. A complete observational view of different scales provides key constraints on galaxy formation models.

A long-lasting obstacle in modeling the processes of gas flowing through galaxies is the multi-phase nature of gas. Localized instabilities easily develop in a multi-phase gas, and can then propagate and grow, or dissipate and fade in ways that are not well understood and are highly stochastic \citep{Gronke22}. 
Multi-wavelength observations from X-ray to radio are necessary to trace the multi-phase, dynamic picture of baryonic flow all the way from the IGM to the star forming regions. 
The neutral atomic hydrogen gas ($\hi$) is an important part of the multi-phase gas at a wide range of scales. 
In the interstellar medium (ISM) where it is abundant, $\hi$ is the raw material to form molecular hydrogen that eventually leads to star formation \citep{Saintonge22}. But even in CGM and IGM, where the $\hi$ is insignificant in mass budget \citep{Tumlinson17}, its amount, distribution, and kinematics shed light on the complex gas physics there. 
Examples include the $\hi$ tidal structures possibly inducing gas cooling through turbulent mixing \citep{Sparre22, Wang23}, the predicted small $\hi$ cloudlets in the $10^7$ K CGM of elliptical galaxies formed through thermal instabilities \citep{Nelson20}, the observed high-velocity $\hi$ clouds around Milky Way (MW) marking the tip of the huge iceberg of the warm ionized gas being accreted onto the disk \citep{Richter17}, and the observed thick $\hi$ disk representing the interface of CGM condensing into the ISM \citep{Marasco19}. 
Detecting and characterizing the $\hi$ in different gaseous environments provide useful clues to the galaxy evolution. 
It is worth pointing out that, so far most quantifications of $\hi$ with column densities less than or equal to $10^{19}~\cmsq$ are based on Lyman-$\alpha$ absorption or Lyman limit systems. 

Traditional interferometric radio telescopes have been very helpful in imaging the $\hi$ in nearby galaxies through the 21-cm emission lines with a resolution sufficient to resolve star-forming clumps, but they struggle in simultaneously capturing the more diffuse and extended $\hi$. 
This is because interferometers construct images of the sky by filtering structures of specific spatial scales, and are thus limited by uv-coverage of the shortest baselines. 
For the time being, the most promising way to solve the problem is using large single-dish radio telescopes to fill in the``short-spacings'', in order to recover the large-scale missing $\hi$ \citep{Stanimirovic02}. Recent technical advances include \citet{Kurono09, Koda11, Rau19}, and recent scientific applications in this direction include \citet{Hess17, deBlok18, Richter18, Das20, Eibensteiner23, Wang23}. Most of these scientific studies confirm that interferometers tend to miss $\hi$ flux, though the exact amounts depend on source structures and telescope details.

We have been conducting an observing program FEASTS (FAST Extended Atlas of Selected Targets Survey, PIs: Jing Wang \& Jie Wang) at the Five-hundred-meter Aperture Spherical radio Telescope (FAST) since 2021, to map the $\hi$ in the disk as well as the $\sim$100 kpc surroundings of nearby galaxies. 
With FEASTS, the $\hi$ gas is mapped with 21-cm emission lines down to a few times $10^{17}~\cmsq$ column densities, approaching regimes typically detected by Lyman limit systems.
The first FEASTS science was conducted on the $\hi$ in/around the interacting galaxy NGC 4631 \citep[\citetalias{Wang23} hereafter]{Wang23}. That work identified a component of excess (by 40\%) $\hi$ detected by FAST but missed by WSRT (Westbork Synthesis Radio Telescope) as part of the HALOGAS (Hydrogen Accretion in LOcal GAlaxieS) survey \citep{Heald11}, which, until recently, was the deepest interferometric $\hi$ survey. 
We refer to this excess $\hi$ as the diffuse $\hi$, because it has a large scale ($\sim$30 kpc), moderate column density ($\lesssim10^{20}~ {\rm cm}^{-2}$), and high velocity dispersion ($\sim$50 $\kms$). 
The diffuse $\hi$ is more closely related to the warm dense $\hi$ that has a velocity dispersion larger than 8 $\kms$ than to the kinematically cooler dense $\hi$, and has a column density high enough to induce cooling flows of the hot CGM. 
Putting together, the diffuse $\hi$ seems to serve as an intermediate phase between the dense $\hi$ and warm ionized gas, and thus traces any energy inputting or dissipating processes that transfer gas between the cool (neutral) and hot (ionized) phases. 
In the context of galaxy evolution, the energy input or dissipative processes include tidal interaction, spiral arm and bar dynamics, stellar feedback and black hole feedback. 
The dissipating or cooling processes includes gas accretion into the ISM, condensation of atomic to molecular phase, and fueling of gas to star formation and black hole growth.
These are major physical processes in a typical galaxy formation model driving the baryonic flow and galaxy mass growth and evolution.
The diffuse $\hi$ newly detected thus provides a promising tool to understand both gas physics and galaxy evolution.

Among the many physical problems of galaxy evolution, how galaxies replenish their gas to sustain star formation has been one of the most challenging. 
Theoretically, galactic gas accretion channels include the cosmological hot and cold mode accretion, the stripping from or merger of satellite galaxies, and the fountain circulation \citep{Putman17}. 
The cosmological cold mode accretion through filaments penetrating the CGM is predicted to be significantly weakened at low redshift around massive galaxies with masses close to or above the MW \citep{Dekel06}. 
Cosmological hot mode accretion tends to couple with the latter two channels, as the circulated gas and stripped gas mix with, enrich, and accelerate cooling of the pristine CGM gas in the accretion process \citep{Grand19}. 
The significance of each channel has been only loosely constrained from observations, but it has become observationally feasible to search for and characterize signatures of gas accretion through the fountain and tidal$\slash$merger channels \citep{Fraternali17}. 
The HALOGAS has greatly advanced the frontier of quantifying extraplanar $\hi$ closely related with the fountain mechanism (e.g., \citealt{Marasco19}).
The diffuse $\hi$ detected far from the galactic disk of NGC 4631 suggests that the CGM cooling induced by tidal tails could be very effective \citepalias{Wang23}.

In this work, the first one of a series of studies combining FEASTS with interferometry, we extend the work of \citetalias{Wang23}, and look into a sample of ten nearby galaxies with both single-dish $\hi$ images from FEASTS and interferometric $\hi$ images from The HI Nearby Galaxy Survey (THINGS, \citealt{Walter08}, \citetalias{Walter08} hereafter). 
The major science goal of this paper is to combine the THINGS and FEASTS data to get a first-order measure of the diverse properties of diffuse $\hi$ in general $\hi$-rich galaxies, including its amount, morphology, distribution, and kinematics.  
As elaborated in the literature (e.g., \citealp{Stanimirovic02}), a combined analysis fully exploiting the advantages of both datasets is complex.
In the first place, we need to ensure the flux calibration between the two datasets to be consistent, so a technical goal of this paper is to optimize a simple procedure of flux cross-calibration in the image domain.
After reasonable cross-calibration of fluxes, we simply define the diffuse $\hi$ as the excess $\hi$ detected by FAST in comparison to uniformly thresholded interferometric data to get a quick overview of properties of this gas component. 
The definition will improve based on astrophysical arguments in the future, by filtering linearly combined or jointly de-convolved images and cubes (Wang et al. in prep, Lin et al. in prep).
We also use relatively simple ways to quantify localized properties of the diffuse $\hi$, but keep in mind that the structure and kinematics of the diffuse $\hi$ can be complex which requires sophisticated decompositions in the future (Lin et al. in prep, Oh et al. in prep). 
We will show that even the coarse measurements already point to interesting and different nature of the diffuse $\hi$ compared to the dense $\hi$; they represent an ensemble measure of complex neutral gas structures, and theoretically may well link to large-scale CGM condensation processes \citep{Gaspari18}.

The paper is organized as follows. 
We introduce the sample and data in Section~\ref{sec:data}.
We describe the observational setting and data reduction procedure of the FEASTS data, and a re-CLEAN of the THINGS data. 
Section~\ref{sec:analysis} describes the data analysis. 
We cross-calibrate the WCS (World Coordinate System) systems and flux intensity levels for the two datasets, which are important steps before scientific combining the two types of data.  
The strategies of the flux cross-calibrating procedure is optimized through mock tests presented in Appendix~\ref{sec:appendix_crosscali}. 
We then make the data cubes and moment images for the diffuse $\hi$. 
The results are presented in Section~\ref{sec:results}. 
We show how much $\hi$ is missed in interferometric observations, and major dependence of the integral diffuse $\hi$ fraction. 
The column density distributions of different types of $\hi$ are displayed.
The moment images of the diffuse $\hi$ are inspected, and radial profiles of properties measured from them are investigated and compared with those of the dense $\hi$. 
The possible physical meanings of the diffuse $\hi$ measurements are discussed in Section~\ref{sec:discussion}. 
We speculate the possible origins of the diffuse $\hi$ including fountains and tidal interactions, the possible link of diffuse $\hi$ to gas accretion, and the difference and similarities between diffuse $\hi$ detected here and that around NGC 4631 in \citetalias{Wang23}. 
Conclusion are made in Section~\ref{sec:conclusion}.
All image data used and radial distribution measurements discussed will be published online\footnote{ \url{https://github.com/FEASTS/LVgal/wiki}}.

\section{Data}
\label{sec:data}

\subsection{The Sample}
The sample targeted in this study is a subset from THINGS (\citetalias{Walter08}). 
We start from the sub-sample of 17 spiral galaxies used by \citet{Bigiel10} to study the star forming efficiency of $\hi$ gas in galactic outer disks. 
We exclude four galaxies at sky locations beyond an efficient observation of FAST, which have a declination above $60\degree$ or below $-5\degree$.
We further exclude three galaxies that have relatively small $\hi$ disks ($R_{\rm HI}<4'$) compared to the FAST beam FWHM of 3.24$'$. 
The $R_{\rm HI}$ is the semi-major axis of the isophote where the $\hi$ surface density reaches 1 $\Msun~{\rm pc}^{-2}$.

The remaining 10 galaxies are the target sample of this study (Table~\ref{tab:tab_sample}). 
These galaxies all have high-quality and relatively uniform interferometric 21-cm $\hi$ data taken at the Very Large Array (VLA) by the THINGS team, which is convenient for combining with the FAST $\hi$ image data. 
Scientific analysis of the combined $\hi$ data in this and future studies will benefit from the rich multi-wavelength data, including deep mid- and far infrared images from Spitzer SINGS (Spitzer Infrared Nearby Galaxies Survey, \citealt{Kennicutt03}) and Herschel KINGFISH (Key Insights on Nearby Galaxies: A Far-Infrared Survey with Herschel, \citealt{Kennicutt11}), deep ultraviolet images from GALEX (Galaxy Evolution Explorer) NGS (Nearby Galaxy Survey, \citealt{GildePaz07}), images of total CO (2-1) from HERACLES (HERA CO-Line Extragalactic Survey, \citealt{Leroy09}), images of dense CO (2-1) from PHANGS (Physics at High Angular resolution in Nearby GalaxieS)-ALMA (Atacama Large Millimeter/submillimeter Array) \citep{Leroy21}, IFS (integral field spectroscopy) data from PHANGS-MUSE (Multi Unit Spectroscopic Explorer) \citep{Emsellem22}, etc. 

 We utilize measurements of the central velocity, distance, inclination, and optical radius $R_{25}$ of the galaxies from \citetalias{Walter08}. 
 $R_{25}$ is the semi-major axis of the 25 mag arcsec$^{-2}$ isophote in the $B$ band.
 We measure the characteristic radius $R_{\rm HI}$ of $\hi$ disks from the THINGS column density images, using the procedure of \citet{Wang16}. 
We take the WISE (Wide-field Infrared Survey Explorer) W1, W2, and W4 fluxes, and GALEX FUV (far ultraviolet) fluxes from the $z=0$ Multiwavelength Galaxy Synthesis (z0MGS, \citealt{Leroy19}). 
We estimate the stellar mass ($M_*$) with the W1 and W2 luminosities, using equation of \citet{Querejeta15}. 
We estimate the star formation rate (SFR) with the FUV and W4 luminosities, using equation of \citet{Calzetti13}. 
We list these properties in Table~\ref{tab:tab_sample}.

    \begin{table*}
    \centering
            \caption{The galaxy sample. }
            \begin{tabular}{c c c c c c c c}
              \hline
              Galaxy  & $v$ & Dis &  $i$ & $R_{25}$ & $R_{\rm HI}$ & log $(M_*/\Msun)$ & log (SFR$/(\Msun/$yr) ) \\
                      &  (km$/$s) & (Mpc) & (degree)& (arcmin) & (arcmin) &   &  \\
              (1)     & (2)     & (3)     & (4)  & (5)    & (6)  & (7)  & (8)\\
              \hline
            NGC 628 & 659.1 & 7.3  &  7 & 4.67 & 8.18     & 10.24 & 0.23 \\
            NGC 925 & 552.5 & 9.2  & 66& 5.24 & 6.81     & 9.75 & -0.17 \\
            NGC 2841& 635.2 &14.1& 74& 4.07 & 7.64     & 10.93 & -0.07  \\
            NGC 2903& 556.6 & 8.9 & 65&5.74 & 8.20      & 10.42 & 0.32  \\
            NGC 3198& 661.2 &13.8& 72& 3.20 & 8.02     & 10.05 & 0.01 \\
            NGC 3521& 798.2 &10.7& 73& 4.20 & 7.60     &10.83 & 0.42 \\
            NGC 5055& 499.3 &10.1& 59&6.59 & 6.88      & 10.72& 0.28 \\
            NGC 5194& 456.2 &8.0  & 42& 5.00 & 6.00     & 10.73 & 0.65 \\
            NGC 5457& 226.5 &7.4  & 18& 10.94 & 15.15 & 10.39 & 0.54 \\
            NGC 7331& 815.6 &14.7& 76& 5.24 & 6.07    & 11.00 & 0.53\\
              
            \hline
            \end{tabular} 
            \\
              {\raggedright
                Column~(1): Galaxy name.
                Column~(2): Central velocity of $\hi$, from \citetalias{Walter08}.
                Column~(3): Luminosity distance, from \citetalias{Walter08}.
                Column~(4): The average kinematical inclination of the $\hi$ disk, from \citetalias{Walter08}.
                Column~(5): The optical radius, From \citetalias{Walter08}.
                Column~(6): The $\hi$ disk radius (this study).
                Column~(7): The stellar mass (this study).
                Column~(8): The SFR (this study). }
        \label{tab:tab_sample}
    \end{table*}

\subsection{The FEASTS Data}
\label{sec:feasts}
The FAST $\hi$ image observations were mostly conducted in the years 2022 and 2023, the first and second observing years of the FEASTS. 
The observing project IDs are PT2021\_0071 and PT2022\_0096 respectively.
The observations have been conducted in a mode similar to basket weaving, using horizontal and vertical on-the-fly scans. 
The receiver is rotated by 23.4 and 53.4$\degree$ in horizontal and vertical scans respectively to achieve an effective sampling spacing of 1.15$'$, so to satisfy the Nyquist-Shannon criterion. 
We refer the readers to \citetalias{Wang23} for more details of the observational settings.
We provide observing information for each target in Table~\ref{tab:tab_obs}, including the observing date, the median zenith angle, the south-east and north-west corners of the scanning region, the effective integration time per pointing, and the radio frequency interference (RFI) contaminating fraction.

The median zenith angles are all below 30$\degree$, causing only a low level of gain attenuation, which we use the equation from \citet{Jiang20} to correct for in data reduction. 
The RFI contaminating rate is minimal.
The observation of NGC 5457 was the most strongly affected by RFI, but after inspecting the data, we confirm that the RFI did not significantly contaminate the region of the galaxy.

    \begin{table*}
            \caption{Observing information of targets studied in this paper. }
            \begin{tabular}{|c|c|c|c|c|c|c|}
              \hline
              Galaxy  & Obs date & $\zeta$ & SE corner & NW corner & t$_{\rm eff}$ & f$_{\rm RFI}$ \\
              & (yy-mm-dd) & (degree) & & & (s) & (\%) \\
              (1) & (2)& (3) & (4) & (5) & (6) & (7) \\
              \hline
              NGC 628  & 2022-11-06 & 11.15 & 01h39m37.1s,+15d04m50.06s & 01h33m46.4s,+16d29m12.14s & 210.8 & 0.00 \\
              NGC 925  & 2021-11-15 & 8.84  & 02h28m34.9s,+33d12m20.66s & 02h25m53.1s,+33d57m03.14s & 250.1 & 0.00 \\
              NGC 2841 & 2023-03-09 & 25.78 & 09h25m13.3s,+50d28m35.76s & 09h18m52.1s,+51d28m35.74s & 269.6 & 0.00 \\
              NGC 2903 & 2022-02-02 & 5.61  & 09h34m16.4s,+21d00m32.85s & 09h30m05.1s,+22d00m29.48s & 216.2 & 0.28 \\
              NGC 3198 & 2023-03-19 & 20.37 & 10h22m46.4s,+45d02m58.86s & 10h17m03.6s,+46d02m58.84s & 252.9 & 0.00 \\
              NGC 3521 & 2022-03-26 & 26.06 & 11h08m01.0s,-00d34m52.23s & 11h03m37.2s,+00d31m04.40s & 208.4 & 0.43 \\
              NGC 5055 & 2022-05-01 & 17.12 & 13h18m16.1s,+41d31m48.81s & 13h13m22.6s,+42d31m45.44s & 243.6 & 0.00 \\
              NGC 5194 & 2022-04-28 & 22.17 & 13h32m49.0s,+46d46m47.01s & 13h27m35.8s,+47d46m43.64s & 256.9 & 0.00 \\
              NGC 5457 & 2023-04-19 & 29.78 & 14h09m13.2s,+53d28m23.16s & 13h57m12.0s,+55d13m27.84s & 354.7 & 10.65 \\
              NGC 7331 & 2022-09-20 & 16.87 & 22h39m29.5s,+33d54m56.76s & 22h34m38.6s,+34d54m56.84s & 230.5 & 0.04 \\
              \hline
            \end{tabular} 
            \\
              {\raggedright
              Column~(1): Galaxy name.
    Column~(2): The starting date of the observation .
    Column~(3): The median zenith angle ($\zeta$) during the observation.
    Column~(4): The South-East corner of the scanning region.
    Column~(5): The North-West corner of the scanning region.
    Column~(6): Effective integration time per pointing of beam size.
    Column~(7): RFI contaminating fraction in the 10-MHz wide low-redshift slice of the data.}
        \label{tab:tab_obs}
    \end{table*}

The data are reduced by a pipeline developed by the FEASTS team (\citetalias{Wang23}). 
The pipeline includes standard steps of RFI flagging, calibration, gridding, and continuum subtraction, and is optimized for the FEASTS data. 
We extract from the raw data a frequency slice of 10 MHz (equivalent to $\sim$2000 $\kms$) to focus on the surrounding environment of each target galaxy. 
As in \citetalias{Wang23}, for each observed rectangular region, we slightly remove data on the four sides in the gridding procedure, to ensure a relatively uniform sampling density throughout the image. 
The cut-out has a typical size of 1 deg$^2$. 
Other details of the procedures and parameter settings can be found in \citetalias{Wang23}. 
The product for each target is a data cube with a spatial resolution of 3.24$'$, pixel size of 30$''$, and channel width of 1.61 $\kms$. 
A flux detection mask is generated for each data cube through SoFiA \citep{Serra15}. 
When running SoFiA on FEASTS images, we use the Smooth \& Clip Finder with a smoothing kernel of 0, 3, 5 and 7 pixels along the $x$ and $y$ direction, and 0, 3, 7 and 13 channels along the $z$ direction. The large extent of smoothing along the z-axis is motivated by the detection of diffuse $\hi$ with high levels of velocity dispersion around the galaxy NGC 4631 in \citetalias{Wang23}. 
We set the detection threshold to be 4-$\sigma$, and the reliability threshold to be 0.99.

\begin{figure*} 
\centering
\includegraphics[width=15cm]{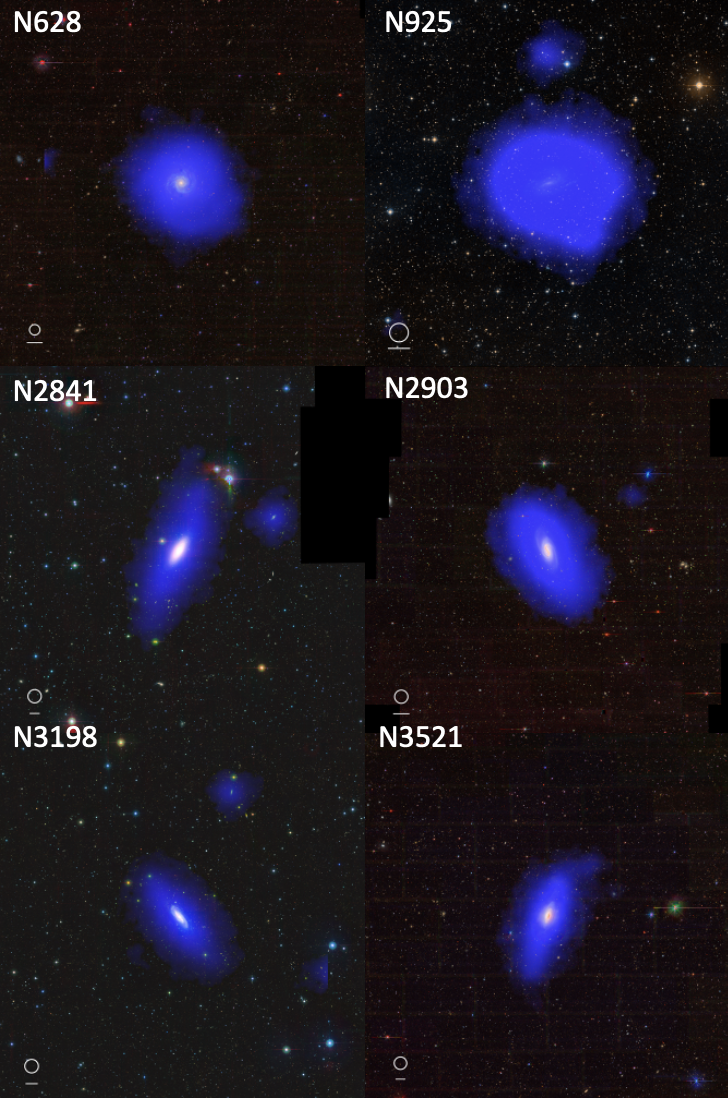}
\caption{ False-color images of FEASTS-detected $\hi$ in the optical image of galaxies. The FoV and coordinates of these plots are similar to those in Figure~\ref{fig:mom_n628} to \ref{fig:mom_n7331} in Appendix F. The small circle and horizontal bar shown at the bottom-left corner of each panel show the beam size of the FEASTS observation and a length of 10 kpc. The optical images are from the Legacy Survey \citep{Dey19}. To be continued.}
\label{fig:atlas}
\end{figure*}

We estimate the rms level of each data cube from the blank regions beyond the SoFiA masks.
We list the rms level and the related 3-$\sigma$ column density limit of $\hi$ by assuming a line width of 20 $\kms$ in Table 2. 
We make moment 0, 1, and 2 maps based on the SoFiA mask of each target source. 
We show the column density, velocity, and velocity dispersion images based on these maps in the first row of Figures~\ref{fig:mom_n628} to \ref{fig:mom_n7331} in Appendix~\ref{appendix:mom}.
We also present an atlas of false-color images in Figure~\ref{fig:atlas}, to give a quick impression of how the FEASTS detected $\hi$ distributes with respect to optical part of the galaxies and surroundings.

   \begin{table*}
    \centering
            \caption{Data properties }
            \begin{tabular}{c| c c| c c c c c c}
              \hline
                     & \multicolumn{2}{c|}{FEASTS} & \multicolumn{6}{c}{THINGS} \\
              Galaxy  & $\sigma$ &  N$_{\rm HI,lim}$ & $\sigma$ & W$_{\rm ch}$ & b$_{\rm maj}$ & b$_{\rm min}$ & N$_{\rm HI,lim}$ & (SNR)$_{\rm mom0}$  \\
                      &  ($\mjyb$) & ($10^{18}~\cmsq$) & ($\mjyb$)& ($\kms$) & (arcsec) & (arcsec) & ($10^{18}~\cmsq$)  & \\
              (1)     & (2)     & (3)     & (4)  & (5)    & (6)  & (7) & (8) & (9)\\
              \hline
NGC 628 & 0.88 & 0.43 & 0.58 & 2.58 & 11.0 & 8.6 & 146 & 3.85 \\
NGC 925 & 0.88 & 0.43 & 0.63 & 2.58 & 5.9 & 5.7 & 444 & 2.98 \\
NGC 2841 & 0.87 & 0.43 & 0.34 & 5.15 & 10.5 & 9.3 & 119 & 3.08 \\
NGC 2903 & 0.85 & 0.42 & 0.38 & 5.15 & 16.5 & 14.3 & 54 & 15.56 \\
NGC 3198 & 0.81 & 0.40 & 0.36 & 5.15 & 12.1 & 10.8 & 93 & 21.91 \\
NGC 3521 & 1.18 & 0.59 & 0.40 & 5.15 & 15.9 & 11.1 & 76 & 21.58 \\
NGC 5055 & 0.83 & 0.41 & 0.35 & 5.15 & 9.0 & 8.2 & 159 & 2.29 \\
NGC 5194 & 0.80 & 0.40 & 0.37 & 5.15 & 11.0 & 9.3 & 123 & 1.85 \\
NGC 5457 & 0.94 & 0.46 & 0.40 & 5.15 & 13.3 & 11.8 & 87 & 15.97 \\
NGC 7331 & 0.74 & 0.37 & 0.52 & 5.15 & 7.5 & 7.3 & 319 & 14.88 \\
             
            \hline
            \end{tabular} 
            \\
              {\raggedright
                Column~(1): Galaxy name. 
                Column~(2): rms level of the FEASTS cube, in unit of mJy per FAST beam.
                Column~(3): Projection corrected $\hi$ column density limit of the FEASTS data, assuming 3-$\sigma$ detection and 20 $\kms$ line widths. The projection correction is conducted by assuming an infinitely thin disk with the inclination listed in Table~\ref{tab:tab_sample}.
                Column~(4): rms level of the THINGS cube, in unit of mJy per VLA beam.
                Column~(5): Channel width of the THINGS data.
                Column~(6): Major axis of the THINGS beam.
                Column~(7): Minor axis of the THINGS beam.
                Column~(8): $\hi$ column density limit of the THINGS data, assuming 3-$\sigma$ detection and 20 $\kms$ line widths. 
                Column~(8): Median SNR of the $\hi$ column density image. }
        \label{tab:tab_dataprop}
    \end{table*}

\addtocounter{figure}{-1}
\begin{figure*} 
\centering
\includegraphics[width=15cm]{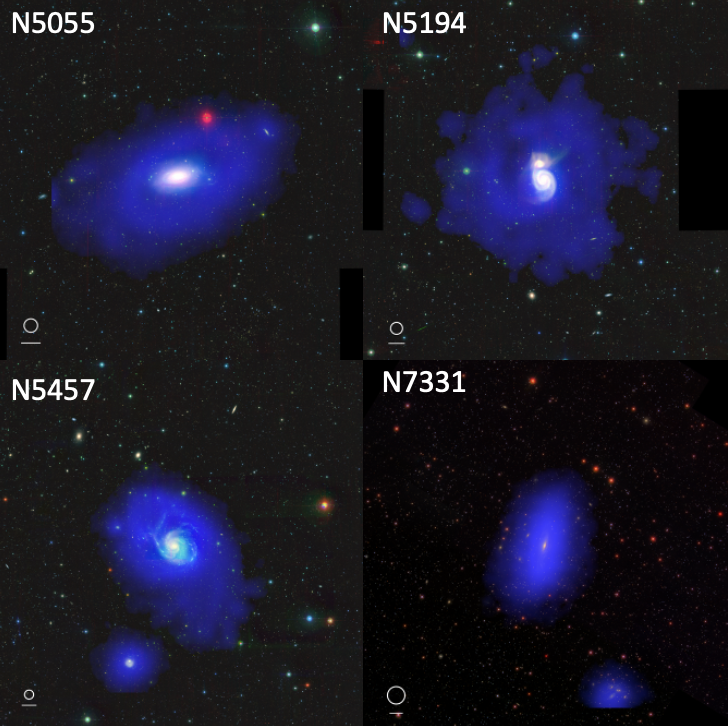}
\caption{False-color images of FEASTS-detected $\hi$ in the optical image of galaxies.  Continued.}
\end{figure*}

\subsection{The THINGS Data}
\label{sec:things}
THINGS $\hi$ data were taken at the VLA in its B, C and D array configurations, achieving baselines that range from 35 to 11.4 km, corresponding to a spatial resolution of 25.2$'$ to 4.64$''$.  
The publicly available cubes of \citetalias{Walter08} were reduced by the THINGS team with the Astronomical Image Processing System (AIPS). 
Conventionally, {\it standard cubes} are produced by directly adding residual cubes to the clean-beam convolved model cubes ({\it convolved model cubes} for short hereafter).
But in \citetalias{Walter08}, the residual maps are rescaled to account for the difference of areas between the dirty and clean beams.
This algorithm has been discussed in detail in \citet{Jorsater95} and \citet{Walter99}. 
The rescaled residual cubes are added to the convolved model cubes to produce the {\it rescaled cubes}. 

We do not use the cubes and related products from \citetalias{Walter08}, including moment images, spectra, and integral fluxes, for all the targets in our sample.  
We re-do the CLEAN procedure, starting from the calibrated and continuum removed visibilities obtained in \citetalias{Walter08}, for three major reasons. 
Firstly, the outer regions of a few galaxies with extended $\hi$ structures (NGC 628, NGC 5055, NGC 5194, NGC 5457) have been excluded in the \citetalias{Walter08} cube creation to avoid dealing with regions with significant primary beam attenuation, and to achieve a relatively uniform noise level throughout the image. 
For a complete comparison (and feathering in the future) with the FEASTS data, we require better coverage of the outer regions.
Secondly, compared to the classical CLEAN algorithm used in \citetalias{Walter08}, the multiscale CLEAN may work better at recovering the large-scale fluxes. 
Thirdly, it may be useful to test the possibility that the convolved model cubes work better than the rescaled cubes in various analyses combined with single-dish data.

We use the CASA script {\it tclean} to conduct the imaging and CLEAN procedures for all the galaxies except for NGC 7331.
We use a pixel size of 1.5$''$ following the data release of \citetalias{Walter08}, `natural' weighting to maximize sensitivity, a cleaning threshold of 1.5-$\sigma$, and a maximum cleaning iteration of 500,000.
We do not use the robust weighting that usually produces better-shaped dirty beams because the sensitivity will be worse while the analysis combined with the FEASTS data is strongly limited by the interferometric depth (section~\ref{sec:NHIdistr}, \ref{sec:prof} and Appendix~\ref{sec:appendix_crosscali}).
We use the `multiscale' deconvolver with a small scale bias of 0.2,  and scales of roughly 0, 1, 2, 4, and 8 times the beam major FWHM. 

Through inspecting convolved models and residual images, we find that the multiscale CLEAN may not be the best deconvolver for NGC 7331, possibly because extended fluxes are hidden by the relatively high noise level and high inclination angle of the galaxy. Instead, the hybrid Hogbom$\slash$Clark$\slash$Steer CLEAN algorithm of Miriad script {\it clean} produces a reasonably good cube for this galaxy. The CLEAN threshold and iterations are set as for other galaxies.

We derive the source masks from standard cubes using SoFiA. 
We use the Threshold Finder, with smoothing kernels of sizes of 30$''$ along the $x$ and $y$ directions, and 3 channels along the $z$ direction.
We use a detection threshold of 2-$\sigma$, and reliability threshold of 0.99.
These settings largely follow those of \citetalias{Walter08}.
For each target, we use the region within the SoFiA mask, to derive the scaling factor to correct for difference between dirty and clean beams, following the method of \citetalias{Walter08}.
We rescale the residual cubes before adding them to the convolved model cubes, and produce the rescaled cubes.  
We then use the SoFiA masks to derive moment maps from the rescaled cubes. 
The related column density, velocity, and velocity dispersion images are displayed in the second row of Figures~\ref{fig:mom_n628} to \ref{fig:mom_n7331} in Appendix~\ref{appendix:mom}.

We list in Table~\ref{tab:tab_dataprop} the rms level, the beam size, and $\hi$ column density limit of each standard cube. 
There might be systematic differences between the new cubes and the \citetalias{Walter08} cubes, due to different reduction processes. 
We will compare measurements from the two sets of cubes to assess the systematic difference in most analyses. 

We have used the W08 method to scale the CLEAN residuals before adding them to the convolved models.
The alternative method to deal with residual fluxes of the CLEAN process  is to deeply clean the data with a ``clean mask''. 
This method firstly identifies regions where deconvolution is need, and then within these regions cleans down very deeply to a fraction of the rms (e.g. 0.1-$\sigma$). 
This avoids the scaling of residuals.
We have adopted the W08 method instead of deep-clean, in consideration of the low-SNR (signal-to-noise ratio) and large-angular-scale nature of the THINGS data.
The detailed justification can be found in Appendix~\ref{sec:appendix_deep_clean}.

\section{Analysis}
\label{sec:analysis}
In the following, we cross-calibrate the geometries and fluxes between the THINGS and FEASTS data, and analyze the possible excess $\hi$ detected by FEASTS in comparison to the THINGS data. 
We refer to the directly measured excess of $\hi$ from FEASTS compared to the THINGS data as the missed $\hi$.
We set a uniform threshold to select dense $\hi$ from the THINGS data, and quantify the respective excess $\hi$ from FEASTS data as the diffuse $\hi$.

\subsection{Aligning the FEASTS and THINGS Images}
The FAST observations have a typical pointing uncertainty of 10$''$ to 20$''$, so we align the FAST images to the THINGS ones before conducting the flux cross-calibration.
The pointing accuracy should not influence the calibration by comparing amplitudes of Fourier transformed images, but affect the sanity check using the imaginary and real parts (Section~\ref{sec:fluxcali}). 
It also significantly influence the deviation of moment images for the diffuse $\hi$ (Section~\ref{sec:creat_dmom}).

We conduct the alignments by deriving the shifts needed in $x$ and $y$ directions for the FEASTS and THINGS moment-0 images to be closest near the galaxy center.
In practice, we modify the header of the FEASTS moment-0 image by adding values ranging from $-0.5$ to 0.5 pixels ($-15$ to 15$''$) with a step of 0.1 pixel (3$''$) to the CRPIX1 and CRPIX2 keywords separately. 
We reproject each shifted FEASTS moment-0 image to the THINGS moment-0 image through comparing their WCS keywords. 
We convolve the THINGS moment-0 image with the FAST beam and subtract it from the reprojected FEASTS moment-0 image. 
We calculate the standard deviation of fluxes within a box with widths of 300$''$ around the galaxy center from the difference image.
The set of $x$ and $y$ shifting values that result in the lowest standard deviation are taken as the best correction, and is applied to the header of the FEASTS data.

We list the derived best corrections in Table~\ref{tab:tab_crosscali}.
The derived corrections in the $x$ and $y$ directions have a median value of $-0.3\pm$0.1 pixel ($-9.0\pm3.3''$) and $-0.25\pm$0.2 pixel ($-7.5\pm6.4''$) respectively.

 \begin{table}
    \centering
            \caption{Cross-calibration result}
            \begin{tabular}{c c c c c}
              \hline
              Galaxy  & $\delta x$  & $\delta y$ & $f_{\rm F/T}$  & $f_{\rm F/T, W08}$  \\
              		  & pixel  & pixel  & & \\
              (1)     & (2)     & (3)  & (4) & (5)     \\
                            \hline
                NGC 628  & -0.4 & 0.1  & 1.08 $\pm$ 0.08 & 1.04 $\pm$ 0.16 \\
                NGC 925  & -0.1 & -0.3 & 0.98 $\pm$ 0.16 & 1.04 $\pm$ 0.14 \\
                NGC 2841 & -0.2 & -0.5 & 1.00 $\pm$ 0.25 & 1.11 $\pm$ 0.10\\
                NGC 2903 & -0.3 & -0.2 & 0.91 $\pm$ 0.12 & 0.96 $\pm$ 0.11\\
                NGC 3198 & -0.3   & -0.5& 1.09$\pm$0.12   & 1.07 $\pm$ 0.05 \\
                NGC 3521 & -0.2   &  0.0& 1.01$\pm$0.15   & 0.91 $\pm$ 0.21 \\
                NGC 5055 & -0.3 & -0.3& 1.08 $\pm$ 0.11 & 1.10 $\pm$ 0.13 \\
                NGC 5194 & -0.3 & -0.1& 1.15 $\pm$ 0.05 & 1.12 $\pm$ 0.08 \\
                NGC 5457 & 0.0 & 0.0 & 0.99 $\pm$ 0.22 & 1.07 $\pm$ 0.31\\
                NGC 7331 & -0.3 & -0.5& 1.13$\pm$0.06 & 1.26 $\pm$ 0.07\\
              \hline
            \end{tabular} 
            
              \raggedright
                Column~(1): Galaxy name.
                Column~(2): Correction to the WCS keyword CRPIX1 in the FEASTS header ($\delta x$).
                Column~(3): Correction to the WCS keyword CRPIX2 in the FEASTS header ($\delta y$).
                Column~(4): Cross calibrated scaling factor $f_{\rm F/T}$, derived from the moment-0 image of the rescaled cube reduced in this study.
                Column~(5): Cross calibrated scaling factor $f_{\rm F/T}$, derived from the moment-0 image of the rescaled cube from \citetalias{Walter08}

        \label{tab:tab_crosscali}
    \end{table}

\subsection{Cross-calibrating the Fluxes between THINGS and FEASTS}
\label{sec:fluxcali}
An important step before we discuss excess or diffuse $\hi$ is to achieve accurate flux calibration of the FEASTS data using the THINGS data as the reference. 
In principle, the systematic difference of fluxes between different telescope observations should be at a relatively small level of a few percent. 
We conduct the flux cross-calibration through comparing amplitudes of Fast Fourier Transformed (FFT) images in overlapping uv space of the two types of data.
We present details of setting up the cross-calibration procedure in Appendix~\ref{sec:procedure_crosscali}, which is improved based on the version of \citetalias{Wang23}, and is adjusted to suit the THINGS data. 

We present the derived scaling factor $f_{\rm F/T}$ in Table~\ref{tab:tab_crosscali}, by which the original FEASTS fluxes should divide in order to be consistent with the THINGS ones.
The median $f_{\rm F/T}$ is 1.07$\pm$0.07, consistent with the fact that systematic uncertainties of the flux calibration in $\hi$ observations are typically a few percents (\citetalias{Walter08}).
We also derive the scaling factors with the moment-0 images of \citetalias{Walter08} for all ten galaxies (also in Table~\ref{tab:tab_crosscali}). 
Because only the inner regions (covered by both types of cubes) are used for cross-calibration, the scaling factors from using the two sets of VLA cubes are expected to be close.
We can see that the two types of $f_{\rm F/T}$ are indeed largely consistent, with a median difference of 0.02$\pm$0.06. 
The consistency in $f_{\rm F/T}$ supports the robustness of our method of deriving the scaling factor between the FAST and VLA fluxes.
All FEASTS $\hi$ fluxes are divided by the corresponding $f_{\rm F/T}$ values from the newly reduced THINGS cubes, and all THINGS data are corrected for the primary beam attenuation hereafter. 

In Appendix~\ref{sec:procedure_crosscali}, the comparison in amplitudes of FFT images also reveals that the FAST observations tend to reveal excess fluxes when the angular scale is larger than 8.82$'$, corresponding to 18.7 to 37.7 kpc in the sample. 
It is thus reasonable to quantify the spatial distribution and kinematics of the missed $\hi$ and the closely related diffuse $\hi$ (Section~\ref{sec:creat_dmom}) with the FAST resolution of 3.2$'$. 

\subsection{Extracting the Missed HI and Diffuse HI}
\label{sec:creat_dmom}

We obtain the cube of the missed $\hi$ for each galaxy as the FEASTS cube minus the resolution degraded THINGS cube. 
The resolution degraded THINGS cube is produced by reprojecting the original THINGS cube to the WCS coordinates of the corresponding FEASTS cube, and then convolving the reprojected cube with the FEASTS beam. 
When the spatially resolved properties of the THINGS detected $\hi$ are compared to those of the missed $\hi$ or total $\hi$, they are always measured from the resolution degraded THINGS data in this study. 

The amount of missed $\hi$ sensitively depends on the varying detection limits of the THINGS data. 
So we extract the diffuse $\hi$ by firstly separating the dense $\hi$ with a relatively uniform flux density threshold from the THINGS data, and then subtracting it from the total $\hi$.
For each THINGS cube we set a uniform flux density threshold as $0.005b_{maj} b_{min}\, {\rm arcsec}^{-2} \mjyb$ ($0.0071 b_{maj} b_{min} \, {\rm arcsec}^{-2} \mjyb$) when the channel width is 5.2 $\kms$ (2.6 $\kms$), where the $b_{maj}$ and $b_{min}$ are the major and minor axes of the THINGS synthesis beam (Table~\ref{tab:tab_sample}). 
We select voxels with flux densities above this threshold in each THINGS cube, and produce a region mask.
We expand the mask by two pixels and blank all THINGS cube voxels beyond the mask. 
By doing so, we have produced a data cube for the relatively dense $\hi$, the dense $\hi$ cubes for short hereafter.
The threshold used corresponds to a 3-$\sigma$ $\hi$ column density limit of 1.83$\times10^{20}~\cmsq$ assuming a line width of 20 $\kms$, which is higher than the column density limits of original THINGS data for most galaxies except for NGC 925 and NGC 7331 (Table~\ref{tab:tab_dataprop}).
But because source finding for the original THINGS cubes is based on smoothed data, the relatively low threshold used in these two galaxies do not significantly bias the detection of dense $\hi$, which is later supported by the lower values of dense $\hi$ fluxes than the THINGS total $\hi$ fluxes for these two galaxies (section~\ref{sec:amount_diffuseHI}).
The dense $\hi$ cubes are reprojected to the WCS coordinates of the corresponding FEASTS cubes, and then convolved with the FEASTS beam. 
The column density, velocity, and velocity dispersion images from the degraded cubes of dense $\hi$ are displayed in the third row in Figure~\ref{fig:mom_n628} to \ref{fig:mom_n7331} in Appendix~\ref{appendix:mom}.
Hereafter, all spatially properties of the dense $\hi$ are measured from the resolution degraded data.

The cube of diffuse $\hi$ for each galaxy is obtained as the FEASTS cube minus the degraded dense $\hi$ cube.
The diffuse $\hi$ defined in this way is roughly the missed $\hi$ plus the $\hi$ that lies between the THINGS detection limits and the selection criteria described above from the THINGS data.
It should be a mixture of low-density $\hi$ and large-scale $\hi$.
We derive moment 0, 1 and 2 images of the diffuse $\hi$ with the SoFiA masks of the FEASTS cubes.
Only line-of-sights where the column densities of diffuse $\hi$ are above $10^{18}~\cmsq$, and the ratios of diffuse $\hi$ over dense $\hi$ higher than 1\% are considered in the deviation of the moment 1 and 2 images.
We display the related column density, velocity, and velocity dispersion images of the diffuse $\hi$ in the fourth row in Figures~\ref{fig:mom_n628} to \ref{fig:mom_n7331} in Appendix~\ref{appendix:mom}.
Obtaining the moment images of diffuse $\hi$ through projecting the diffuse $\hi$ cubes is totally equivalent to deriving them through mathematical combinations of moment images of the dense $\hi$ and total $\hi$. We put the relations in Appendix~\ref{appendix:mom_relation}, which helps cross-check patterns from images of different $\hi$ components, and assess errors propagated from the THINGS data. 

We also derive the {\it difference velocity image} as the degraded dense $\hi$ moment-1 image minus the diffuse $\hi$ moment-1 image. 
The difference velocity images show how much slower or faster the rotational$\slash$orbital motion of the diffuse $\hi$ is compared to that of the dense $\hi$.
 The difference velocity image for each galaxy is displayed in the fifth row of Figures~\ref{fig:mom_n628} to \ref{fig:mom_n7331} in Appendix~\ref{appendix:mom}.

Except in Section~\ref{sec:missed_HI} and \ref{sec:NHIdistr}, we present results mostly on the diffuse $\hi$ in the main part of the paper.

\subsubsection{Deriving Radial Profiles of the Diffuse $\hi$ Emission}
\label{sec:derive_prof}
We derive radial distribution of properties for the total $\hi$, the dense $\hi$ (degraded to FAST resolution), and the diffuse $\hi$, which are denoted by the subscripts tot, dens, and diff, respectively.  

For the moment-0 and 2 images of different $\hi$ types, we derive radial profiles using annuli which have the same position angles and axis ratios as the optical disks (Table~\ref{tab:tab_sample}).  
All profile measurements start from the FAST beam FWHM to minimize the spatial smoothing effects of the beam.
We do not correct for projection effects, as the geometry of the diffuse $\hi$ in outer disks is highly uncertain. 

We also derive radial profiles of the normalized velocity difference between the dense and diffuse $\hi$.
It measures the lagged extents of rotational$\slash$orbital velocities of the diffuse $\hi$ in comparison to the dense $\hi$.
For each galaxy, a slit is put along the major-axis of the disk, with a width equal to the FAST beam FWHM. 
The averaged values are obtained along this slit from the difference velocity image ($v_{{\rm r,dens}}-v_{{\rm r,diff}}$), and from the velocity image of the dense $\hi$ ($v_{{\rm r,dens}}$). 
The former is divided by the latter to obtain the normalized velocity difference. 
The normalized velocity difference is positive and within unity when the diffuse $\hi$ is co-rotating (or orbiting in the same direction) with but more slowly than the dense $\hi$, is negative when the former is co-rotating (or orbiting in the same direction) with but faster than the latter, and above unity when the former is counter-rotating (or orbiting in the opposite direction) with respect to the latter. 
The infalling gas from a companion coming in in the same direction as the disk or from a well aligned CGM should still orbit in the same direction; but if the interaction geometry is inclined or the CGM mis-aligned, the infall gas forms a warped outer disk or tail, which has a chance to show (projected) faster or opposite orbiting.

\section{Results}
\label{sec:results}
In the following, we present results of this paper.
We briefly discuss the amount of missed $\hi$, which is the $\hi$ detected by FEASTS but not by the THINGS data.
We focus more on the diffuse $\hi$, which includes the missed $\hi$ but is more uniformly selected by flux densities than the missed $\hi$. 

\subsection{Global Measures of the Missed $\hi$}
\label{sec:missed_HI}

The integrated $\hi$ fluxes of FEASTS, and the fraction of missed $\hi$ ($f_{\rm missed}$) in the THINGS data estimated from both newly reduced cubes and \citetalias{Walter08} cubes are listed in Table~\ref{tab:tab_flux}.
The total $\hi$ fluxes, and amount of flux underestimation in this THINGS subset is for the first time uniformly quantified with a homogeneous and deep single-dish image dataset. 

\begin{figure*} 
\centering
\includegraphics[width=5.4cm]{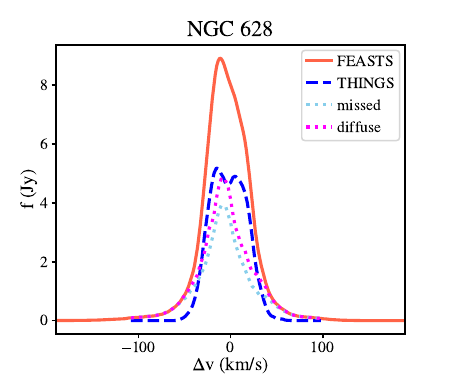}
\includegraphics[width=5.4cm]{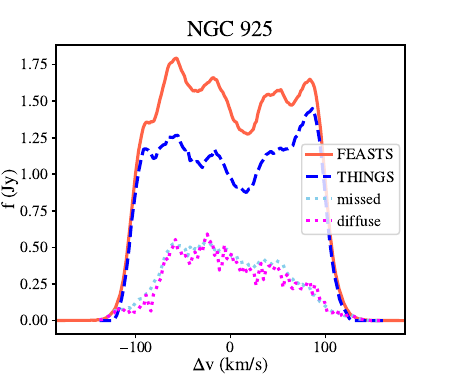}
\includegraphics[width=5.4cm]{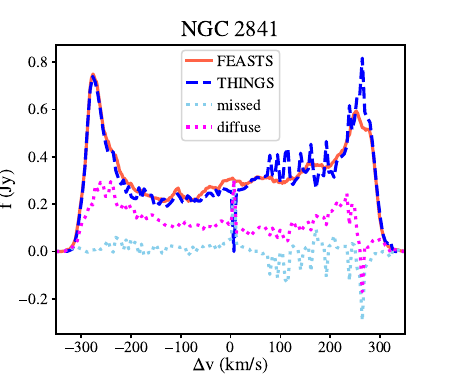}

\includegraphics[width=5.4cm]{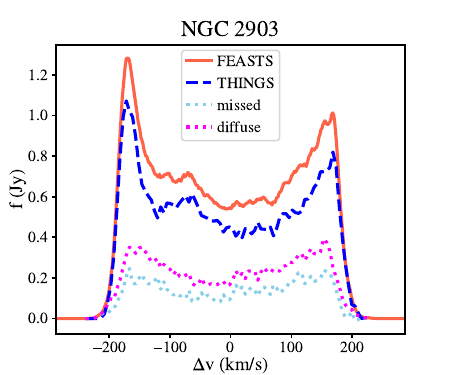}
\includegraphics[width=5.4cm]{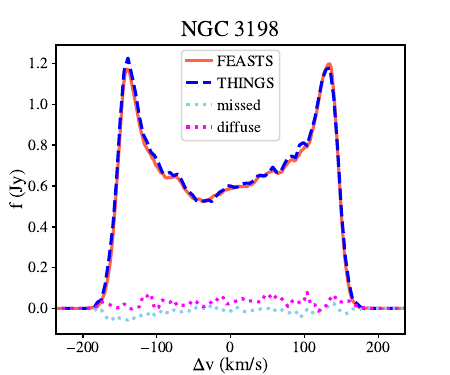}
\includegraphics[width=5.4cm]{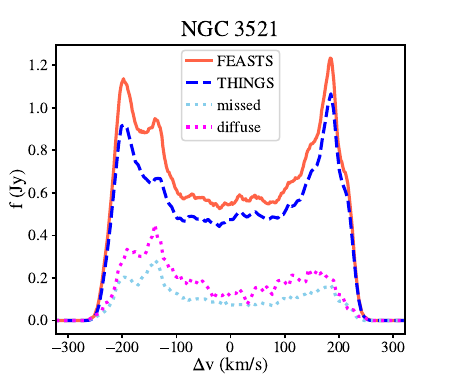}

\includegraphics[width=5.4cm]{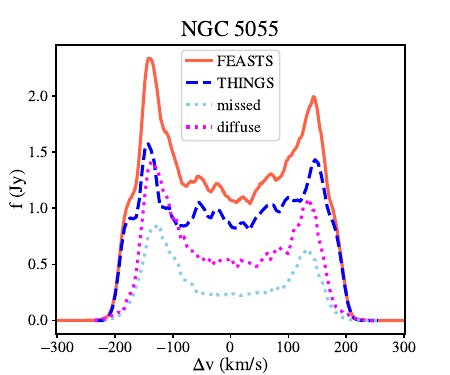}
\includegraphics[width=5.4cm]{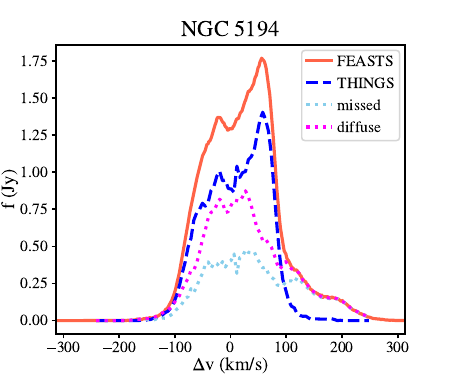}
\includegraphics[width=5.4cm]{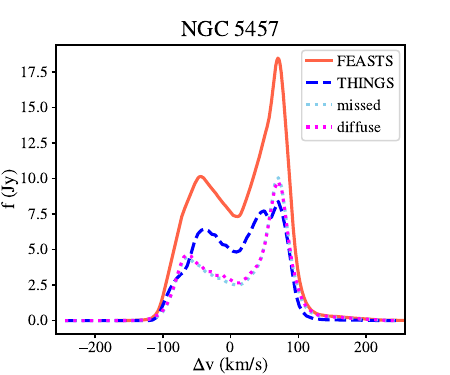}

\includegraphics[width=5.4cm]{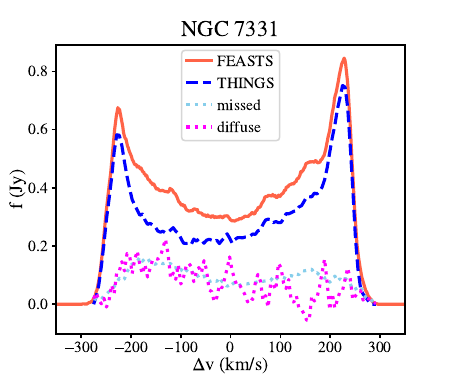}

\caption{The integrated $\hi$ spectra from FEASTS and THINGS data. The red solid lines and blue dashed curves are for the FEASTS and THINGS spectra, respectively. The cyan dotted curves plot the spectra of the missed $\hi$, which is the FEASTS spectrum minus the THINGS spectrum for each galaxy. 
The magenta dotted curves plot the spectra of the diffuse $\hi$, which is the FEASTS spectrum minus the dense $\hi$ spectrum for each galaxy. 
}
\label{fig:spec}
\end{figure*}

Among the whole sample, $f_{\rm missed}$ has a median value of 23\%. 
A significant fraction of missed $\hi$ is detected in almost all galaxies except for NGC 2841 and NGC 3198 which have high inclinations. 
The negative $f_{\rm missed}$ of NGC 3198 should reflect noise, and the low $f_{\rm missed}$ of NGC 2841 should also indicate non-detection. 
The $f_{\rm missed}$ ranges from 17 to 44\% in the eight galaxies with detectable missed $\hi$. 
These detections are robust against the systematic uncertainty of 5\% related to the flux cross-calibration (Appendix~\ref{sec:appendix_mock}).
Comparisons of $\hi$ spectra from the FEASTS and THINGS data, and for the missed $\hi$ are displayed in Figure~\ref{fig:spec}.
They confirm the $f_{\rm missed}$ values discussed above.

\begin{figure} 
\centering
\includegraphics[width=9cm]{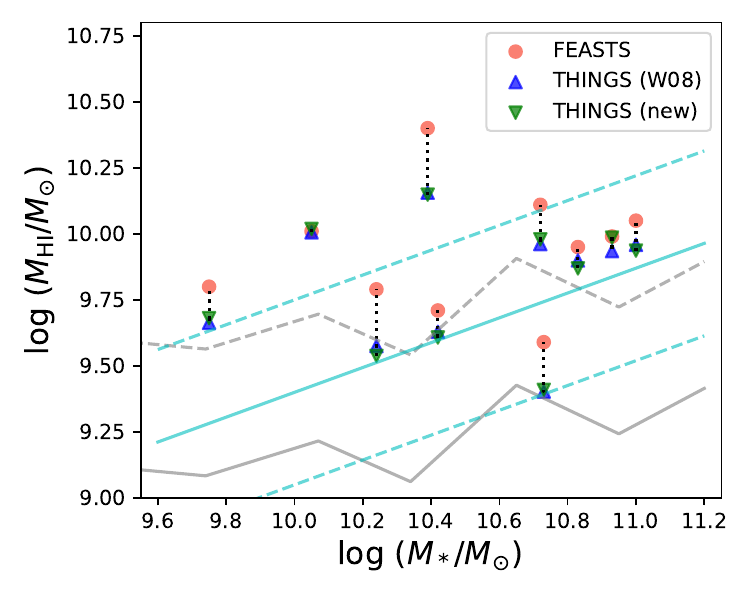}
\caption{The change of M$_{\rm HI}$ due to the inclusion of missed $\hi$ in the $\hi$ mass versus stellar mass space. The red, green and blue symbols mark the M$_{\rm HI}$ derived from the FEASTS data, the new THINGS cubes, and \citetalias{Walter08} THINGS cubes. We link the measurements for the same galaxy with dotted lines. The solid and dashed grey lines show the median and 75 percentile values of $M_{\rm HI}$ respectively as a function of $M_*$ for general galaxies \citep{Catinella18}; the solid and dashed cyan lines show the mean relation of $M_{\rm HI}$ versus $M_*$ and its scatter for star-forming galaxies \citep{Janowiecki20}.  }
\label{fig:mhi}
\end{figure}

In Table~\ref{tab:tab_flux} (and in Figure~\ref{fig:mhi}), we also see that the $f_{\rm missed}$ values based on newly reduced VLA cubes are comparable with those based on the \citetalias{Walter08} cubes.
The two types of $f_{\rm missed}$ on the median differ by 0.0$\pm$4.3\%. 
It reflects that the multi-scale CLEAN used for most galaxies in this paper do not recover significantly more $\hi$ flux than those of \citetalias{Walter08}, possibly because the data are more limited by high noise level than by limited baseline short-spacing.
This result is different from a few studies focusing on other subsets of the THINGS sample \citep{Rich08}, which almost always obtain more fluxes with multiscale CLEAN than with classical CLEAN. 
In addition to possibly different structures of galaxies and rms levels of data, we rescaled the residual cubes before adding them to the convolved model cubes while many of the other studies do not, which may also be relevant in explaining the apparent discrepancy.  

We show in Figure~\ref{fig:mhi}, the relation between $\hi$ masses and stellar masses, with the $\hi$ masses derived from both THINGS and FEASTS data. 
For half of the sample, the difference in $\hi$ masses from the two types of data are comparable to the scatter (1-$\sigma$) of the $M_{\rm HI}$ versus $M_*$ mean relation of star-forming galaxies \citep{Janowiecki20}. 
They are also comparable to half the scatter (50 to 75 percentiles) of the $M_{\rm HI}$ versus $M_*$ relation of general galaxies (xGASS, \citealt{Catinella18}). 
The deviations of $M_{\rm HI}$ from these scaling relations are related to the physical processes that replenish, consume, or remove the $\hi$ reservoir in galaxies. 
We can expect a similar shift of data points around the scaling relation of $M_{\rm HI}$ versus SFR, the deviation of which is commonly interpreted as $\hi$ amount fluctuations due to star formation fueling and feedbacks \citep{Ostriker22} or the lagged response of star formation to abrupt gas removal or accretion \citep{Cortese21}.
Thus, recovering the $\hi$ missed in the interferometric observation is essential to assess the evolutionary stage, and link global and local properties of these nearby galaxies.

We expect that the face-on, large galaxies, which have large angular scales along the minor axes in $\hi$, tend to have higher f$_{\rm missed}$, as confirmed by the trend of red triangles in Figure~\ref{fig:fdiffuse}. 
Several observational effects play a role. 
When the $\hi$ distribution minor axis is small, the angular scales of $\hi$ column density variations are shifted to low values even for a large disk, so the interferometric short-spacing problem becomes less severe. 
Meanwhile, the column densities are higher in more inclined systems, so the sensitivity problem becomes less severe.
These two effects are confirmed by the positive trend of f$_{\rm missed,PBa}$ with the galaxy minor axis (green triangles in Figure~\ref{fig:fdiffuse}), where f$_{\rm missed,PBa}$ is similar to f$_{\rm missed}$ but derived after we apply the VLA primary beam response to both FEASTS and THINGS $\hi$ images to control for the primary beam attenuation effect. 
Furthermore, when both major and minor axes are large, the $\hi$ in the outer region is hard to detect in the interferometry data, due to the strong primary beam attenuation and rms-level-based source finding. 
This effect is reflected by the offsets between the missed $\hi$ fraction with and without primary beam attenuation applied (f$_{\rm missed}$-f$_{\rm missed, PBa}$, vertical distance between green and red triangles for each galaxy in Figure~\ref{fig:fdiffuse}). 
The offsets tend to be larger for the four largest galaxies.
Primary beam attenuation only accounts for a moderate fraction ($<1/6$) of f$_{\rm missed}$ for the two largest galaxies, and never exceeds half f$_{\rm missed}$ for the whole sample.
Due to these effects, it is more likely to have high values of $f_{\rm missed}$ in face-on large galaxies.

Finally, we note that five galaxies from the sample are also included in the ALFALFA (Arecibo Legacy Fast ALFA) Catalog for Extended Sources \citep{Hoffman19}. 
The ALFALFA flux over FEASTS flux ratios range from 0.90 to 1.02, with a median value of 0.92. 
The on average slightly lower ALFALFA fluxes are consistent with the conservative source masks put by hand by the ALFALFA team.

 \begin{table*}
    \centering
            \caption{$\hi$ fluxes}
            \begin{tabular}{c c c c c c}
              \hline
              Galaxy   & $F_{\rm HI}$  & log ($M_{\rm HI}/\Msun$)  & $f_{\rm missed, W08}$ &$f_{\rm missed}$  & $f_{\rm diffuse}$ \\
                     & ($\jykms$)  &   &  & & \\
              (1)     & (2)     & (3)   & (4)   & (5) & (6) \\
              \hline
NGC 628 & 493.97 & 9.79 & 0.39 & 0.44 & 0.54 \\
NGC 925 & 319.29 & 9.80 & 0.27 & 0.24 & 0.22 \\
NGC 2841 & 208.48 & 9.99 & 0.12 & 0.01 & 0.38 \\
NGC 2903 & 278.94 & 9.71 & 0.17 & 0.21 & 0.34 \\
NGC 3198 & 229.93 & 10.01 & 0.01 & -0.02$^a$ & 0.05 \\
NGC 3521 & 332.89 & 9.95 & 0.11 & 0.17 & 0.27 \\
NGC 5055 & 533.05 & 10.11 & 0.29 & 0.26 & 0.48 \\
NGC 5194 & 261.47 & 9.59 & 0.35 & 0.34 & 0.55 \\
NGC 5457 & 1948.54 & 10.40 & 0.43 & 0.44 & 0.45 \\
NGC 7331 & 220.68 & 10.05 & 0.19 & 0.23 & 0.19 \\ 
            \hline
            \end{tabular} 
      
              \raggedright
                Column~(1): Galaxy name.
                Column~(2): $\hi$ flux from the FEASTS data. We do not add error bars because the flux uncertainties are less than 0.01\%. 
                Column~(3): $\hi$ mass from the FEASTS data.
                Column~(4): The fraction of interferometry missed $\hi$ flux over the total $\hi$ flux ($f_{\rm missed}$) based on \citetalias{Walter08} cubes.
                Column~(5): The fraction of interferometry missed $\hi$ flux over the total $\hi$ flux ($f_{\rm missed}$) based on newly reduced THINGS cubes. 
                Column~(6): The fraction of diffuse $\hi$ over the total $\hi$ flux ($f_{\rm diffuse}$) based on newly reduced THINGS cubes.

      Note {\rm a:} the negative value should reflect noise.          
        \label{tab:tab_flux}
    \end{table*}

\subsection{The Amount of Diffuse $\hi$}
\label{sec:amount_diffuseHI}
The ratios of diffuse $\hi$ over total $\hi$ fluxes, $f_{\rm diffuse}$, are also listed in Table~\ref{tab:tab_flux}.
The values of $f_{\rm diffuse}$ range from 5 to 55\%, with a median value of 34\%. 
The galaxies with a significant $f_{\rm diffuse}$ ($\gtrsim40$\%) include NGC628, NGC 5055, NGC 5194, and NGC 5457. 
These four are the most face-on galaxies in the sample, with inclinations below 60$\degree$. 
The $\hi$ disk of NGC 5457 further has the largest $R_{\rm HI}$ in our sample, while NGC 5194 also has a wide-spreading $\hi$ distribution due to its on-going close interaction with the companion galaxy NGC 5195.

\begin{figure} 
\centering
\includegraphics[width=9cm]{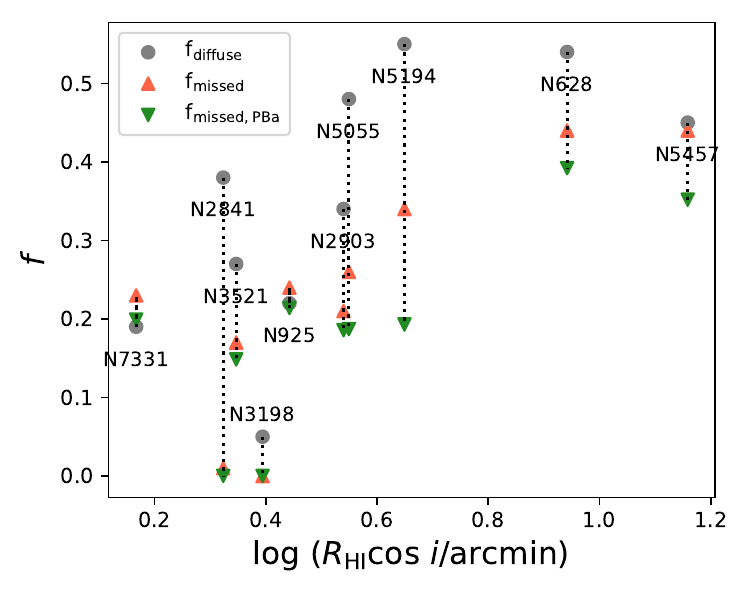}
\caption{The change of missed or diffuse $\hi$ fraction as a function of angular sizes of galactic $\hi$ minor axes. The diffuse $\hi$ fraction f$_{\rm diffuse}$ are plotted in grey dots. The missed $\hi$ fraction without (f$_{\rm missed}$) and with (f$_{\rm missed,PBa}$) VLA primary beam attenuation applied to both single-dish and interferometric data are plotted in red and green triangles respectively. Galaxy names are given. }
\label{fig:fdiffuse}
\end{figure}

The correlations with inclination and disk size reflect a more intrinsic dependence of $f_{\rm diffuse}$ on the minor-axis disk angular size in the $\hi$, as shown in Figure~\ref{fig:fdiffuse} (grey dots) \footnote{There is also a significant correlation of $f_{\rm diffuse}$ with the kpc-unit minor-axis $\hi$ disk size. 
Because the sample dynamic range of luminosity distances is small, we cannot rule out the possibility that the $f_{\rm diffuse}$ depends more on physical scales than angular scales.}.
By definition the diffuse $\hi$ includes and is slightly more than the missed $\hi$, so any instrumental dependence of $f_{\rm missed}$ as discussed in Section~\ref{sec:missed_HI} may propagate into the diffuse $\hi$. 
This dependence indicates that the current definition of diffuse $\hi$ may be rather dependent on observations (e.g. the physical size, the inclination, the distance of galaxies). 
In the following, when studying the relation of $f_{\rm diffuse}$ with other galactic properties, we need consider possible mutual dependence on the inclination or minor axial size.

Despite the possible bias, the diffuse $\hi$ in its current definition provides a first-order indicator of the actual diffuse $\hi$ that may have a physically clean definition. 
We can use the current measurements to begin exploring the nature of diffuse $\hi$.
In each individual galaxy, it is a relatively consistent measure of the diffuse part of $\hi$ in regions throughout the disk and extending to the CGM.
We show the spectra of the diffuse $\hi$ in Figure~\ref{fig:spec}.
The diffuse $\hi$ is found throughout the velocity range of each galaxy.  
Meanwhile, the diffuse $\hi$ does not extend beyond the spectral line edges of the dense $\hi$, except for NGC 5194 which is a strongly interacting galaxy.
It indicates that most of the diffuse $\hi$ is well confined in the potential well of each galaxy, and possibly the bulk motion of the diffuse $\hi$ tends to follow that of the dense $\hi$.

\subsection{Column Density Distributions of the Different Types of $\hi$}
\label{sec:NHIdistr}
\begin{figure} 
\centering
\includegraphics[width=9cm]{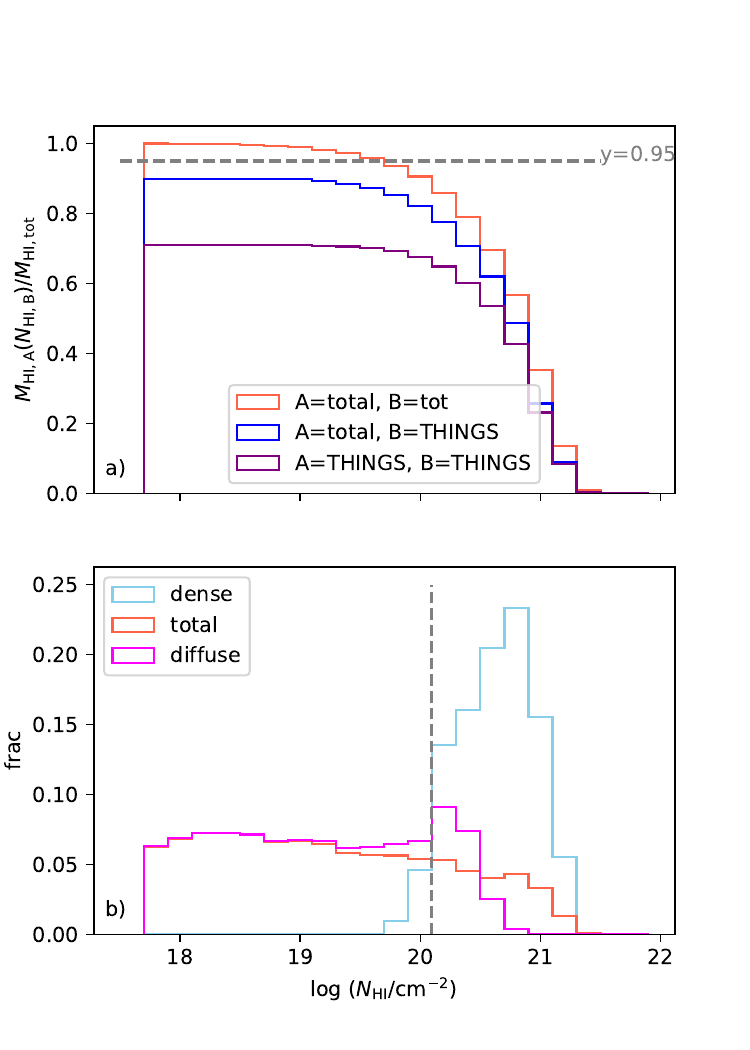}
\caption{The mass and number distributions of column densities for the different types of $\hi$. 
All measurements are made at the same resolution of the FEASTS (9.1$\pm$2.5 kpc).
Each galaxy has an equal weight in both panels. 
{\bf a:} Cumulative $\hi$ mass distribution as a function of $\hi$ column densities. 
Each set of histogram shows a cumulative $\hi$ mass distribution in $\hi$ images of type A related by pixels to column densities of type B greater than given values. 
The cumulation of a distribution starts from the high column density end, and is normalized by the total $\hi$. 
The A and B types are denoted in the plot, with their images being registered to the same WCS system. 
The grey dashed line marks a y axis value of 0.95.
{\bf b:} Differential $\hi$ column density distributions. The blue, red and magenta histograms are for the dense, total, and diffuse $\hi$ respectively. The areas of histograms are normalized to unity. The dashed line marks a rough division between the two types of $\hi$ at a column density of $10^{20.1}~\cmsq$. 
}
\label{fig:hist_NHI}
\end{figure}

We show in panel a of Figure~\ref{fig:hist_NHI} the cumulative $\hi$ mass distribution in total $\hi$ images with column densities greater than given values. 
The resolution area over which the column densities are averaged is the FAST FWHM of 3.24$'$, or 9.1$\pm$2.5 kpc. 
It shows that, only 5\% of the total $\hi$ mass is found below a total $\hi$ column density of $10^{19.8}~\cmsq$.
The fraction is largely consistent with the column density function of \citet{Zwaan05}, indicating that most of the $\hi$ masses are found at high column densities even with image depth reaching $10^{17.7}~\cmsq$.
Because this fraction is lower than the median $f_{\rm missed}$ of the sample, while $10^{19.8}~\cmsq$ is close to the detection limits of THINGS data, the missed $\hi$ and the diffuse $\hi$ are not equivalent to $\hi$ found at the low end of the column density distribution. 
Instead, the missed $\hi$ is possibly a large-scale, low-density, and more diffuse part of $\hi$, which is co-spatial with the THINGS detected $\hi$.
This implication is confirmed by comparing the conditional cumulative distribution of FEASTS (denoted by tot) and THINGS detected $\hi$ masses in relation to THINGS detected $\hi$ column densities (blue and purple histograms). 
All measurements are made at the same FEASTS resolution.
These two cumulative distributions flatten below unity when the THINGS detected column densities are below the detection limits. 
These flattened parts show the FEASTS  and THINGS detected $\hi$ masses distributed in THINGS detected regions, which are normalized by the total $\hi$ masses throughout the galaxies.
The THINGS detected $\hi$ missed the total $\hi$ by an average of $\sim$30\%. 
On average only $\sim1/3$ of this missed $\hi$ is found in regions extending beyond the THINGS detected regions. 
Because the missed $\hi$ is a subset of the diffuse $\hi$ in the THINGS detected regions, we expect that most diffuse $\hi$ is found co-spatial with the dense $\hi$.

We also show in panel b of Figure~\ref{fig:hist_NHI}, the distribution of column densities for the dense and diffuse $\hi$ at the same resolution of 3.24$'$.
Because the dense $\hi$ is not well resolved at this resolution, its column densities are shifted to lower values than their original ones.
 This plot shows the division in column densities between the dense and diffuse $\hi$ when measured at the same resolution. 
The two types of $\hi$ have a rough division at $10^{20.1}~\cmsq$, close to the 2-$\sigma$ threshold estimated with the flux density criteria for selecting the dense $\hi$ (section~\ref{sec:creat_dmom}). 
Moreover, the total and diffuse $\hi$ has a log N$_{\rm HI}$ dynamic range around four times as wide as the dense $\hi$, hinting significantly more diverse physical conditions traced by them.

\subsection{ Localized Distribution and Kinematics of the Diffuse HI}
\label{sec:result_mom}

\subsubsection{Inspecting Moment Images of the Diffuse HI}
\label{sec:inspect_mom}
We firstly inspect the column density images of the diffuse $\hi$ (the left panel in the fourth row of Figure~\ref{fig:mom_n628} to \ref{fig:mom_n7331}).
In most of the relatively face-on galaxies, there is a smooth distribution of diffuse $\hi$ throughout the disks. 
Around the strong interacting galaxy NGC 5194, there is a significantly enhanced amount of irregularly distributed diffuse $\hi$ beyond the disk and tidal tails detected in dense $\hi$.
We study this system in detail in another project (Lin et al. in prep).
For moderately interacting galaxies (NGC 5055 and NGC 5457), there is clear evidence of a large amount of diffuse $\hi$ linking to the surrounding satellites.
The $\hi$ bridges have no counterparts in stellar light (Figure~\ref{fig:atlas}, also see mid-infrared and violet images in \citealt{Leroy09}). 
The weakly interacting, relatively face-on galaxies (NGC 628 and NGC 925) have neighbors, and show lopsided spiral arms, but neither $\hi$ nor optical arms link to their neighbors. 
In these two systems, the diffuse $\hi$ tends to follow the trailing side of the more open arm.
The weakly interacting, relatively inclined galaxies (NGC 2841 and NGC 3521) show a tail or warp in diffuse $\hi$ on one side, but neither $\hi$ nor optical tails$\slash$warps point to the neighbors.
These two galaxies may be relatively inclined analogs of NGC 628 or NGC 925.
The small neighbors may have closely passed by these four galaxies in the past, inducing the arms, tails, or warps.
The remaining highly inclined galaxies seem to have a close neighbor detected in $\hi$, but do not clearly show perturbed features in $\hi$  (NGC 2903, NGC 3198, NGC 7331) .
They also have relatively small amounts of diffuse $\hi$, mostly in the outer disks.

The velocity images of diffuse $\hi$ (the middle panel in the fourth row of Figure~\ref{fig:mom_n628} to \ref{fig:mom_n7331}) look similar to the velocity images of the total (and dense) $\hi$, indicating the diffuse $\hi$ to follow similarly rotating disks as the total (and dense) $\hi$.
In some galaxies, there seem to be abnormality in the central regions of velocity fields of the diffuse $\hi$, but we refrain from over-interpreting before a more careful modeling to account for projection and beam smearing effects is conducted in the future. 
We focus on the difference velocity images ($v_{\rm dense}-v_{\rm diffuse}$, the fifth row in Figure~\ref{fig:mom_n628} to \ref{fig:mom_n7331}) which show the difference in velocity between dense $\hi$ and diffuse $\hi$.
For most galaxies, the difference velocity fields show a similar pattern of rotation as the velocity fields of total and dense $\hi$ in at least the relatively inner disks.
But the amplitudes of such patterns are systematically lower than the latter.
It indicates that the diffuse $\hi$ in at least the inner disk tends to follow the rotation of the main disk or dense $\hi$, but in a relatively lagged way.
The lag of rotation is seen almost throughout the disks of the relatively inclined galaxies NGC 2841, NGC 3198, NGC 3521, and NGC 7331. 
But it is limited to a relatively small radial range from the center in the relatively face-on and moderately$\slash$highly interacting galaxies NGC 5055 and NGC 5457. 
The velocity difference is less uniform in outer disks possibly due to the on-going or past tidal interactions and existence of warps.
It is particularly interesting to point out the case of NGC 5457, which seems to have a perfectly lagged rotation pattern in the inner disk, but actually the pattern center is off-centered with respect to the main disk and dense $\hi$ rotation. 
This feature of shifted center can be seen more clearly in the radial profile of rotational lagging in the next section.

We finally inspect the velocity dispersion images of the diffuse $\hi$ (the right panel in the fourth row in Figure~\ref{fig:mom_n628} to \ref{fig:mom_n7331}).
High velocity dispersions of the diffuse $\hi$ seem a ubiquitous feature observed throughout the sample. 
In relatively face-on galaxies, velocity dispersions of the diffuse $\hi$ are typically higher than the dense $\hi$ throughout the disks.
Their inner disks have higher velocity dispersions in the diffuse $\hi$ than the outer disks.
In the inner disks of highly inclined galaxies, we often see X shaped structures, which are possibly caused by projection and beam smearing.

\subsubsection{Radial Distribution of Diffuse $\hi$ Properties}
\label{sec:prof}

\begin{figure*} 
\centering
\includegraphics[width=17cm]{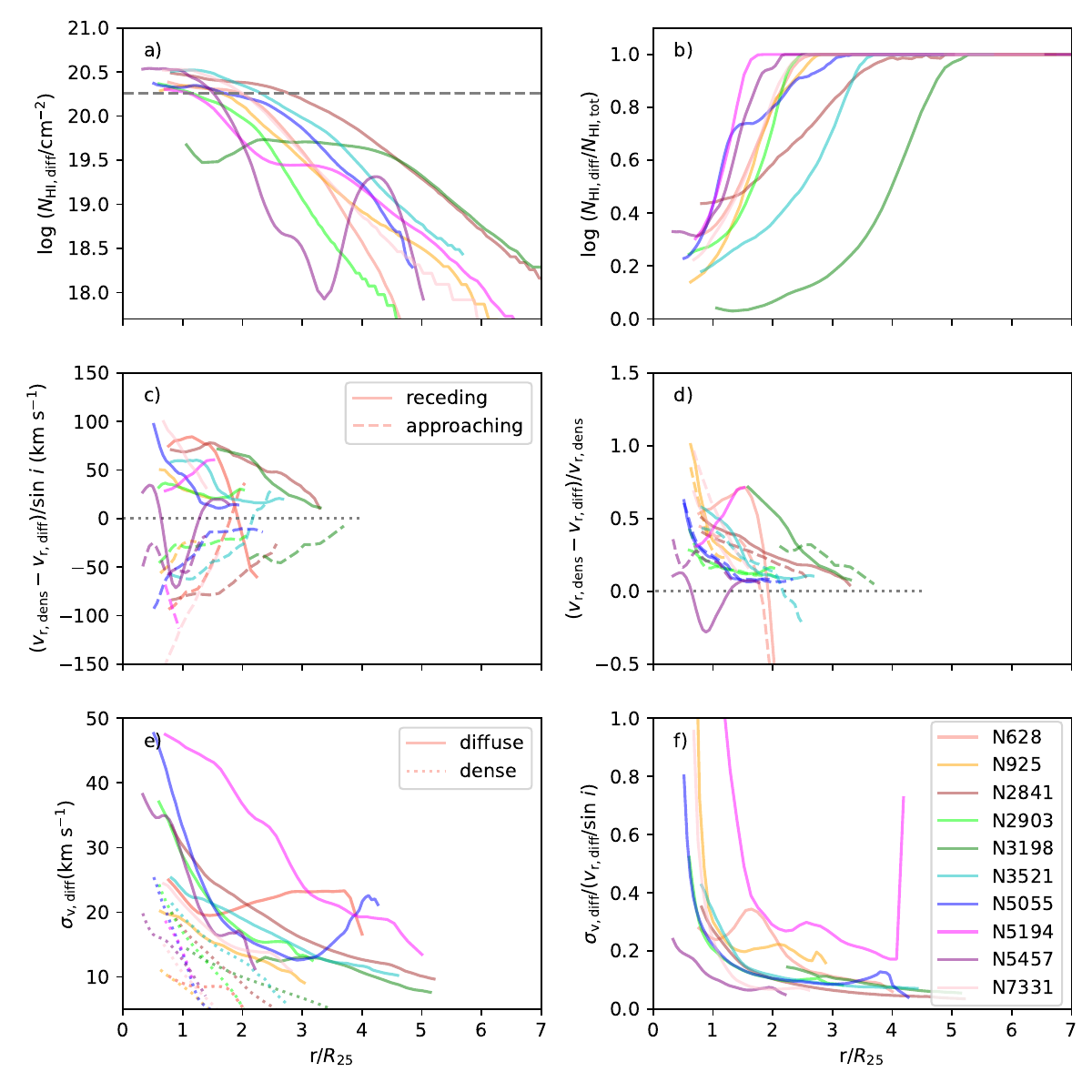}
\caption{Radial profiles of diffuse $\hi$ properties. 
The profiles are color coded by galaxy names as denoted in each panel, and the radius are normalized by the optical radius $R_{25}$. All profiles start from radius equal to beam FWHM of FAST. 
All measurements are made at the same resolution of FEASTS  (9.1$\pm$2.5 kpc). 
{\bf a:} the column density profile. 
The column densities are not corrected for projection because the inclinations in outer disks where diffuse $\hi$ dominates the total $\hi$ are highly uncertain. 
The dashed horizontal line shows the rough column density threshold of selecting the dense $\hi$.
{\bf b:} the profile of diffuse $\hi$ column densities over the total $\hi$ column densities.
{\bf c:} the profile of projection corrected radial velocity difference between the dense and diffuse $\hi$ (former minus latter) measured along major axis of galaxies (see Section~\ref{sec:derive_prof}). 
The projection-corrected radial velocities along the major axis approximate rotational velocities, but we warn of their uncertainties.
The solid and dashed curves are for the receding and approaching side of the disks respectively. 
The dotted line marks the position of y$=$0.
{\bf d:} similar as in Panel c, but the radial velocity difference is normalized by the radial velocity of the dense $\hi$ along the major axis. 
Values between zero and unity indicate lagged rotational$\slash$orbital velocity (see Section~\ref{sec:derive_prof}).
{\bf e:} the profile of velocity dispersion. The solid and dashed lines are for the diffuse and dense $\hi$ respectively. 
{\bf f:} the profile of velocity dispersion over the projection corrected radial velocity along major axis for the diffuse $\hi$.
}
\label{fig:prof}
\end{figure*}

We present the radial distribution of diffuse $\hi$ properties, including the column density, extent of lagged rotation, velocity dispersion, and kinematical hotness (i.e. the extents of motion being dominated by random instead of rotational motions), in Figure~\ref{fig:prof}.  
We compare the diffuse $\hi$ to dense $\hi$ in most properties, which are measured at the same FAST resolution of 3.24$'$. 
The plots summarize our visual impression on individual moment images in a more quantitive and statistical way.
We caution that all the results below are based on relatively coarse and first-order measure, and we will need careful kinematical modeling to better account for possible beam smearing and projection in the future (Randriamampandry et al. in prep.). 
Meanwhile, all profiles start from radius equal to the FWHM of the FAST beam, while the extended nature of the diffuse $\hi$ makes it less susceptible to beam smearing or smoothing than the dense $\hi$.
The beam effects are further mitigated when we focus on the relative ratio instead of absolute differences between the diffuse and dense $\hi$.

In panel a of Figure~\ref{fig:prof}, we show the radial profiles of diffuse $\hi$ column densities.
Most of the diffuse $\hi$ has column densities below the threshold of the dense $\hi$.
They generally tend to flatten near the center and decrease with radius, similar to radial profile shapes of normal $\hi$ column densities \citep{Wang16}.
They do not exhibit obviously abrupt truncation at large radius, implying a smooth transition to the CGM at least down to a column density limit of $10^{18}~\cmsq$, though future modeling will be needed to fully rule out spatial smoothing effects of the beam. 

\begin{figure} 
\centering
\includegraphics[width=9cm]{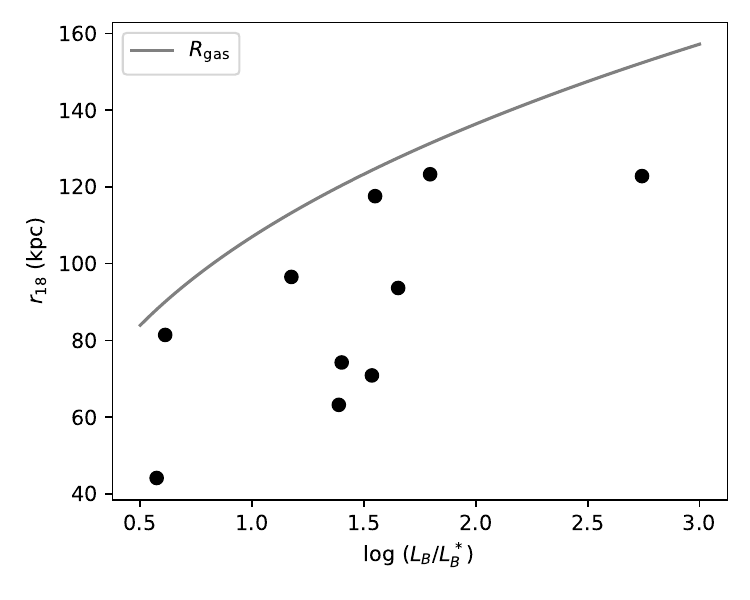}
\caption{The relation between $r_{18}$ and normalized $B$ band luminosity. $r_{18}$ is the characteristic radii where the azimuthally averaged column density of diffuse $\hi$ reaches $10^{18}~\cmsq$ in a galaxy. The curve shows the relation between $R_{\rm gas}$ and the normalized $B$ band luminosity from \citet{Chen10}, where $R_{\rm gas}$ is the typical border radius for cool gas to be found in the CGM.  }
\label{fig:r18_Rgas}
\end{figure}

The diffuse $\hi$ extends out to 4 to 7 times the optical radius.  
We obtain from the radial profiles the characteristic radius $r_{18}$, where the column density of diffuse $\hi$ reaches $10^{18}~\cmsq$. 
We plot in Figure~\ref{fig:r18_Rgas} the $r_{18}$ as a function of the $B$ band luminosity. 
We compare such a relation to that of fiducial boundary radius ($R_{\rm gas}$) for Mg {\textsc i}{\textsc i}-traced cool gas to be detected the CGM, which is $R_{\rm gas}=(107~ {\rm kpc})~(L_B/L_B^*)^{0.35}$ \citep{Chen10}, where $L_B$ is the $B$ band luminosity (from \citetalias{Walter08}), and $L_B^*$ is the characteristic luminosity of luminosity function in the $B$ band from \citet{Faber07} . 
The $r_{18}$ values are systematically lower than, and thus consistent with the cool gas boundary radius $R_{\rm gas}$ at a given $B$ band luminosity.
The ratios $r_{18}/R_{\rm gas}$ have an average value of $0.74\pm0.15$. 
It indicates that around $\hi$-rich galaxies like those in the THINGS sample, the $\hi$ or cool gas detected in the CGM is possibly linked to the disk instead of being floating clouds.

In panel b of Figure~\ref{fig:prof}, we show the radial distribution of ratio of diffuse $\hi$ over the total $\hi$. 
The diffuse $\hi$ contributes at a low level to the total $\hi$ within $R_{25}$ in most galaxies, except for the most strongly interacting galaxy NGC 5194.
For the remaining less interacting galaxies, the contribution of diffuse $\hi$ to the total $\hi$ increases toward large radii, roughly with an exponential profile.
For the three galaxies with the lowest $f_{\rm diffuse}$ (NGC 2841, NGC 3198, NGC 3521), the ratio of diffuse $\hi$ rises above half only when $r>4~R_{25}$. 
It confirms the more extended nature of the diffuse $\hi$. 
A detailed comparison in radial profile shapes between the total, dense, and diffuse $\hi$ will be presented in a future paper.

In panel c of Figure~\ref{fig:prof}, we show the velocity difference between the dense $\hi$ and diffuse $\hi$ ( $v_{{\rm r,dens}}-v_{{\rm r,diff}}$) divided by sin $i$ for both the receding (solid curves) and approaching (dashed curves) sides of galaxies.
The sin $i$ term uses the optical inclination angle, and corrects the radial velocities for projection assuming rotational motions. 
We can see that for most galaxies, on the receding (approaching) sides of disks, the velocity differences tend to be positive (negative). 
Because both lagged rotation and counter-rotation lead to positive (negative) velocity differences on the receding (approaching) sides, it can be more informative to check the normalized velocity difference between the dense $\hi$ and diffuse $\hi$ ($(v_{{\rm r,dens}}-v_{{\rm r,diff}})/v_{{\rm r,dens}}$, see Section~\ref{sec:derive_prof}). 

In panel d, we can see that the normalized velocity difference profiles are largely positive and within unity, for both the receding and approaching disk sides of most galaxies.
The behavior of most galaxies indicates that the majority of diffuse $\hi$ in the sample is rotating in the same direction but in a lagged way in comparison to the dense $\hi$.
The rotations of the diffuse $\hi$ lag behind the dense $\hi$ by a median of 25\% in velocity in the radial range of 1 to 2 $R_{25}$, but by as high as $\sim$50\% within $R_{25}$. 
On the whole the extents of lagging drop from small to large radius. 
This lagging is unlikely due to beam smearing: because the diffuse $\hi$ is less (centrally and locally) concentrated than the dense $\hi$, its observed velocity is also less artificially ``lagged'' by beam smearing. 
Because the dense $\hi$ is possibly more strongly smeared and artificially ``lagged'' when smoothed to the FEASTS resolution, the actually extents of lagging of the diffuse $\hi$ may be even higher than shown here.
On the outskirts ($r>1.5~R_{25}$), some profiles become negative possibly due to tidal perturbations (i.e. the gas rotates faster or deviates from the circular orbits). 
NGC 5194 is an outlier in this plot, with normalized velocity difference on the receding side rising as a function of radius from the center, and reaching values above unity; its profile for the approaching side is not shown due to values larger than unity, implying orbiting in the opposite direction with respect to the dense $\hi$.

Panel e of Figure~\ref{fig:prof} shows the velocity dispersion radial profiles for diffuse $\hi$.
These profiles drop from the galaxy center toward outer disks, a trend was seen for the dense $\hi$ \citep{Tamburro09}, and was interpreted as a result of energy input from stellar feedback and gas inflow activities. 
The velocity dispersions of diffuse $\hi$ have values roughly twice those of dense $\hi$ (dotted curves), the latter showing values close to 10 $\kms$ typically observed for high-column density $\hi$ \citep{Tamburro09}.

Panel f shows the kinematical hotness of the diffuse $\hi$ disk, measured as the ratio of velocity dispersion over the rotational velocity approximated as the projection-corrected radial velocity, $\sigma_{\rm v,diff}/(v_{\rm r,diff}/{\rm sin}~i)$. 
In the outer disks ($r>2R_{25}$), the galaxies largely have rotationally dominated, kinematically cool disks, with $\sigma_{\rm v,diff}/(v_{\rm r,diff}/{\rm sin}~i)<0.25$.
In the inner disks ($r<2R_{25}$), many galaxies have $\sigma_{\rm v,diff}/(v_{\rm r,diff}/{\rm sin}~i)$ reaching values above 0.2. 
It is a bit surprising to see that the strongly interacting galaxy NGC 5457 has the kinematically coldest diffuse $\hi$.

Similar results based on analysis of the missed $\hi$ is presented in Figure~\ref{fig:miss_prof} in Appendix~\ref{appendix:prof}.
The plots show that distribution and kinematical properties of the missed $\hi$ are largely consistent with those of the diffuse $\hi$. 
But we highlight that, from panel a of Figure~\ref{fig:miss_prof}, most of the missed $\hi$ has column densities below the detection limit of THINGS.
It implies that most of the THINGS missed $\hi$ is due to its low-density and not necessarily large-angular scale.
It is different from the missed $\hi$ in \citetalias{Wang23} comparing the FEASTS and HALOGAS data, where 90\% of the diffuse $\hi$ is missed by HALOGAS due to its large-angular scale instead of low-density.

\section{Discussion }
\label{sec:discussion} 
 In this section, we discuss possible physical scenarios related to the properties observed for the diffuse $\hi$.

\subsection{ Kinematically Warm Disks of Diffuse $\hi$} 
Keeping in mind that the projection and beam smearing effects are not perfectly accounted for, we find that for most galaxies, the diffuse $\hi$ is organized in rotating disks that resemble but are kinematically warmer than the dense $\hi$ disks. 
Its kinematics is largely rotation-dominated, and tends to follow the rotating direction of the dense $\hi$ throughout the disks.
But its rotation is usually slower than the dense $\hi$ by around 25\%, and its velocity dispersions (typically 10 to 25 $\kms$) are twice that of the dense $\hi$. 
These kinematical features are unlikely dominantly caused by beam smearing effects, as the diffuse $\hi$ distributes in less concentrated way than the dense $\hi$. 
The ``lag in rotation'' is closely related to the disk asymmetric drift, suggesting kinematics contributed partly by random instead of purely rotational motions. 
But it can further involve dynamic mixing with the CGM, and possible inflow motions due to cooling. 
The diffuse $\hi$ is thus likely a kinematically hotter extension of the dense $\hi$ into the hotter and more ionized regime of CGM. 
Because the diffuse $\hi$ dominates the total $\hi$ at large radius, while the dense $\hi$ is usually related to star formation at small radius, we separately discuss the possible scenarios in relatively inner and outer disks, roughly divided by 1.5 $R_{25}$. 

\subsubsection{The Similarity of Diffuse and Thick $\hi$ Disks}
The kinematics of diffuse $\hi$ disks except for the very outskirts in a few galaxies resembles that of the thick $\hi$ disk identified in deep interferometric $\hi$ observations (e.g., \citealt{Oosterloo07, Heald11}), as they both show lagged rotation and high velocity dispersions within roughly the optical radius \citep{Marasco19}. 
Despite different technical details, there is no obvious reason to consider the two types of disks being intrinsically different.

Galactic fountains which have been suggested to be an important channel to produce the thick $\hi$ disks \citep{Fraternali06} should also contribute to producing the diffuse $\hi$ detected in this work. 
Future efforts on kinematical modeling of the diffuse $\hi$, revealing multiphase gaseous structures using multiwavelength data, and linking them to the localized star formation rates will be needed to quantitatively confirm the role of fountains.
For now, if we take observations of the HVC consisted thick $\hi$ disk of the MW (the properties of which are highly consistent with being fountain driven) as the reference, the lagged extents of rotation at the level of $\sim$20 and 10\% at a radius around 1.5 and 2 $R_{25}$ in our sample correspond to a height of $\gtrsim$3 and 1 kpc above the disk plane \citep{Marasco12}, respectively. 

\subsubsection{The Connection of Diffuse $\hi$ to CGM Properties in the Outer Regions}
\begin{figure} 
\centering
\includegraphics[width=9cm]{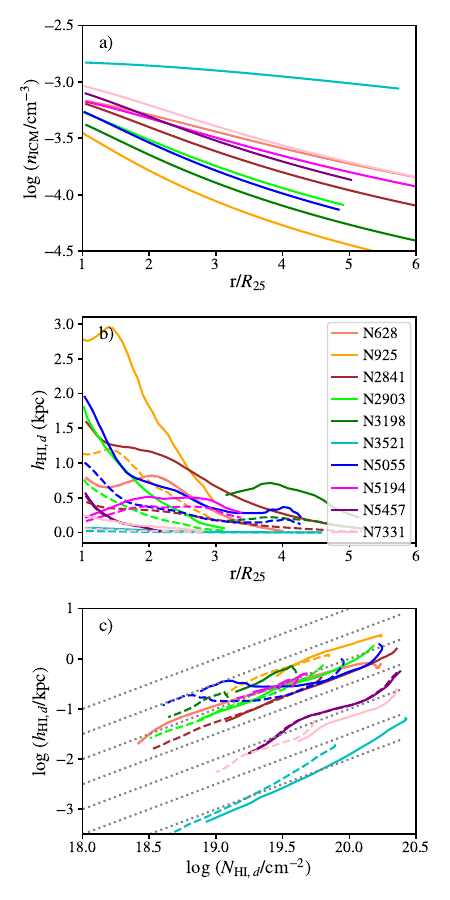}
\caption{ Relation between radially distributed $\hi$ properties and CGM properties. Different colors are for different galaxies. 
{\bf a:} CGM density as a function of radius, estimated based on scale relations of group CGM properties.
{\bf b:} the thickness ($h_{\rm HI}$) profile derived assuming the diffuse $\hi$ is in pressure equilibrium with the CGM.  
{\bf c:} relation between the diffuse $\hi$ thickness and the column density. All dotted lines have a slope of unity.
In panel b and c, the estimation based on column densities which are (not) corrected for projection effects are plotted with (solid) dashed curves.}
\label{fig:thickness}
\end{figure}

We perform an order-of-magnitude estimation on the thickness of the diffuse $\hi$ in outer disks, largely following the method described in \citetalias{Wang23}.
We assume the diffuse $\hi$ to be internally supported by turbulent pressure (mass per particle $\mu=1.273~m_{\rm H}$), and externally in pressure equilibrium with the hot CGM ($\mu=0.609~m_{\rm H}$).
We take the group mass of each galaxy from \citet{Kourkchi17}, and estimate the viral temperature.
We estimate the CGM gas mass from the group mass using the equation of \citet{Ettori15}, and assume the CGM gas distribution following an isothermal $\beta$-model with $\beta$=0.64 and core radius equal to 0.19 r$_{500}$, where r$_{500}$ is the radius of group where the average density is 500 times the critical density of the universe.
The estimated CGM density radial profile is present in panel a of Figure~\ref{fig:thickness}. 
We divide the column density of diffuse $\hi$ by the volume density and obtain the thickness. 
Because the diffuse $\hi$ seems to largely follow the rotation$\slash$orbits of the dense $\hi$, the column densities have disk-like projection effects, and the derived thickness should be viewed as upper limits.
We also multiply the column density of diffuse $\hi$ by cos $i$ to obtain the ``projection corrected'' column density before deriving the thickness, where $i$ is the inclination of the dense $\hi$ disk (Table~\ref{tab:tab_sample}).
This projection correction assumes the $\hi$ to be in a thin disk, and the corrected column density and thickness should be lower limits. 

As we show in panel b of Figure~\ref{fig:thickness}, the thickness profiles of all galaxies decrease with radius. 
Except for in NGC 925, the directly derived thicknesses of diffuse $\hi$ of all galaxies are less than 3 (1.5) kpc when the radius are larger than 1 (1.5) $R_{25}$ \footnote{These thickness values can increase by roughly twice if the localized ionizing rate is above 0}.
The projection corrected diffuse $\hi$ thicknesses of all galaxies are below 3 kpc outside $R_{25}$.
The real thicknesses are possibly between the projection corrected and uncorrected values.
The typical thickness of diffuse $\hi$ in outer disk regions in this study is thus much smaller than that inferred in a similar way for NGC 4631 in \citetalias{Wang23}, which is 4.5 and 10.0 kpc at a galactocentric distance of 30 and 55 kpc. 
The reason is that most column densities and velocity dispersions of diffuse $\hi$ derived in this study are slightly and much lower than the typical values of diffuse $\hi$ detected around NGC 4631 ($10^{19.7}~\cmsq$ and $50~\kms$ respectively). 

In panel c of Figure~\ref{fig:thickness}, we plot the relation of thickness against column density for the diffuse $\hi$ in each galaxy.
The different intercepts of relations probably reflect projection effects, but it is surprising that each relation is nearly linear, regardless of applying projection corrections or not.
Such linear trends imply a constant volume density for the diffuse $\hi$ in each individual galaxy while the CGM volume densities vary for more than 0.5 dex in the same radial range.
Because the gas cooling rate is correlated with the gas volume density, the diffuse $\hi$ possibly marks a shell of roughly constant cooling rate.
It implies that the diffuse $\hi$ detected here may indeed be a cooling interface between the ISM and CGM.
We emphasize that the thickness values derived in this way have large uncertainties, due to the simplified assumptions of the CGM model and pressure equilibrium status, and beam smoothing$\slash$smearing and projection effects. 
These estimates should be revisited in the future when the diffuse $\hi$ distribution and kinematics are better modeled and direct measurements of CGM densities and temperatures are available.

To sum up, while the diffuse $\hi$ detected here has spatial scales ($\gtrsim$18 kpc, section~\ref{sec:fluxcali}) similar to that of NGC 4631, its thickness is much smaller. 
It is more layer like, while the latter is more halo-like.
Nevertheless, they may both have close relations to CGM cooling.
The constant volume density beyond $R_{25}$ tentatively supports its role as a cooling interface between the ISM and the CGM for the galaxies in this study, while the $\gtrsim10^{19.7}~\cmsq$ level high-column density of the diffuse $\hi$ in NGC 4631 indicates an efficient cooling flow driven by thermal conductions.

\subsection{The Possibly Important Role of Tidal Interactions in Producing the Diffuse HI}
\begin{figure*} 
\centering
\includegraphics[width=14cm]{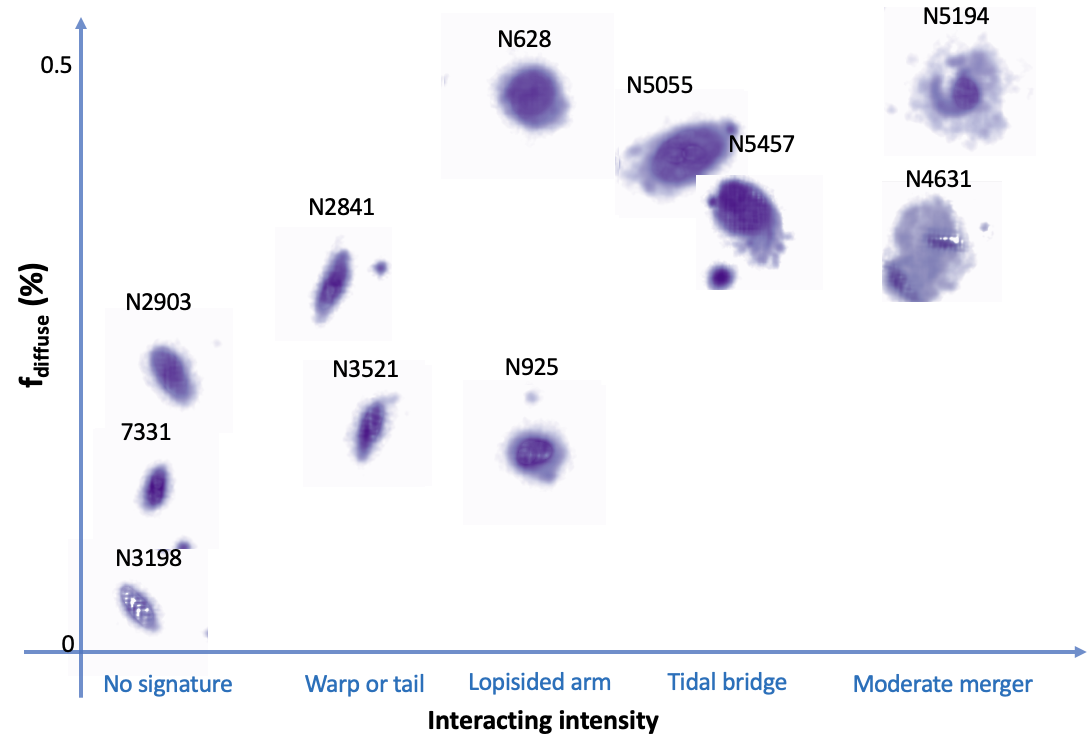}
\vspace{1cm}

\includegraphics[width=14cm]{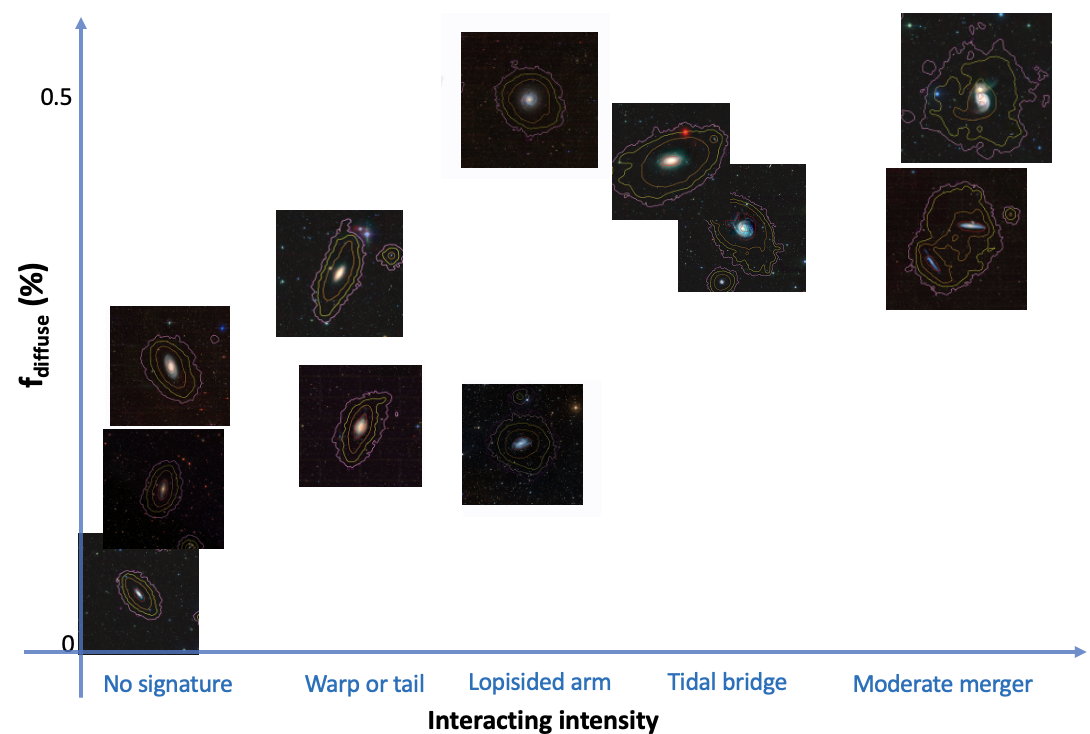}
\caption{The variation of galaxy $\hi$ morphology in the space of tidal interaction strength and diffuse $\hi$ fraction. {\bf Top:} the stamps are moment-0 images of the diffuse $\hi$. 
The galaxy names are denoted. 
The fraction and column density map of the diffuse $\hi$ in galaxy NGC 4631 have been re-derived using the method of this study, but the $f_{\rm diffuse}$ of NGC 4631 should be taken as a lower limit, because its HALOGAS data has better sensitivity on short baselines than the THINGS data. 
{\bf Bottom:} the stamps are contours of total $\hi$ column density levels overlaid on top of the optical images. The outmost contour starts from the detection limit and is always dominated by the diffuse $\hi$. The outer regions of the total $\hi$  are always dominated by the diffuse $\hi$, and because the diffuse $\hi$ distribution is quite flat, its contours will largely be the outmost contours of the total $\hi$.
}
\label{fig:fdiffuse_stid}
\end{figure*}

Keeping in mind that the sample is relatively small, we find that the fraction of diffuse $\hi$ ($f_{\rm diffuse}$) seems to be related to the morphology and kinematical abnormalities in the diffuse $\hi$ which imply intensities of tidal interaction (Section~\ref{sec:result_mom}). 
We summarize this trend in Figure~\ref{fig:fdiffuse_stid}.
Such a relation between diffuse $\hi$ and tidal interactions has been speculated in \citetalias{Wang23}, based on detailed study of the interacting system NGC 4631, but here studied in a statistical way.
We also include NGC 4631 in this figure.
The fraction and column density map of the diffuse $\hi$ in NGC 4631 have been derived using the method of this study, but they both should be taken as lower limits, for the HALOGAS data has better sensitivity on short baselines than the THINGS data.  

As in the top panel of Figure~\ref{fig:fdiffuse_stid}, the interaction intensities can be roughly ordered by the $\hi$ morphologies as having no significant $\hi$ tidal features, having $\hi$ tails or warps possibly induced by a past encounter, having tidal bridge linking to a small companion galaxy, and being in the matured stage of a major merger. 
The highest level of interacting intensity is further supported by kinematic abnormalities in difference velocity fields in the outer disks (NGC 5194 and NGC 5457, Section~\ref{sec:inspect_mom}). 
This relation is likely robust against the systematic dependence of $f_{\rm diffuse}$ on galaxy minor-axis angular sizes, because it exists among both relatively face-on and inclined galaxies.
Moreover, the diffuse $\hi$ becomes more dominating the total $\hi$ along tidal features typically found at large galactocentric radius, which also supports its link to tidal interactions. 
The bottom panel of Figure~\ref{fig:fdiffuse_stid} shows the corresponding morphologies in the optical, but most tidal features in the $\hi$ are not seen in the optical. 
In previous observational studies, it is common for interferometers to detect extraplanar $\hi$ \citep{deBlok14}, and single-dish telescopes to detect diffuse $\hi$ in tidal interactions \citep{Borthakur10}. 
The main progress here is that the diffuse $\hi$ is mapped to a relatively low column density limit at a moderate resolution for a relatively large sample (thanks to the large diameter and 19-beam receiver of FAST), so that the distribution and kinematics of diffuse $\hi$ directly show tidal features.

Besides stellar feedbacks (fountains) strengthened due to tidally induced star formation in gas-rich galaxies \citep{Moreno15}, theoretical studies predict that tidal interactions may produce the diffuse $\hi$ in at least the following three ways.
 Firstly, tidal interactions strip $\hi$ initially in disks into tails and bridges, and enhance turbulent mixing that may produce the diffuse $\hi$. 
Simulations predict that, when neutral gas moves subsonically in the CGM, a cooling-dominating turbulent mixing layer develops with the column density correlated with the initial traveling speed, and the velocity dispersion one or two magnitudes lower than the initial traveling speed \citep{Ji19, Yang23}.
The velocities of tidal tails and bridges in our sample rarely exceed the maximum rotational velocity of the dark matter halo, and thus should be mostly sub-sonic; the velocity dispersions of the diffuse $\hi$ are on the level of 20 $\kms$, consistent with predictions for gas in the turbulent mixing layer.  
The diffuse $\hi$ near the tidal bridges and tails of NGC 5194, NGC 5457, and NGC 5055 may be related to the turbulent mixing.
Secondly, tidal interactions drive spiral arms in galaxies, and produce the diffuse $\hi$ through shocks of gas running into arms. 
Possible candidates in support of this scenario include NGC 628 and NGC 925. 
They have lopsided spiral arms, and the diffuse $\hi$ tends to follow the trailing side of the more open arm which is likely rotating faster toward large radius than the other arm.  
Spiral arms in NGC 628 and NGC 925 have a pattern speed of 41.8 and 7.7 $\kms$ kpc$^{-1}$ respectively derived in the literature \citep{MartinezGarcia14, Elmegreen98}.
It only takes a time lag of 66 and 12 Myr respectively to fill an angle of 30$\degree$ behind the arms, and produce the observed morphology in these two galaxies.
Thirdly, interaction with companion galaxies may directly heat the $\hi$ disk, with gravitational and hydrodynamic effects.
This mechanism may contribute to the diffuse $\hi$ in the inner disks of NGC 628, NGC 925, NGC 2841, and NGC 3521.

\subsection{The Abnormal Diffuse $\hi$ in NGC 4631}
A question after comparing the sample studied here with NGC 4631 (\citetalias{Wang23}) is why NGC 4631 stands out?
The HALOGAS data is an order of magnitude deeper in column density than the THINGS data, and their uv coverage on short baselines by design is much denser than the latter, yet they missed 40\% ($>$70\%) of the $\hi$ fluxes detected by FEASTS in the NGC 4631 system (CGM region), the central galaxy of which is edge-on  \citepalias{Wang23}. 
In order to clearly demonstrate this contrast with the THINGS sample, we conduct the same analysis as done in this study for the THINGS data to the HALOGAS data of three overlapping galaxies, NGC 925, NGC 3198, and NGC 5055.
The derived fraction of missed $\hi$ from the HALOGAS images is 12, 0, and 8\% for these three galaxies, respectively, in comparison to 24, 0, and 26\% from the THINGS images. 
These HALOGAS missed-$\hi$ fractions decrease to 10, 0, and 4\% when the WSRT primary beam is applied to both HALOGAS and FEASTS images (i.e., when the contribution from primary beam attenuation is removed). 
These comparisons suggest that, the HALOGAS images have smaller short-spacing and sensitivity limitations than the THINGS data in detecting the diffuse $\hi$. 
NGC 4631 is indeed a special system for which even the HALOGAS observation misses a large fraction of $\hi$ in comparison to the FEASTS data. 
If it were observed by THINGS, the missed $\hi$ fraction would have been even higher, possibly higher than all the THINGS galaxies studied in this paper.

The diffuse $\hi$ of NGC 4631 has distinct column densities ($\gtrsim$3 times higher), velocity dispersions ($\sim$2.5 times higher), and inferred thicknesses ($\sim$10 times higher) in comparison to all other galaxies studied here.
Most of the diffuse $\hi$ is distributed perpendicular to instead of along the disk of NGC 4631, again in contrast to the sample here.
NGC 4631 is indeed an extreme case in term of its halo properties, not only in diffuse $\hi$, but also in radio continuum, H$\alpha$, X-ray, etc \citep{Martin01, Wang01, Irwin12}. 
All these may be related to the extensive tidal tails.
Then it is particularly interesting to compare between NGC 4631 and NGC 5194, both of which have grand-design tidal structures in the dense $\hi$ and particularly in the diffuse $\hi$.  

Do these two galaxies have different degrees of galaxy-galaxy interactions?
The strength of galaxy-galaxy interactions depend the mass ratio, orbits, number of involved galaxies, times of encounters, etc.
NGC 4631 is 2.5 times less massive than NGC 5194, but its major companion NGC 4656 is 16 times less massive than NGC 5195 which interacts with NGC 5194.
The lower mass ratio in the NGC 4631 system, which implies lower degree of tidal interaction, is unlikely to explain its more extreme diffuse $\hi$. 
But NGC 4631 is simultaneously interacting with four $\hi$-rich neighbors \citep{Rand94}, and strongly with at least 2 of them \citep{Combes78}, while NGC 5194 seems mainly interacting with NGC 5195 \citep{Dobbs10}.
The interactions with NGC 4631 mainly occur in the polar direction \citep{Combes78}, while those with NGC 5194 are rather co-planar \citep{Dobbs10}.
Hydrodynamic simulations demonstrate that highly inclined orbits sometimes lead to stronger perturbations than co-planar orbits \citep{Bustamante18}.  
It is possible the large numbers of involved galaxies, and the polar obits result in higher degrees of tidal interactions, leading to more efficient diffuse $\hi$ production around NGC 4631. 

The stellar feedback may also play a role. 
Though the two galaxies have similar SFRs (4.2 vs 4.4 M$_{\odot}$ yr$^{-1}$), NGC 4631 has a 2.5 times lower stellar mass than NGC 5194, and its SFR distributes in more clustered ways, producing the super shells observed in $\hi$ and molecular gas \citep{Rand94, Rand00}. 
Such concentrated, bursty star formation in a lower gravitational potential produces more effective stellar feedbacks.

The intensive stellar feedbacks and extensive tidal interactions in NGC 4631 may produce its abnormal, halo-like diffuse $\hi$. 
These two effects may together drive large-scale CGM turbulence followed by a top-down cascade in the multi-phase gas \citep{Gaspari18};
warm ionized and neutral gas condense out of the turbulent eddies, producing diffuse $\hi$ in the half-way. 
The difference from the turbulent mixing at the neutral gas-CGM interface is that it is a top-down condensation starting from the large-scale CGM.  
One interesting prediction of this scenario is that the ensemble velocity dispersions of $\hi$ on scales of tens kilo-parsecs trace the cascading turbulent velocity of the CGM where the $\hi$ condenses out \citep{Gaspari18}. 
The turbulent diffuse $\hi$ of NGC 4631, which has a velocity dispersion of 50 $\kms$ in an ensemble resolution element of 8 kpc (\citetalias{Wang23}), can be a promising candidate to support this theoretical scenario.

\subsection{Implications on Gas Accretion}
The diffuse $\hi$ may trace a transition between the dense $\hi$ and the CGM, possibly serving as an interface of gas accretion from the CGM.
Theoretically, channels for gas accretion onto massive nearby galaxies can be roughly divided into the cosmological hot accretion, fountains and mergers \citep{Putman17, Grand19}. 
In practice, the first channel is hard to directly observe, but is likely mixed with the latter two channels \citep{Grand19}.
Through comparing the star formation rate, and mass and kinematics of the thick disk $\hi$ and associated warm$\slash$hot ionized gas, circulation of gas through fountains seems to be an important gas accretion channel in the nearby star-forming galaxies including the MW \citep{Richter17, Fraternali17, Marasco22}.
Since the diffuse $\hi$ in the inner disk region resemble the thick disk, it may contribute to the gas accretion through the fountain channel.
The importance of the last channel, tidal interactions or mergers, seems observable but most inconclusive. 

First of all, whether mergers lead to increase of $\hi$ richness (at a given stellar mass) has been inconclusive. 
Theories and simulations suggest its feasibility \citep{Gaspari18, Sparre22}, while recent observations suggest that the $\hi$ gas and total neutral gas masses in post-mergers and merger-induced LIRGs seem to be similar to normal star-forming galaxies \citep{Ellison18, Shangguan19}. 
The cooling of hot gas should not immediately lead to an increase in the dust content of a galaxy so one may not use dust emission to trace an increase in total neutral gas content. 
On the other hand, the conversion of $\hi$ to molecular gas possibly becomes faster and more important in LIRGs (Luminous Infrared Galaxies) \citep{Bellocchi22}, which limits the use of total $\hi$ measurements. 
The nearby galaxies of more normal star formation may be a better laboratory to study the physical process of tidally induced gas condensation, and the diffuse $\hi$ may provide a unique tracer. 

Secondly, whether merger related gas accretion is important for general star-forming galaxies has been questionable.
Through counting the $\hi$ masses in satellite galaxies and deriving the merger timescale, it was largely ruled out that direct merger or stripping of low-mass $\hi$-rich satellites can sustain the star formation in central galaxies \citep{Kauffmann10, DiTeodoro14}. 
However, the ubiquitous existence of diffuse $\hi$ around $\hi$-rich galaxies and the link to tidal interaction strengths, together with the theoretical possibility of large-scale gas condensation due to turbulence cascade and small-scale cooling flow due to turbulent mixing, seem to imply that tidal interaction with neighbor galaxies may bring more cool gas into a galactic disk than just adding the neutral ISM of satellites.
As almost all galaxies in our sample have $\hi$-bearing small companions, wet tidal interactions of nearby $\hi$ rich galaxies can be more frequent than reported in optical flux limited observations or in resolution limited simulations. 
The role of tidal interactions in gas accretion and star formation fueling may thus worth reconsideration.

Nevertheless, linking the diffuse $\hi$ to cooling in the CGM is a complicated problem, because the diffuse $\hi$ is only an intermediate phase. 
The fate of diffuse $\hi$ depends on the interplay between many physical processes including UV background ionization, Rayleigh-Taylor instabilities, Kelvin-Helmholtz instabilities, turbulent mixing and cascading, radiative cooling, and thermal conduction \citep{Tumlinson17}. 
These processes can be better characterized by observationally dissecting the multi-phase gas, as well as obtaining information on the metallicity, magnetic fields, and cosmic rays.
We hope our statistical characterization of the diffuse $\hi$ may provide motivation for observational efforts on the multi-phase gas and useful constraints on the theoretical efforts.

\section{Summary and Conclusion}
\label{sec:conclusion}
We have analyzed in combination the FAST FEASTS images and VLA THINGS images of $\hi$ for ten nearby galaxies. 
Special care has been taken to ensure flux density scale alignment between the two observations, so that we can robustly quantify the excess $\hi$ detected by FEASTS in comparison with the THINGS data. 
Missing $\hi$ from the THINGS data is robustly detected in at least 8 out of 10 galaxies.
The integral $\hi$ mass from interferometric observations of nearby large galaxies should thus be used with caution.

The diffuse $\hi$ is defined as the difference between total and dense $\hi$, while the dense $\hi$ is selected from the THINGS cubes with a relatively uniform flux density threshold. 
The fractions of diffuse $\hi$ over the total $\hi$ range from 5\% to more than half with a median value of $\sim$34\% among the sample.
From analyzing the distribution, velocity and velocity dispersion of the diffuse $\hi$, we find that most of the diffuse $\hi$ is likely organized in a rotating disk which is kinematically hotter than the dense $\hi$. 
The diffuse $\hi$ detected here is more like a thin layer with around 1-kpc thickness, either by itself in the outer galactic region or as a shielding layer at a height of 1 to 3 kpc above the thin disk of dense $\hi$ in the inner region, instead of halo-like structures detected in NGC 4631.
Thus, the diffuse $\hi$ may have different categories, and possibly different producing mechanisms. 
Tidal interaction intensities seem to be closely linked to the fraction, morphology, and kinematics of diffuse $\hi$ detected in galaxies. 
While fountains may contribute more to the inner star-forming disks, tidal interactions may be an efficient producer of diffuse $\hi$ throughout the disks of general $\hi$-rich galaxies. 

The possible role of diffuse $\hi$ as an intermediate phase between the dense $\hi$ and the CGM, and the theoretical links of fountains and tidal interactions to gas condensation indicates a promising role of diffuse $\hi$ in dealing with the gas accretion problem. 
Separating these different mechanisms, and better understanding the fate and role of different types of diffuse $\hi$ in galaxy evolution, will be major tasks in up-coming work, as more data (for a planned sample of in total of 118 galaxies) are obtained and analyzed in the FEASTS program.

\section*{acknowledgments}
We thank Thijs van der Hulst, Yong Shi, Guilin Liu, Ivy Wong, Xi Kang for useful discussions. 
JW thanks support of the research grants from Ministry of Science and Technology of the People's Republic of China (NO. 2022YFA1602902),  the National Science Foundation of China (NO. 12073002), and the science research grants from the China Manned Space Project (NO. CMS-CSST-2021-B02).  
LCH was supported by the National Science Foundation of China (11721303, 11991052, 12011540375, 12233001), the National Key R\&D Program of China (2022YFF0503401), and the China Manned Space Project (CMS-CSST-2021-A04, CMS-CSST-2021-A06).
L.C. acknowledges support from the Australian Research Council Discovery Project funding scheme (DP210100337). Parts of this research were supported by the Australian Research Council Centre of Excellence for All Sky Astrophysics in 3 Dimensions (ASTRO 3D), through project number CE170100013.
Parts of this research were supported by High-performance Computing Platform of Peking University.

This work made use of the data from FAST (Five-hundred-meter Aperture Spherical radio Telescope). FAST is a Chinese national mega-science facility, operated by National Astronomical Observatories, Chinese Academy of Sciences.

The DESI Legacy Imaging Surveys consist of three individual and complementary projects: the Dark Energy Camera Legacy Survey (DECaLS), the Beijing-Arizona Sky Survey (BASS), and the Mayall z-band Legacy Survey (MzLS). DECaLS, BASS and MzLS together include data obtained, respectively, at the Blanco telescope, Cerro Tololo Inter-American Observatory, NSF’s NOIRLab; the Bok telescope, Steward Observatory, University of Arizona; and the Mayall telescope, Kitt Peak National Observatory, NOIRLab. NOIRLab is operated by the Association of Universities for Research in Astronomy (AURA) under a cooperative agreement with the National Science Foundation. Pipeline processing and analyses of the data were supported by NOIRLab and the Lawrence Berkeley National Laboratory (LBNL). Legacy Surveys also uses data products from the Near-Earth Object Wide-field Infrared Survey Explorer (NEOWISE), a project of the Jet Propulsion Laboratory/California Institute of Technology, funded by the National Aeronautics and Space Administration. Legacy Surveys was supported by: the Director, Office of Science, Office of High Energy Physics of the U.S. Department of Energy; the National Energy Research Scientific Computing Center, a DOE Office of Science User Facility; the U.S. National Science Foundation, Division of Astronomical Sciences; the National Astronomical Observatories of China, the Chinese Academy of Sciences and the Chinese National Natural Science Foundation. LBNL is managed by the Regents of the University of California under contract to the U.S. Department of Energy. The complete acknowledgments can be found at https://www.legacysurvey.org/acknowledgment/.

\facilities{FAST: 500 m, VLA }
\software{Astropy \citep{astropy:2013, astropy:2018, astropy:2022}, 
numpy \citep[v1.21.4]{vanderWalt11}, photutils \citep[v1.2.0]{Bradley19}, Python \citep[v3.9.13]{Perez07}, scipy \citep[1.8.0]{Virtanen20} }


\appendix
\section{ Justification on Not Using Deep-Clean with Clean masks}
\label{sec:appendix_deep_clean}
To account for the remaining uncleaned pedestal, we have used the W08 method to scale the residual image before adding it to the convolved CLEAN model for each THINGS data cube.
To solve the same problem, many observations and surveys over the last few decades have instead made use of "clean masks".
They identify the regions where deconvolution is needed (i.e. where signal is present), and within these masks clean down very deeply, for example, to $\sim$0.1-$\sigma$. 
This deep-clean method avoids residual scaling and limits the number of clean components needed.
However, we do not adopt the deep-clean method, as it does not seem to perform compatibly with the multi-scale CLEAN deconvolver for the THINGS data, as detailed below.
The following justification is supported by a test (test-1) of combining deep-clean with multi-scale CLEAN, which is applied to two galaxies (NGC 2901 and NGC 3521) and one mock image ($\kappa=$2.2, SNR=2.8, see Appendix~\ref{sec:appendix_crosscali}). 
For each data set, the clean mask is generated by dilating the SoFiA mask of the THINGS cube used in the main part of this paper.
The dilating width is 24 pixels ($\sim$4 FWHM of the clean beam).

Firstly, deep-cleaning regions with sufficient flux and recovering large-scale faint structures are possibly contradictory criteria for low SNR data. 
Low-amplitude but large-scale structures play an important role in recovering the flux structure in the CLEAN process, by restricting the power attributed to the small-scale structures \citep{Rich08}.
But they tend to be buried in noise (even after extensive smoothing) in the final (residual added) cube, and not to be captured by source finding or properly enclosed in clean masks.
On the other hand, blindly dilating source masks to large scales tend to include unnecessary noise peaks.
In test-1, the warning of ``MSCLAN minor cycle stopped at large scale negative or diverging'' appears much more frequently than when clean masks are not used. 

Secondly, (possibly partly caused by the first reason) deep-clean is computationally expensive for large galaxies like the THINGS sample. 
For each data set in test-1, after 1,000,000-1,500,000 iterations (30 to 45 hours on a 6-core, 12-thread, end-of-2010s work station), the absolute peak residual within the mask still does not reach the criteria threshold of 0.1-$\sigma$.
Instead, it shows hints of converging to (or fluctuating around) $\sim$0.3-$\sigma$ without further decreasing.
For comparison, we conduct a second test (test-2) of combining deep-clean with the ``hogbom'' deconvolver. 
For each data set (real or mock), the absolute peak residual within the mask does decrease, and reaches the 0.1-$\sigma$ threshold after 1,500,000 to 4,000,000 iterations (15 to 48 hours). 

For these considerations, we have adopted the W08 method to scale residual images when producing the THINGS data cubes. 
Combining deep-cleaned images from a high-SNR interferometric dataset with the FEASTS data remains to be investigated in the future.

\section{The Procedure of Cross-calibrating the FEASTS and THINGS Fluxes}
\label{sec:procedure_crosscali}
\subsection{A Review of the Procedure in \citetalias{Wang23}}
We use a similar procedure as described in \citetalias{Wang23} to cross-calibrate the fluxes. 
The procedure of \citetalias{Wang23} was built on basis of the package {\it uvcombine}\footnote{https://github.com/radio-astro-tools/uvcombine/}, and the steps are summarized below.
We reproject the single-dish image to the same WCS system as the interferometric image of the same source, and then attenuate the reprojected single-dish image with the same primary beam response as the interferometric image. 
The product is called the {\it single-dish PBA image} for short.
We convolve the single-dish PBA image with the clean beam of the interferometric data, and convolve the interferometric data with the beam of FAST, so that these two data have the same resolution.
We divide both the convolved single-dish PBA image and the convolved interferometric image by the beam area to let the pixel values have the unit of Jy per pixel.
We Fourier transform these two images, and obtain the single-dish FFT image and the interferometric FFT image, respectively. 
We select the FFT pixels that are in a overlapping frequency range with the lower and upper limits determined by the largest and smallest spatial scales detected in the interferometric and single-dish data, respectively.
We further select the pixels that have a SNR above 4 in both FFT images.
The cross-calibration factor of fluxes $f_{\rm F/T}$ is derived as the 3-sigma-clipped mean of the ratio between the single-dish FFT amplitudes and the interferometric FFT amplitudes of the selected pixels.

\subsection{Updated Beam Shape and Beam area}
\begin{figure} 
\centering
\includegraphics[width=9cm]{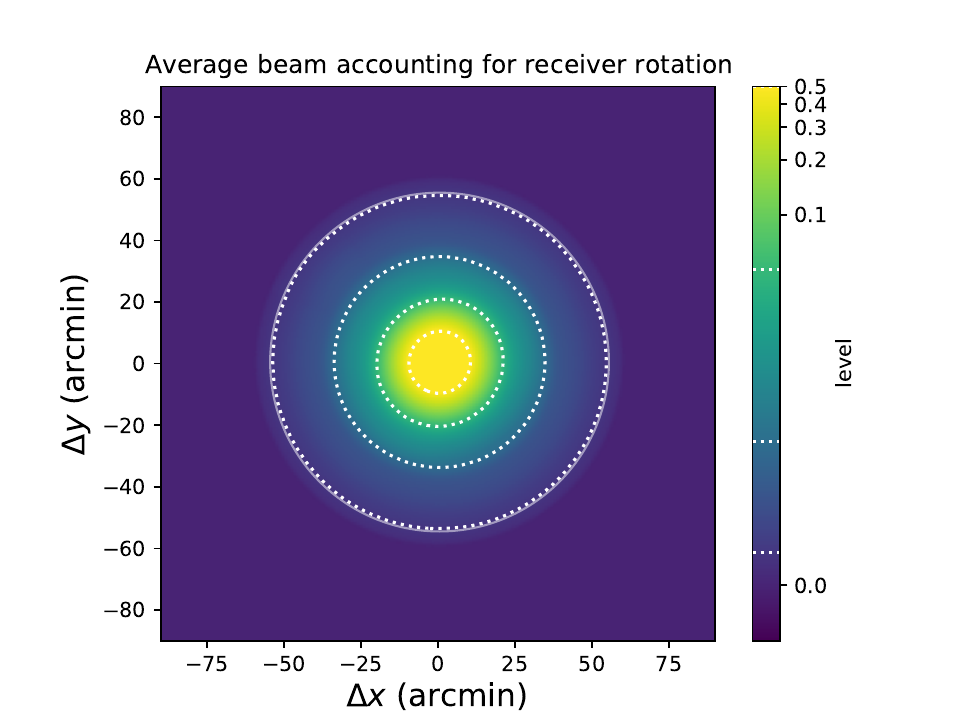}
\caption{The average beam image accounting for the rotated receivers. The white, dotted contours are at levels of 5e-4, 5e-3, 5e-2, and 0.5. The levels are also marked in the color bar. A solid white circle is plotted near the outmost contour for reference, to highlight that the averaged beam is circular in the outer region.  }
\label{fig:beam}
\end{figure}

Compared to \citetalias{Wang23}, we have updated the average beam image of FAST.
The average beam shape when the receiver has a position angle of zero has been used in \citetalias{Wang23}, but in FEASTS observation, the receiver was rotated to achieve the Nyquist-Shannon sampling rate. 
The unrotated receiver of FAST has a $m=6$ harmonic pattern in the outer region of the beam image. 
As the rotated receiver of the horizontal and vertical scans have position angles differing by 30$\degree$, the $m=6$ harmonic patterns of the two rotated beams cancel out. 
As a result, the effective average beam of FEASTS has an almost azimuthally symmetric shape in the outer region, as shown in Figure~\ref{fig:beam}. 

The radial profile of the beam does not change in comparison to the one showed in \citetalias{Wang23}.
We also point out that, as most of the diffuse $\hi$ in \citetalias{Wang23} have high column densities, this correction in beam shape does not significantly affect the results on diffuse $\hi$ in \citetalias{Wang23}.

As in \citetalias{Wang23}, the beam FWHM of 3.24$'$ is confirmed by fitting double gaussian models to the beam image. 
But the actual beam area is larger that of the gaussian beam by a factor of 1.064 due to the existence of side-lobes.
This factor is considered in all cases when converting the unit of $\jyb$ to Jy$/$arcsec$^2$ or to Jy$/$pixel.

\subsection{Adjustment in the Cross-calibrating Procedure}
We adjust and improve the procedure, to optimize it for the THINGS data that are shallower than the HALOGAS data used in \citetalias{Wang23}. 
Many of these adjustments are developed based on tests with mock data that have similar qualities as the FEASTS and THINGS data. 
The mock simulation and the tests are presented in detail in Appendix~\ref{sec:appendix_crosscali}. 
The adjustments are summarized below.
\begin{enumerate}
\item We use the moment-0 image (produced in section~\ref{sec:feasts} and \ref{sec:things}) instead of the cube to conduct the comparison.

\item The THINGS moment-0 image used is derived from the rescaled cube, using SoFiA mask of the standard cube. 
Using this type of moment-0 image to correctly recover as much faint and extended fluxes as possible is important in order to have a consistent comparison in amplitudes with the FEASTS data. We show that the rescaled cubes are much better than the standard cubes, and slightly better than the convolved model cubes for cross-calibration based on tests with mock data in appendix~\ref{sec:appendix_crosscali}. 

\item Only the inner region with a radius of 750$''$ is used for the calibration, as the primary beam of THINGS data quickly drops below 0.5 beyond that.

\item The upper limit of spatial frequency is set to be 1.5 times the FAST beam size as before, while the lower limit of the spatial frequency is set to be 8.82$'$, corresponding to a baseline of 100 m. 
Like in \citetalias{Wang23}, the interferometric data start to miss fluxes in comparison to the FAST data before reaching the largest scale allowed by its shortest baseline. 
The reason is likely a combination of limited sampling density of the shortest baseline, which can be worsened by RFI flagging (like in \citetalias{Wang23}), and the relatively high rms level due to limited integration time (unlike the case of \citetalias{Wang23}). 
\end{enumerate}

\subsubsection{Arguments for Using Moment-0 Images in Cross-calibration}
Among the many adjustments listed above, a most controversial one may be using moment-0 images instead of the more conventional choice of cubes or visibilities for cross-calibration. 
The data cubes already have the shortage of having nonlinear noises introduced by the CLEAN, while the moment-0 images further have the disadvantage of being after noise dependent thresholding of source finding.

A first support of this method can be based on the assumption that interferometric moment-0 images widely used for characterizing $\hi$ distributions should be reasonably good in regimes where the SNR and uv sampling are sufficiently good. 
But more importantly, the data quality requires us to use the moment-0 images instead of cubes for cross-calibrating.
Firstly, a sufficiently high SNR level is required for a robust cross-calibration. 
We show in Appendix~\ref{sec:appendix_crosscali} that the interferometric image should have a median SNR above 2.6, in order for the systematic uncertainty of cross-calibration to be below 5\%. 
Most THINGS moment-0 images have median SNR just above this threshold (Table~\ref{tab:tab_dataprop}), while the related channel maps only have SNRs roughly one tenth that level. 
Increasing the SNR through a sufficient smoothing in the spectral direction is almost equivalent to making the moment-0 image.
Smoothing in the image domain does not help either, as it is already included in the source-finding (smoothed with a gaussian kernel that has FWHM of 30$''$) and cross-calibrating (smoothed with the FAST beam) procedures.
Secondly, a sufficiently wide dynamic range in the uv coverage, or spatial scales of $\hi$ structures, is necessary for a robust cross-calibration \citep{Stanimirovic02, Kurono09}. 
In principle, the lower and upper limits of the overlapping region are determined by the FAST resolution and the shortest baseline of the THINGS observation, but the real flux structures do not always cover the selected region with sufficient power, which is exacerbated by the limited sensitivity. 
In many channel maps of the THINGS cube, the flux detected regions can be smaller than the FAST beam. 
This problem is much mitigated in the moment-0 image. 

A possibly most fundamental and accurate (but see \citealt{Stanimirovic02}), but computational expensive method is to convert the FEASTS images to visibilities and compare the visibilities for cross-calibration. 
This method has advantages of keeping the interferometric data closer to observations, being less affected by filtering and thresholding.
However, this method should also be limited by the noise level of data and spatial spanning of fluxes. 
Moreover, producing simulated visibilities for single-dish images introduces interpolation errors. 
These effects remain to be explored. 

Thus, in this first study of combining FEASTS data with interferometry for a relatively large sample of galaxies, we adopt and optimize the relatively simple and quick method of cross-calibrating with moment-0 images. 
This method can be easily used by non-radio-experts, and can be applied to combining large samples of interferometric images and single-dish images. 
Such images are likely to be available in the near future with WALLABY \citep{Koribalski20}, Apertif \citep{Adams22}, and CRAFTS \citep{Zhang19} wide-field $\hi$ surveys.

\subsection{Results of Cross-calibration}
\begin{figure*} 
\centering
\includegraphics[width=16cm]{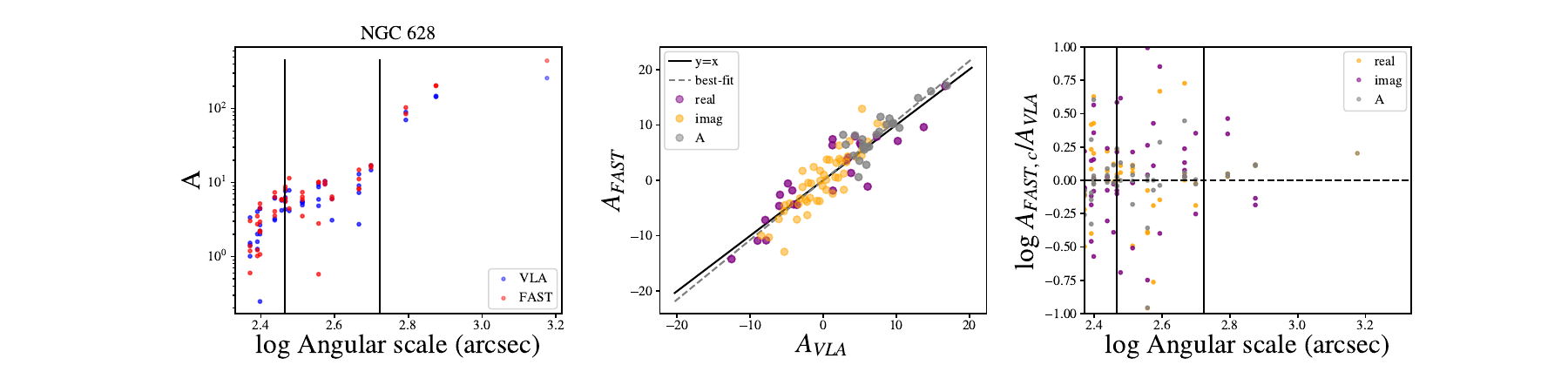}
\includegraphics[width=16cm]{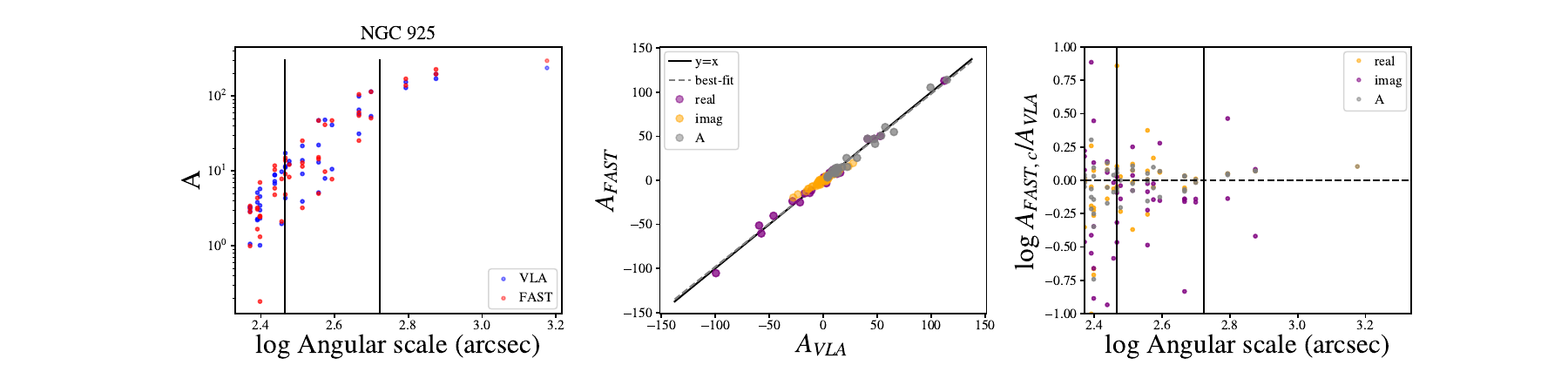}

\includegraphics[width=16cm]{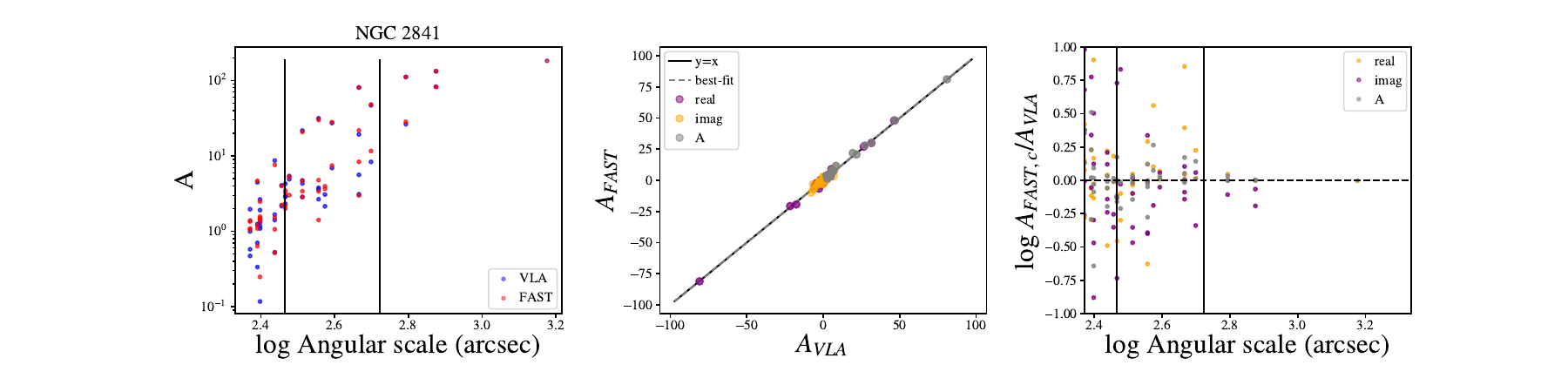}
\includegraphics[width=16cm]{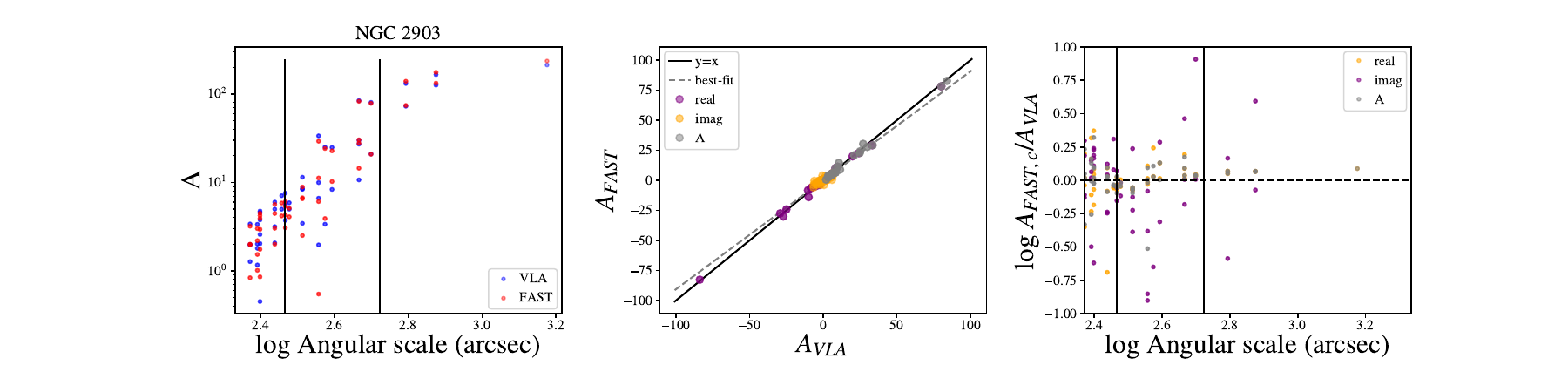}

\includegraphics[width=16cm]{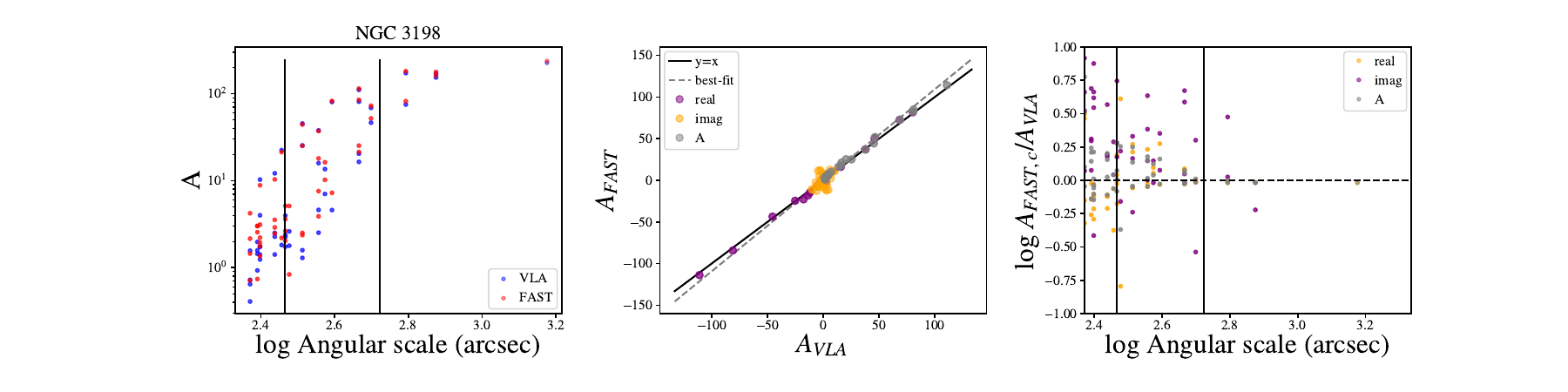}
\caption{The cross-calibration procedure plots for each galaxy. 
{\bf Left:} 
the FFT amplitudes of the FAST and VLA moment-0 images as a function of angular scales (the inverse of spatial frequency). 
The FAST and VLA measurements are plotted in red and blue respectively.
The two vertical lines mark the lower and upper limits of the overlapping angular scales, between which the pixels are selected. 
{\bf Middle:} 
comparing the FFT amplitudes between the two types of data.
The FFT amplitudes of selected pixels from the two types of FFT images are plotted in gray dots, and the gray dashed line has an intercept of 0 and a slope of the derived $f_{\rm F/T}$. 
The imaginary$\slash$real parts of selected pixels from the two types of FFT images are plotted as orange and purple dots to confirm the goodness of the derived $f_{\rm F/T}$. 
The black solid line shows the position of $y=x$.
{\bf Right:} the ratio of corrected FAST amplitudes over VLA amplitudes as a function of angular scales. The two vertical lines from the left panel are repeated. The dashed horizontal line mark the position of $y=0$.
To be continued.}
\label{fig:crosscali}
\end{figure*}

Figure~\ref{fig:crosscali} presents the results from the flux cross-calibration procedure. 
Each row is for a target galaxy.
The left panel of each row shows the FFT amplitudes of the FAST and THINGS moment-0 images as a function of angular scales. 
The two vertical lines mark the lower and upper limits of the overlapping angular scales, between which the pixels are selected. 
We can see that between the two vertical lines the two types of amplitudes often have a similar range of values, but beyond the vertical line of the upper limit the FAST amplitudes are often higher. 
In the middle panel of each row, the gray dots show the relation of FFT amplitudes from selected pixels between the types of data, and the gray dashed line has an intercept of 0 and a slope of the derived $f_{\rm F/T}$. 
The gray dashed line is usually not far away from the black line of unity. 
We also plot the imaginary$\slash$real parts of selected pixels from the two types of FFT images. 
Most of the data points follow the relation of the amplitudes, justifying the robustness of the cross-calibration. 
In the right panel of each row, we plot as a function of angular scales the ratio of corrected FAST FFT amplitudes over VLA FFT amplitudes. 
The scaled FAST FFT amplitudes are now consistent with the VLA FFT amplitudes within the selected angular range, but start to be higher than the VLA amplitudes roughly when the angular scale is larger than the upper limit, which is roughly 8.82$'$, or 18.7 to 37.7 kpc in this sample.

 \addtocounter{figure}{-1}
\begin{figure*} 
\centering
\includegraphics[width=16cm]{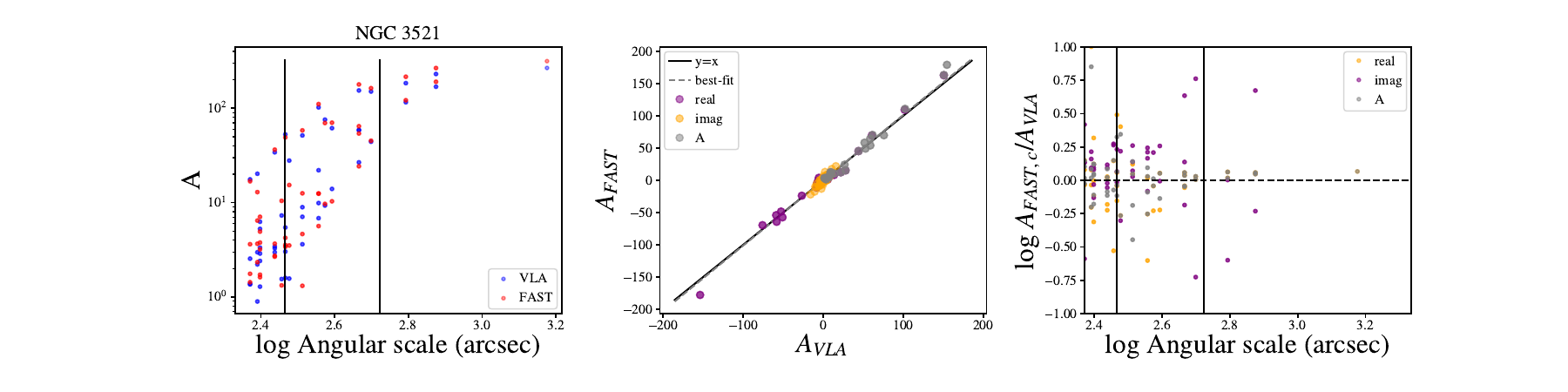}

\includegraphics[width=16cm]{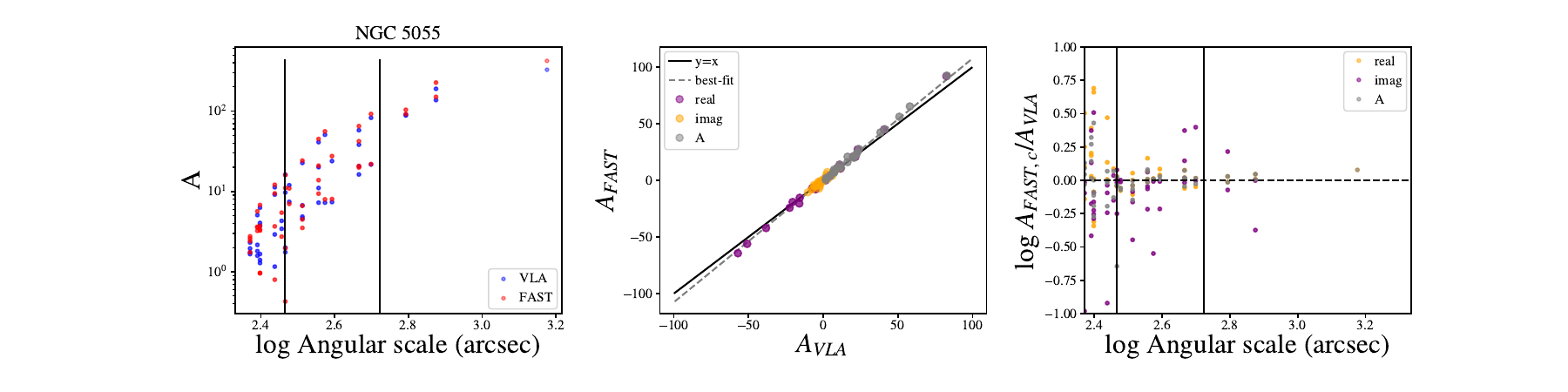}
\includegraphics[width=16cm]{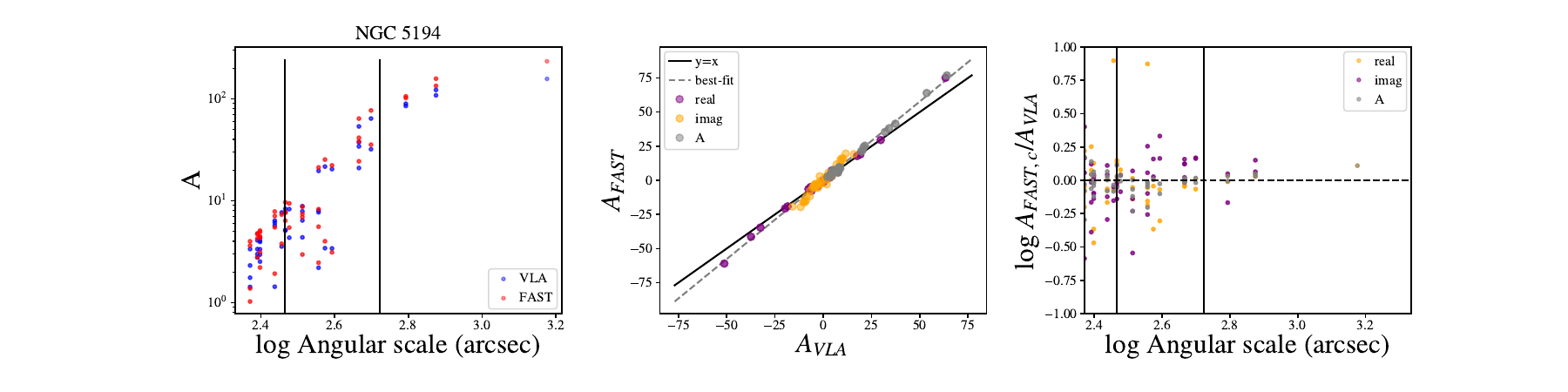}

\includegraphics[width=16cm]{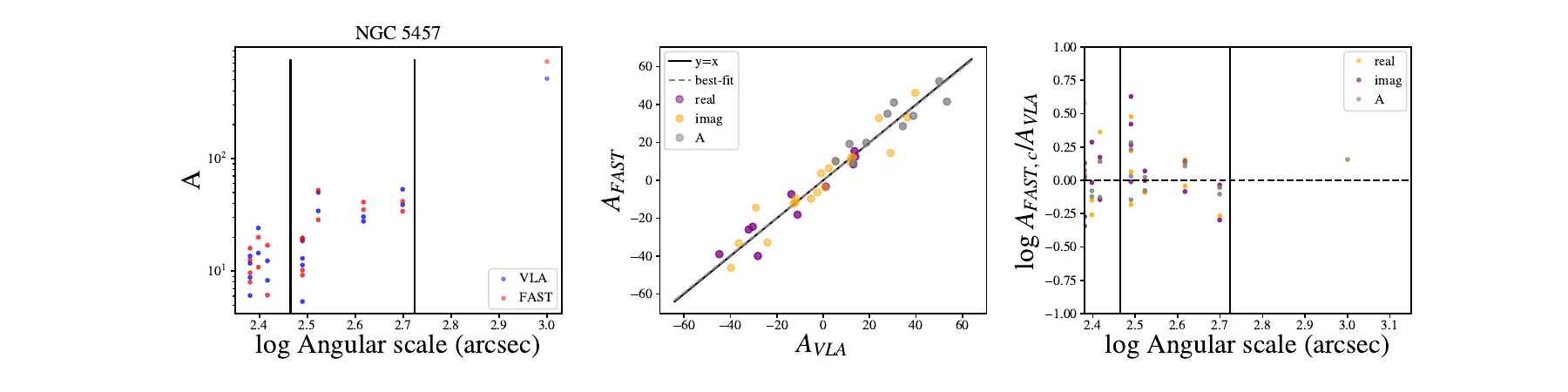}
\includegraphics[width=16cm]{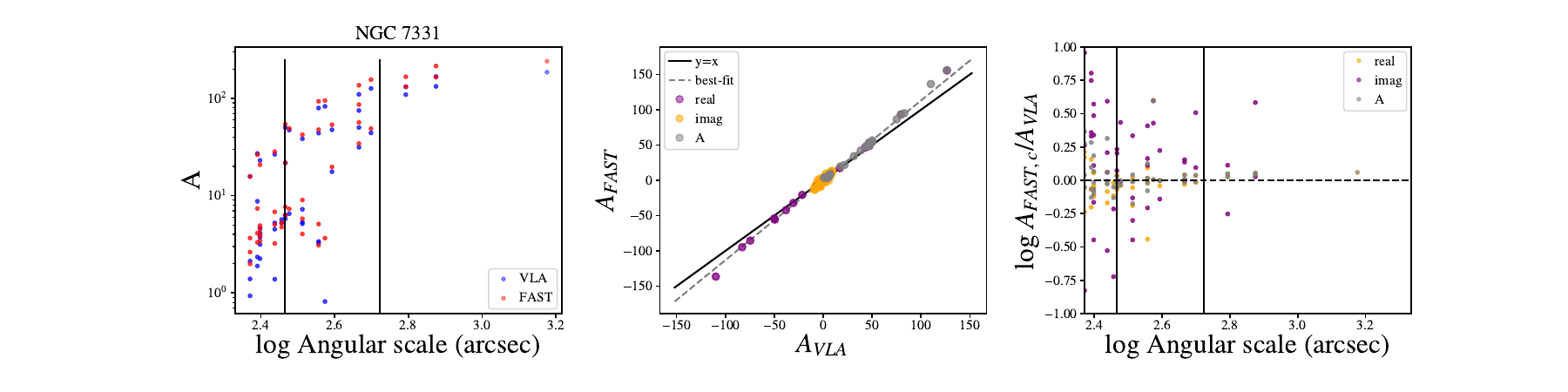}

\caption{The cross-calibration procedure plots for each galaxy. Continued. }
\end{figure*}

\section{ Mock Test to Optimize the Flux Cross-calibrating Procedure }
\label{sec:appendix_crosscali}
The primary questions to address here are whether we can achieve reasonable accuracy in flux cross-calibration between the VLA and the FAST data, and what type of VLA image is best for the cross-calibration. The candidates of VLA maps include the standard clean map, the rescaled clean map, and the convolved model map. 
To address these questions, we simulate mock image of disks which have size, power spectrum slope, and SNR close to the moment 0 images of real galaxies. We analyze them in the same way as we do for real data, and check how the derived scaling $f_{\rm F/T}$ deviates from the actual value of unity. 

\subsection{Mock HI disks}
\label{sec:appendix_mock}
We describe below the step of simulating a mock $\hi$ disk.
We firstly create an image of a face-on disk which has a $R_{HI}$ of 9$'$, and surface density radial profile following the median profile of \citep{Wang16}.
The $\hi$ disk is larger than most target galaxies in our sample, so the mocks represent the most difficult case for VLA observations.
The image size is set to be 2000 pixels on each size, and the pixel size 1.5$'$'.
We use the python package {\it TurbuStat} \citep{Koch19} to create a synthetic image of the same size based on a power-law noise distribution. 
The power spectrum slope is set to a value $-k$ close to real galaxies in $\hi$ moment-0 images (see Appendix~\ref{sec:appendix_ps}).
The flux scale of the synthetic noise image is globally normalized to have values between $-1$ and 1. 
The synthetic noise image multiplying with the face-on disk image is added to the face-on disk image. 
Any pixels below zero is set to be an arbitrarily positively low value.
Finally, the pixel values are globally scaled to have a total flux of 23.22 Jy. 
We have produced a mock $\hi$ disk which have similar small and large-scale structures as real $\hi$ disks.
We produce mock $\hi$ disks with six different $k$ values, ranging from 1.8 to 3.2. 
The peak fluxes of these disks range from 0.3 to 0.14 mJy pixel$^{-1}$. 

\subsection{Mock FAST images}
The mock $\hi$ disk is convolved with the FAST beam image which has been re-sampled to have the same pixel size. 
We generate a noise image following a random gaussian distribution with a $\sigma$-width of 0.57 mJy$/$beam. The $\sigma$ looks lower than the values listed in Table 2 because the FAST data are smoothed along the velocity direction when reprojected to the WCS system of the VLA data. We add the noise image to the mock disk convolved with the FAST beam, and this produces a mock FAST image. 

\subsection{Mock VLA images}
We use the CASA task {\it simobserve} to simulate VLA observations, and produce mock visibilities.
We feed the mock disk to the task as the sky model. We use the coordinate of the galaxy NGC 628 as the direction of pointing. 
To mimic the uv coverage of real THINGS observations, we simulate three sets of observations, in the B, C, and D array configurations, with integration time of 7, 2.5 and 1.5 hours, respectively. 
We follow the lengths of integration time of the THINGS observation of NGC 628 to achieve similar uv coverage.
We only simulate one channel map, and thus can conveniently adjust the rms level by changing the channel width of the observation. 
We use channel widths of 0.25, 0.05, 0.04, 0.03, 0.02, 0.01, and 0.001 MHz, to achieve an rms level of 0.31, 0.50, 0.53, 0.60, 0.71, 0.96 and 2.86 mJy$/$beam.
We note that, the SNR ratio distribution in the whole flux detected region is more informative than the absolute rms level in the problems we discuss.
Later, we find rms levels to correspond to median SNR of 4.4, 2.8, 2.6, 2.4, 2.1, 1.7, and 0.9 in flux detected regions of the cleaned and rescaled images for the $k=2.2$ mock disks.

After creating the mock visibilities, we follow the procedure described in section~2, using the {\it tclean} task of CASA to reduce them. 
We obtain the standard images, the convolved model images, the residual images, and the rescaled images of the mock VLA observation. 
The clean beam of all maps has major and minor FWHM of 9.0$'$ and 8.9$'$.

\subsection{Cross-calibrating the Mock FAST and Mock VLA images}
We run the cross-calibrating procedure on the mock images, to derive $f_{\rm F/T}$ calibrating the flux ratio between mock FAST and mock VLA observations. We compare the $f_{\rm F/T}$ to the true value of unity, and study how the deviations vary.

In Table~\ref{tab:tab_mockcali1}, we list $f_{\rm F/T}$ values derived from mock images generated from the same $k=2.2$ mock disk. 
The mock VLA observations have different rms levels, and the VLA images used for the cross-calibration have different types.
We also test two options of quantity compared between the FAST and VLA data to derive $f_{\rm F/T}$: the amplitudes, and the complex (real and imaginary) parts of FFT images.

We find that when the median SNR$\lesssim$2.4 (1.7), all derived $f_{\rm F/T}$ deviate from the truth of unity by more than 5\% (10\%). 
Because the flux calibration systematic uncertainty between observations is typically $\lesssim$10\%, it is possible more appropriate not to cross-calibrate, or manually set $f_{\rm F/T}=1$, when the VLA data has a median column density SNR below 1.7 (or rms above 0.96 mJy$/$beam).
When the median SNR$\gtrsim$2.6, the derived $f_{\rm F/T}$ are close to unity when using the amplitudes as the quantity, and the rescaled images as the VLA data for comparison. 
We thus take the FFT amplitudes and rescaled images as the fiducial combination for cross-calibration.

We check the cross-calibration quality when using other combinations of measurements and data.
When we replace the FFT amplitudes with FFT real and imaginary parts in the fiducial combination, the derived $f_{\rm F/T}$ values become slightly less accurate, deviating  by $\sim$0.05 from the true value when SNR$\gtrsim$2.6.
When we replace the rescaled images with the convolved model images in the fiducial combination, the derived $f_{\rm F/T}$ values are slightly less accurate.
But when we replace the rescaled images with the standard images in the fiducial combination, the derived $f_{\rm F/T}$ values are noticeably less accurate, deviating by $\sim$0.08 from the true value when SNR$=$2.6.
Therefore, either changing FFT amplitudes to real$\slash$imaginary components, or changing rescaled images to other images, we get worse results in cross-calibration than using the fiducial combination.

As justified in \citetalias{Walter08}, in regions with significant fluxes, the rescaling method is able to properly correct for the difference of dirty beam from the clean beam in the residual map. The rescaled map is able to retain in a relatively complete and accurate way the faint and extended fluxes which would otherwise be left in the residual map or be added back with inconsistent scaling from the convolved model. This might be the reason why tests with the rescaled clean maps lead to best results.  

In Table~\ref{tab:tab_mockcali2}, we conduct a similar test as in Table~\ref{tab:tab_mockcali1}, but we fix rms of VLA observations at a relatively low level of 0.5 mJy$/$beam (SNR$\sim$2.8) and change the $k$ values of mock disks.
Throughout the table, the most reliable $f_{\rm F/T}$ values are obtained when using the fiducial combination of FFT amplitudes and rescaled images.
The largest deviation of 0.05 is found when $k=1.8$, possibly because images with shallower power spectra are more susceptible to noises. 
But 1.8 is below the minimum $k$ value of the THINGS sample (see Appendix~\ref{sec:appendix_ps}), which mitigate this problem.

To summarize, the combination of FFT amplitudes and rescaled VLA images are best for the cross-calibration among all options considered here. 
With such a combination, we can achieve reasonable accuracy in obtaining the flux cross-calibrating factor $f_{\rm F/T}$, with a typical systematic uncertainty less than 0.05 when the median SNR of the image $\gtrsim$2.8.
Because most of the THINGS moment-0 images have median SNR above 2.4, the flux cross-calibration with FEASTS data should be reasonably correct. 
The THINGS channel maps typically has a SNR 6 to 8 times lower than that of moment images, thus, cross-calibrating with the THINGS channel maps is not recommended.
Because of the minimal systematic uncertainty, a detection of diffuse $\hi$ with a mass ratio (over the total $\hi$) above 5\% is likely to be reliable.

   \begin{table*} 
    \centering
            \caption{Flux cross-calibration factors from tests with the $k=2.2$ mock disk.}
            \begin{tabular}{c| c c c c c c c }
              \hline
                 VLA data type    & \multicolumn{7}{|c}{Median SNR} \\
                           &4.4  & 2.8 &  2.6 &  2.4 & 2.1  &  1.7  & 0.9\\ 
              (1)     & (2)     & (3)     & (4)  & (5)  & (6) & (7)  & (8)    \\
             \hline
             &  \multicolumn{7}{c}{Using amplitudes of FFT image} \\
             Rescaled image    & 0.98$\pm$0.01  &  0.99$\pm$0.03 &0.97$\pm$0.03 &0.94$\pm$0.03  & 0.92$\pm$0.03  & 0.84$\pm$0.02 &  0.65$\pm$0.04 \\
             Convolved model   & 0.96$\pm$0.01  &  0.95$\pm$0.04 &0.96$\pm$0.04 &0.95$\pm$0.04 &  0.90$\pm$0.03  & 0.85$\pm$0.02 &  0.67$\pm$0.04  \\
             Standard image    &  0.96$\pm$0.01 &  0.93$\pm$0.01 &0.92$\pm$0.01 &0.95$\pm$0.02  &  0.93$\pm$0.02  & 0.82$\pm$0.02 & 0.48$\pm$0.02   \\
            \hline
            &  \multicolumn{7}{c}{Using real and imaginary parts of FFT image} \\
             Rescaled image    & 0.95$\pm$0.02  &  0.96$\pm$0.02 &1.00$\pm$0.03 &0.92$\pm$0.03  & 0.91$\pm$0.03    & 0.85$\pm$0.03 & 0.57$\pm$0.04\\
             Convolved model   & 0.97$\pm$0.03  &  0.94$\pm$0.03 &0.95$\pm$0.02 &0.94$\pm$0.02 & 0.92$\pm$0.03   & 0.82$\pm$0.04 & 0.58$\pm$0.04 \\
             Standard image    &  0.92$\pm$0.02 &  0.94$\pm$0.02 &0.96$\pm$0.02 &0.95$\pm$0.02  &  0.92$\pm$0.03  & 0.89$\pm$0.03 & 0.47$\pm$0.03 \\
            \hline
            \end{tabular} 
            \\
              {\raggedright
                Column~(1): The type of VLA cube from which the moment image is derived.
                Column~(2)-(8): The cross calibrated scaling factor $f_{\rm F/T}$ at a given SNR level of the mock VLA observation. Two sets of results are provided: the first three rows are for tests comparing the amplitudes of FFT images to derive $f_{\rm F/T}$, and the last three rows are for tests comparing the real \& imaginary parts. 
               }
        \label{tab:tab_mockcali1}
    \end{table*}

   \begin{table*}
    \centering
            \caption{Flux cross-calibration factors from tests with mock disks of different power spectra }
            \begin{tabular}{ c| c c c c c c }
              \hline
              VLA data type    &     \multicolumn{6}{c}{Power spectral slope $k$ of the mock disk} \\
               &    1.8  &2.0 &  2.2 & 2.5 & 2.8 & 3.2 \\
               &    (1) & (2)   & (3) & (4)  & (5) & (6)  \\
             \hline
             &  \multicolumn{6}{c}{Using amplitudes of FFT image} \\
            Rescaled image      &  0.95$\pm$0.03  & 0.96$\pm$0.03  & 0.99$\pm$0.03  &  0.97$\pm$0.02 & 0.98$\pm$0.02 & 1.02$\pm$0.02   \\
            Convolved model   &  0.96$\pm$0.04    & 0.97$\pm$0.02  &0.95$\pm$0.04 &  0.96$\pm$0.03 &  0.99$\pm$0.01 & 1.02$\pm$0.02 \\
            Standard image      &   0.96$\pm$0.03 & 0.95$\pm$0.02  & 0.93$\pm$0.01 & 0.93$\pm$0.01  & 0.98$\pm$0.02  & 1.02$\pm$0.03\\
            
            \hline
             & \multicolumn{6}{c}{Using real and imaginary parts of FFT image} \\
             Rescaled image     & 1.00$\pm$0.03 & 1.00$\pm$0.03  & 0.96$\pm$0.02  & 0.94$\pm$0.03 & 1.01$\pm$0.02  & 1.05$\pm$0.02 \\
             Convolved model  & 0.92$\pm$0.03  & 0.97$\pm$0.02 &0.94$\pm$0.03  & 0.96$\pm$0.03   & 1.01$\pm$0.02  & 1.05$\pm$0.02\\
             Standard image      & 0.88$\pm$0.03&  0.95$\pm$0.02 & 0.94$\pm$0.02  & 0.92$\pm$0.02  & 1.01$\pm$0.02 & 1.05$\pm$0.02 \\
            
            \hline
            \end{tabular} 
            \\
              {\raggedright
                Column~(1)-(6): The cross calibrated scaling factor $f_{\rm F/T}$ for the mock observation of the mock disk with the given power spectral slope $k$. 
               }
        \label{tab:tab_mockcali2}
    \end{table*}

\section{Power Spectrum of the Target Galaxies }
\label{sec:appendix_ps}
We use the {\it PowerSpectrum} task from the python package {\it Turbustat} to derive one-dimensional power spectra from the THINGS $\hi$ moment-0 images of rescaled cubes. 
We divide the power spectrum of each moment-0 image by the power spectrum of its clean beam. 
We select the angular scale range between the largest sensitive scale of 8.82$'$ and 1.5 times the major-axis FWHM of the clean beam, and fit a linear relation between the power and the frequency in the logarithm space. 
The slope of this linear relation is the power law index of the power spectrum. 

\begin{figure*} 
\centering
\includegraphics[width=16cm]{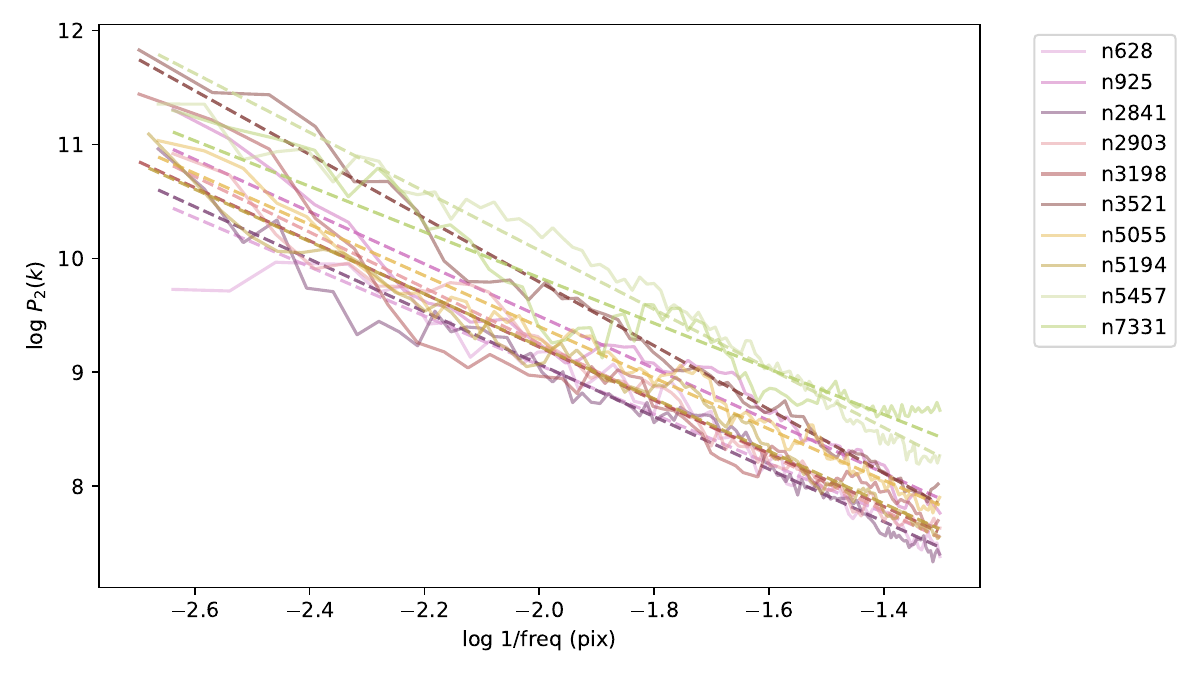}
\caption{The 1 dimensional power spectra of $\hi$ distribution in THINGS moment-0 images. The solid lines are the directly derived power spectra, and dashed lines of the same colors are the best linear fit in the logarithm space. The different colors represent different galaxies as denoted in the figure.}
\label{fig:ps}
\end{figure*}

We show the power spectrum and best-fit linear relations in the logarithm space in Figure~\ref{fig:ps}. 
We list the slopes in Table~\ref{tab:tab_ps}. The slopes have a median value of -2.3, and standard deviation of 0.2. 
We use these measurements as reference to set the power spectrum of mock images created in Appendix~\ref{sec:appendix_mock}.

 \begin{table}
    \centering
            \caption{Power spectral slope}
            \begin{tabular}{c c }
              \hline
              Galaxy  & $k$ \\
              (1)     & (2)       \\
                \hline
                    NGC 628 & 2.17\\
                    NGC 925 & 2.29\\
                    NGC 2841& 2.30\\
                    NGC 2903& 2.44\\
                    NGC 3198& 2.33\\
                    NGC 3521& 2.80\\
                    NGC 5055& 2.24 \\
                    NGC 5194& 2.30 \\
                    NGC 5457& 2.59\\
                    NGC 7331& 2.00\\
              \hline

            \hline
            \end{tabular} 
            \\
              {\raggedright
                Column~(1): Galaxy name.
                Column~(2): The slope of the 1 dimensional power spectrum of $\hi$ distribution. 
                }
        \label{tab:tab_ps}
    \end{table}

\section{Relations between the Moment Images of the Diffuse HI, the Dense HI and the Total HI }
\label{appendix:mom_relation}
We mathematically link the moment images of the total, dense and diffuse $\hi$. 
The moment-0 (column density) images of the diffuse $\hi$ is related to those of the total and dense $\hi$ simply as:
\begin{equation} 
N_{{\rm HI,diff}}=N_{{\rm HI,tot}}-N_{{\rm HI,dens}}.
\label{eq:mom0}
\end{equation}

The moment-1 (velocity field) images of the diffuse $\hi$ is related to those  of the total and dense $\hi$ as: 
\begin{equation} 
v_{\rm diff}=\frac{ N_{{\rm HI,tot}} v_{\rm tot}-N_{{\rm HI, dens}} v_{\rm dens}}{N_{{\rm HI, diff}}}.
\label{eq:mom1}
\end{equation}

If we ignore the beam smearing effects, the moment-2 image of the total $\hi$ has four parts in each pixel: the velocity dispersion of the diffuse (dense) $\hi$ around the central velocity of the diffuse (dense) $\hi$, and the offset of central velocity of the diffuse (dense) $\hi$ from the central velocity of the total $\hi$. 
So, mathematically, the velocity dispersions of the total, dense and diffuse $\hi$ are related by the following equation. 
\begin{align}
  N_{{\rm HI, tot}}\sigma_{\rm tot}^2=  {} & N_{{\rm HI,diff}}( \sigma_{\rm diff}^2 + v_{\rm 1,diff}^2) \notag \\
               & + N_{{\rm HI,dens}}( \sigma_{\rm dens}^2+ v_{\rm 1,dens}^2)   .
  \label{eq:mom2}
\end{align}
where $v_{\rm 1,diff}=v_{\rm diff}-v_{\rm tot}$ and $v_{\rm 1,dens}=v_{\rm dens}-v_{\rm tot}$.
This equation reflects the conservation of kinetic energy in transforming of reference frames and holds regardless of the actual spectral shape, a detailed deduction of which can be found in the appendix~\ref{appendix:dmom2}.

These equations hint expectations for cross-checking patterns from images of different $\hi$ components, and propagation of errors dominated by the THINGS data and dependent on $f_{\rm diffuse}$ (or locally on N$_{\rm HI,diff}$/N$_{\rm HI,tot}$).
Due to the relatively high noise levels of the THINGS data, we should avoid extract information more complex than the moment images from the cubes of the diffuse $\hi$.
A future joint deconvolution to coherently take advantages of both datasets, and derive consistent models for both datasets should enable more sophisticated analysis of structural and kinematical components (Lin et al. in prep).

\subsection{The Equation Linking the Velocity Dispersions of the Dense HI, Diffuse HI and Total HI }
\label{appendix:dmom2}
We assume the $\hi$ to be comprised of cloud elements, each having a volume density $n$ and velocity $v'$.
Obviously, for each line of sight, the column density $N$, average velocity $v$, and velocity dispersion $\sigma$ are related to the statistics of the clouds by
\begin{equation}
\begin{split}
N &=\sum n, \\
v &=\frac{\sum n v'}{\sum n}, \\
\sigma^2 & = \frac{\sum n (v'-v)^2}{\sum n}.
\end{split}
\label{eq:dmom2a}
\end{equation}

For each line of sight, we assume the total $\hi$ (denoted with $s$) to be comprised of the dense $\hi$ and the diffuse $\hi$. 
Their properties are link as:


\begin{equation}
\begin{split}
N_{\rm tot} \sigma_{\rm tot}^2  = {}& \sum n_s(v_{\rm tot}'-v_{\rm tot})^2  \\
                     ={}& \sum n_{\rm dens}(v_{\rm dens}'-v_{\rm tot})^2 \\
                      &+ \sum n_{\rm diff}(v_{\rm diff}'-v_{\rm tot})^2.
\end{split}                  
\label{eq:dmom2b}
\end{equation}

For each gas phase, the average velocity deviate from that of the total $\hi$ by $v_{1,x}=v_x-v_{\rm tot}$, where $x={\rm diff}$ or ${\rm dens}$. So that,


\begin{equation}
\begin{split}
 \sum & n_x(v_x'-v_{\rm tot})^2   \\
 & =  \sum n_x (v_x'-v_x+v_{1,x})^2   \\
                                &   =  \sum n_x [ (v_x'-v_x)^2+ v_{1,x}^2 +2(v_x'-v_x)v_{1,x}]  \\
                                &   =  \sum n_x (v_x'-v_x)^2+ N_x v_{1,x}^2  \\
                                &   = N_x  ( \sigma_x^2+ v_{1,x}^2).
\end{split}                                
 \label{eq:dmom2c}
\end{equation}

 We have used equations~\ref{eq:dmom2a} in the above deduction. 
 Equation~\ref{eq:dmom2b} thus becomes Equation~\ref{eq:mom2}.

\section{Moment Images of the Original and Degraded THINGS Cubes}
\label{appendix:mom}
\begin{figure*} 
\centering
\includegraphics[width=5.5cm]{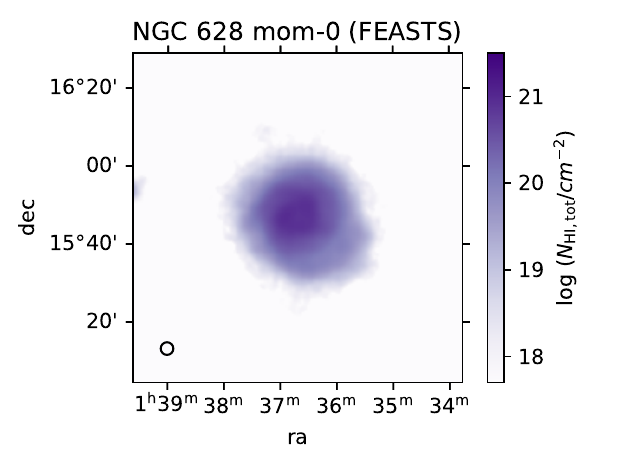}
\includegraphics[width=5.5cm]{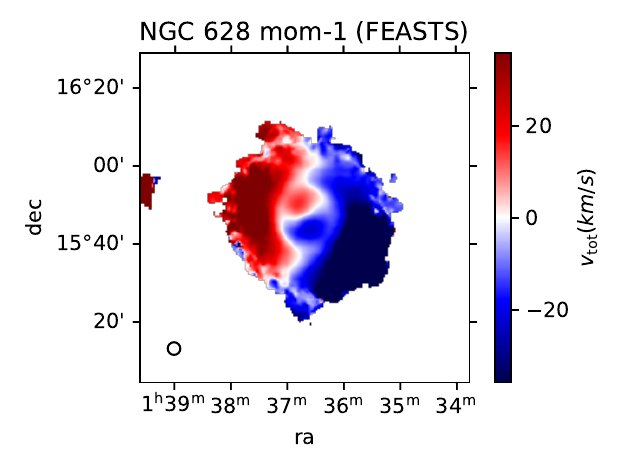}
\includegraphics[width=5.5cm]{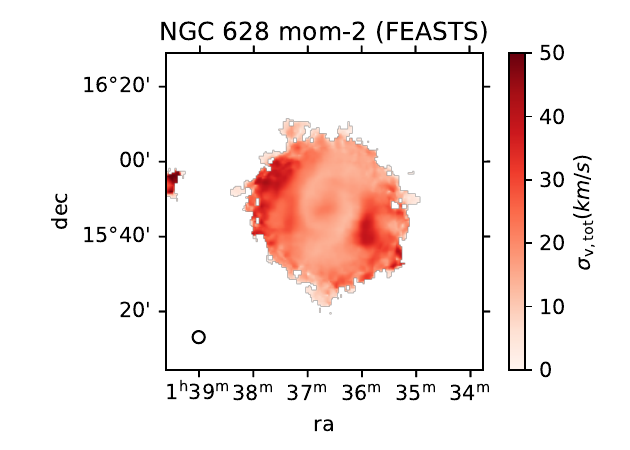}

\centering
\includegraphics[width=5.5cm]{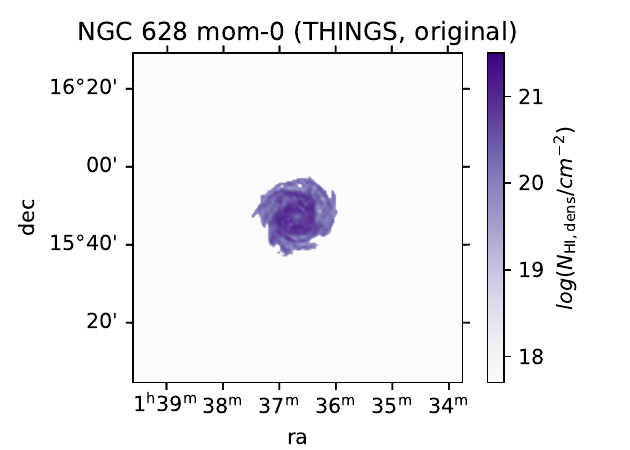}
\includegraphics[width=5.5cm]{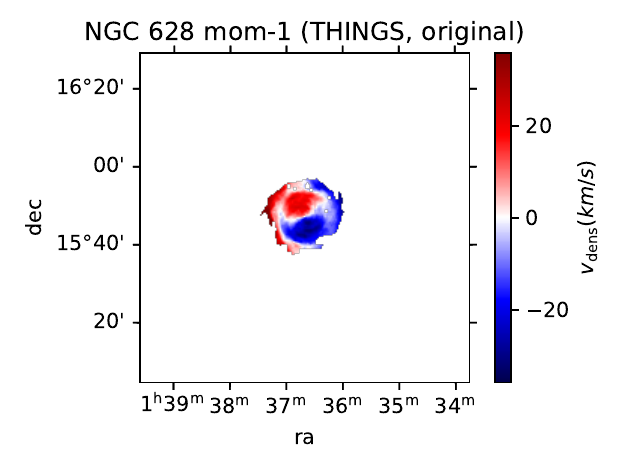}
\includegraphics[width=5.5cm]{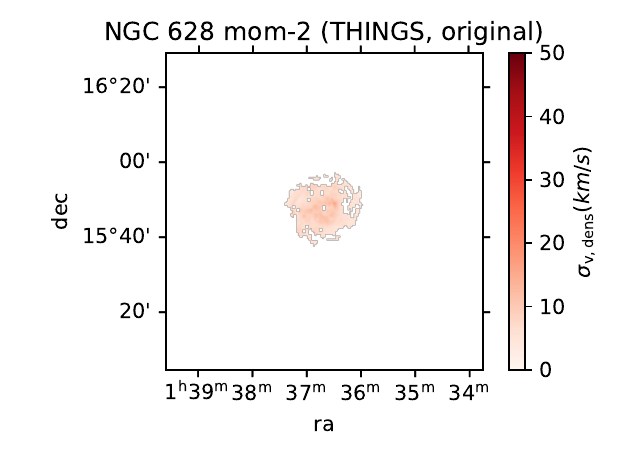}

\centering
\includegraphics[width=5.5cm]{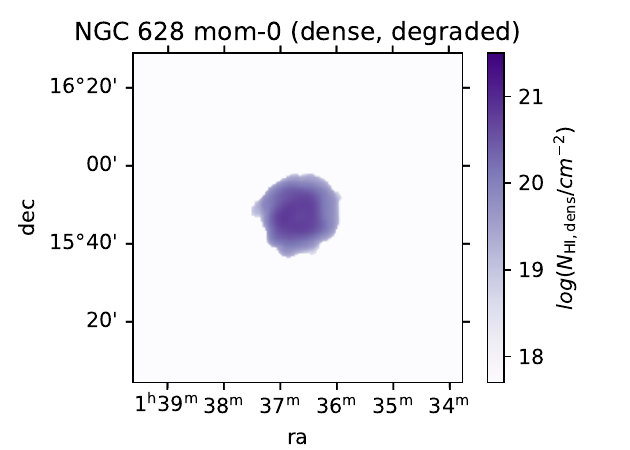}
\includegraphics[width=5.5cm]{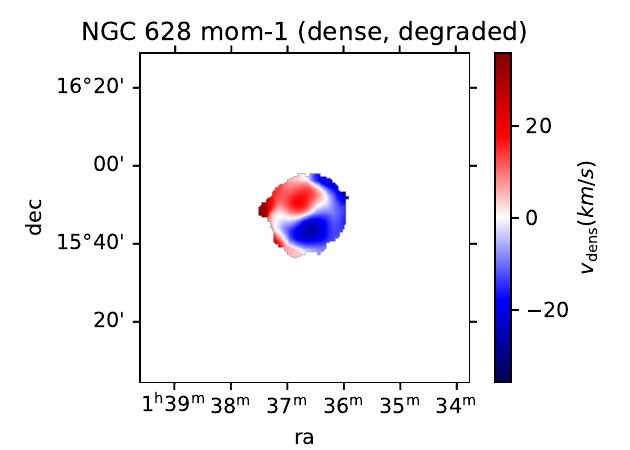}
\includegraphics[width=5.5cm]{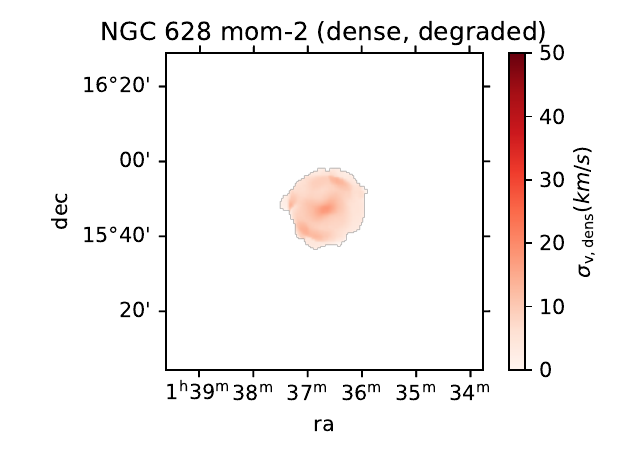}

\centering
\includegraphics[width=5.5cm]{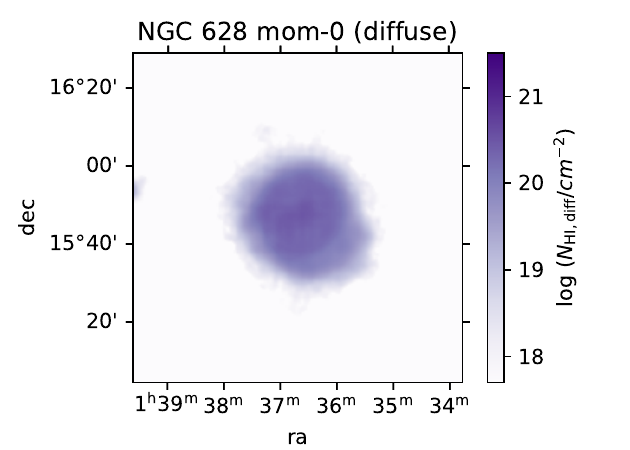}
\includegraphics[width=5.5cm]{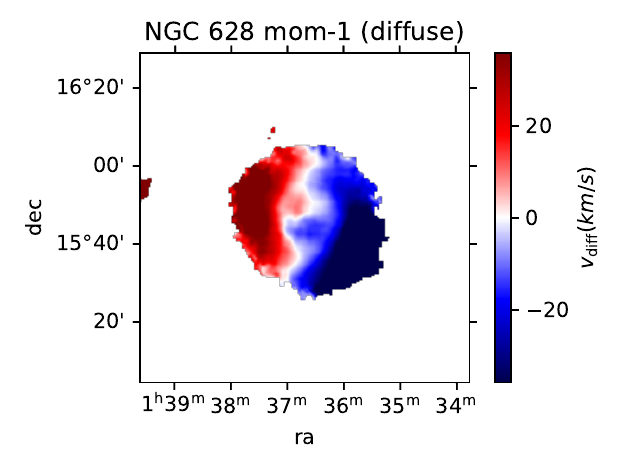}
\includegraphics[width=5.5cm]{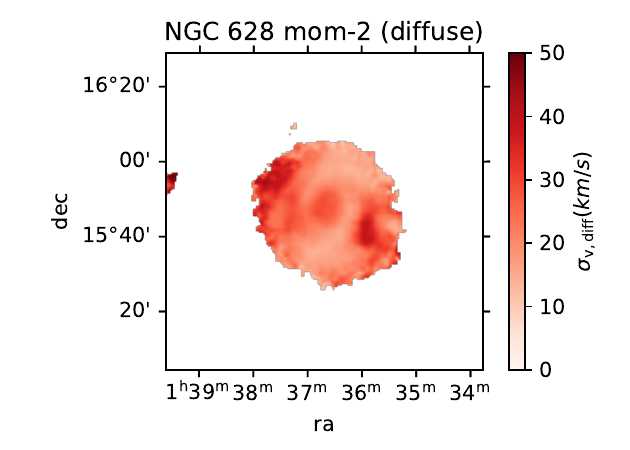}

\centering
\includegraphics[width=5.5cm]{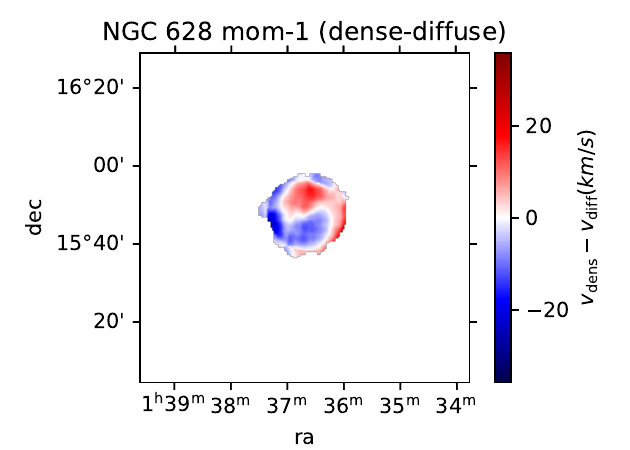}

\caption{The moment images of the galaxy NGC 628. The sky region displayed is the full region of the FEASTS data.
{\bf Row 1:} moment 0-2 images of FEASTS data, representing the total $\hi$ at the FAST resolution. The circle in the bottom left of the shows the size of the FAST beam.)
{\bf Row 2:} moment 0-2 images of THINGS data, representing the dense $\hi$ at the VLA resolution. We do not show the VLA synthesis beam as it is too small compared to the field of view.
{\bf Row 3:} moment 0-2 images of dense $\hi$ at the FAST resolution. 
{\bf Row 4:} moment 0-2 images of the total $\hi$ minus the dense $\hi$, representing the diffuse $\hi$ at the FAST resolution. 
{\bf Row 5:} the difference in $\hi$ moment-1 images between the dense $\hi$ and diffuse $hi$ at the FAST resolution (the former minus the latter).  }
\label{fig:mom_n628}
\end{figure*}

\begin{figure*} 
\centering
\includegraphics[width=5.5cm]{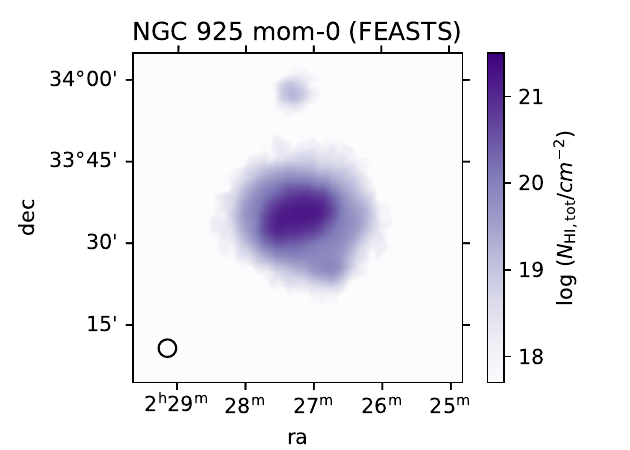}
\includegraphics[width=5.5cm]{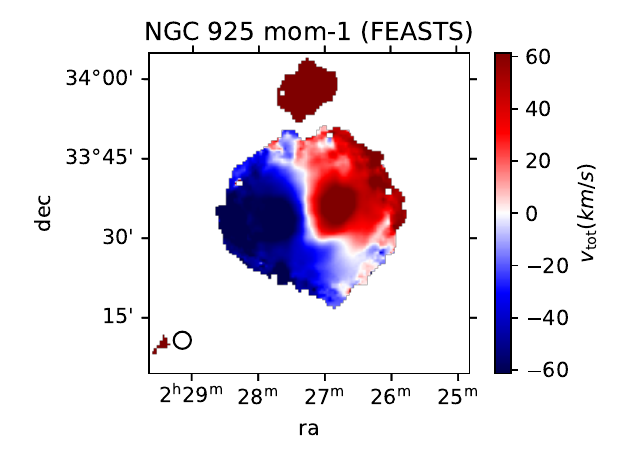}
\includegraphics[width=5.5cm]{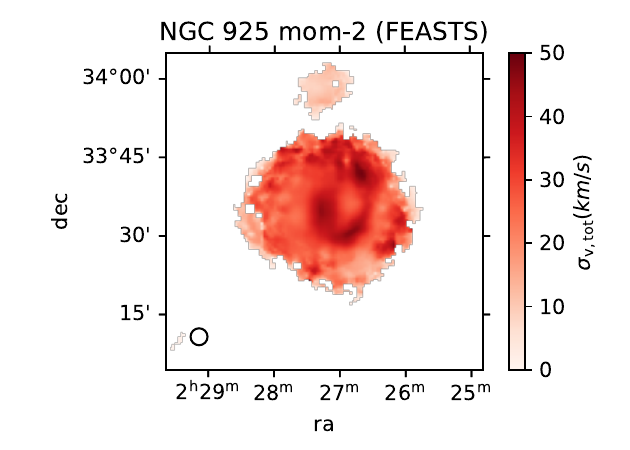}

\centering
\includegraphics[width=5.5cm]{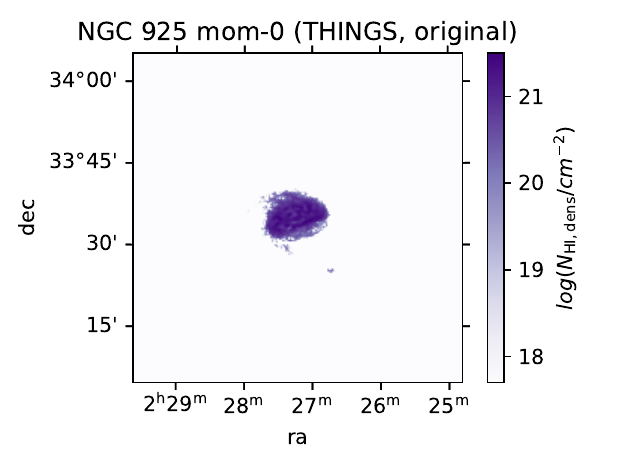}
\includegraphics[width=5.5cm]{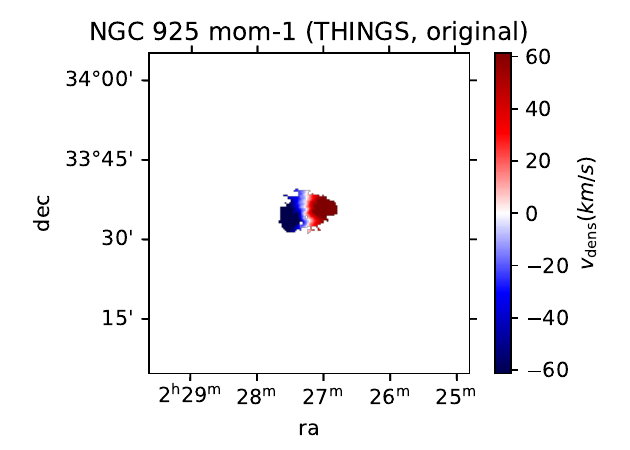}
\includegraphics[width=5.5cm]{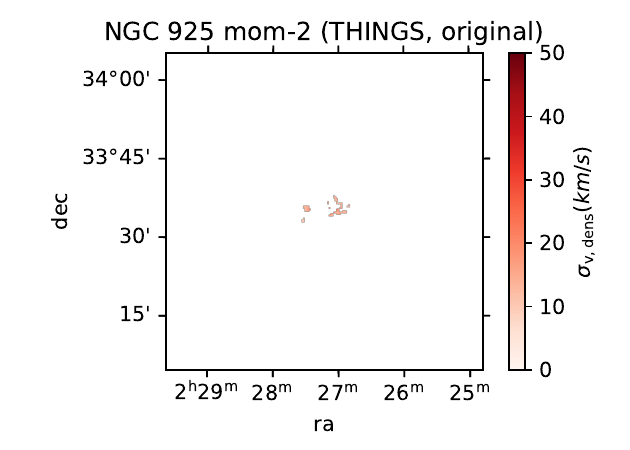}

\centering
\includegraphics[width=5.5cm]{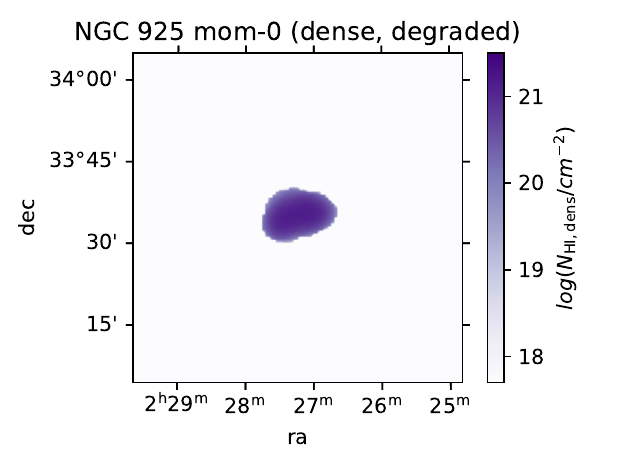}
\includegraphics[width=5.5cm]{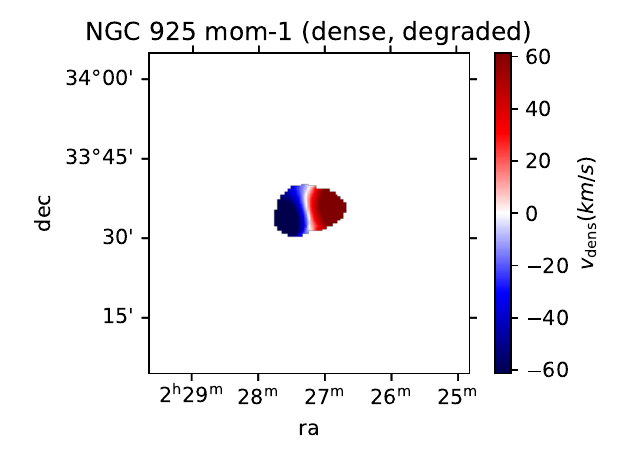}
\includegraphics[width=5.5cm]{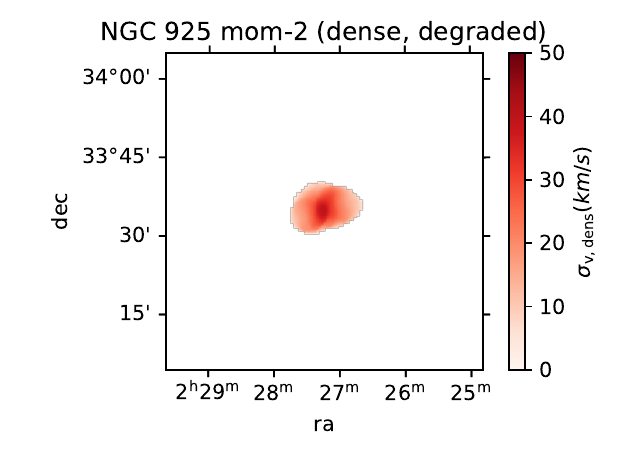}

\centering
\includegraphics[width=5.5cm]{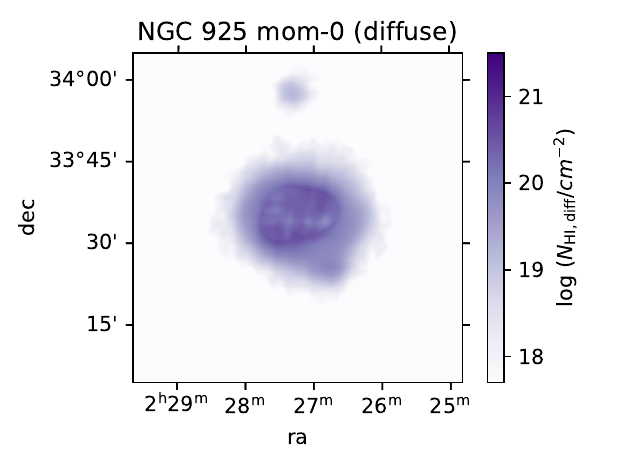}
\includegraphics[width=5.5cm]{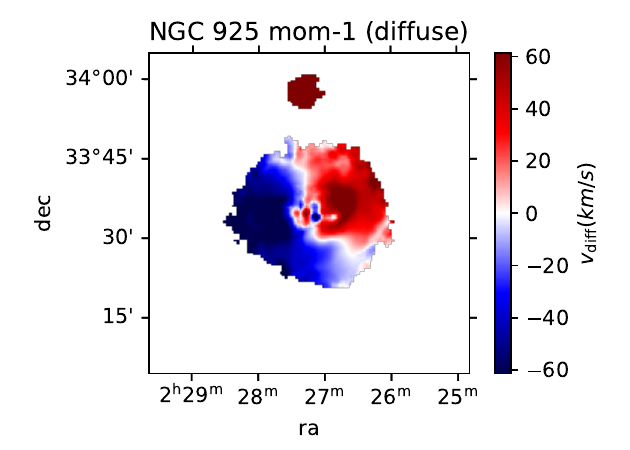}
\includegraphics[width=5.5cm]{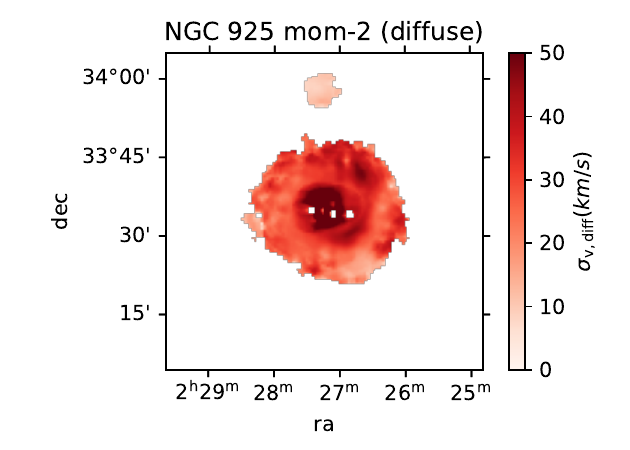}

\centering
\includegraphics[width=5.5cm]{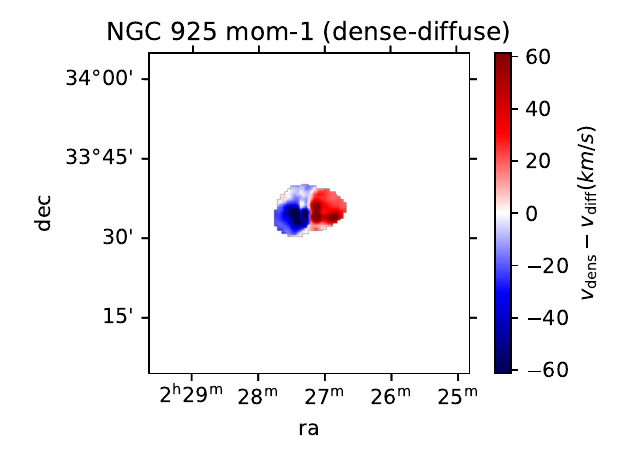}

\caption{Same as Figure~\ref{fig:mom_n628}, but for the galaxy NGC 925. }
\label{fig:mom_n925}
\end{figure*}

\begin{figure*} 
\centering
\includegraphics[width=5.5cm]{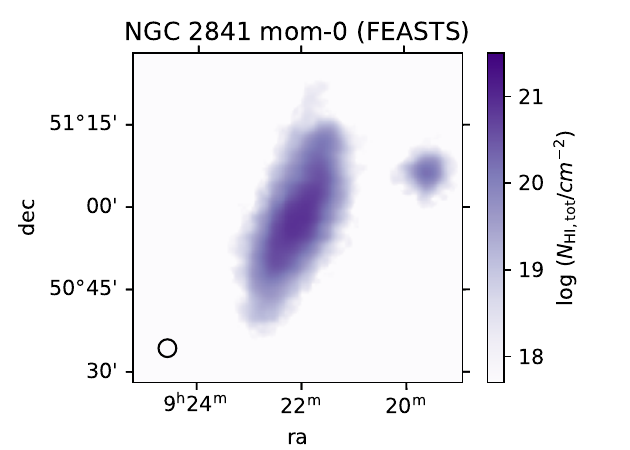}
\includegraphics[width=5.5cm]{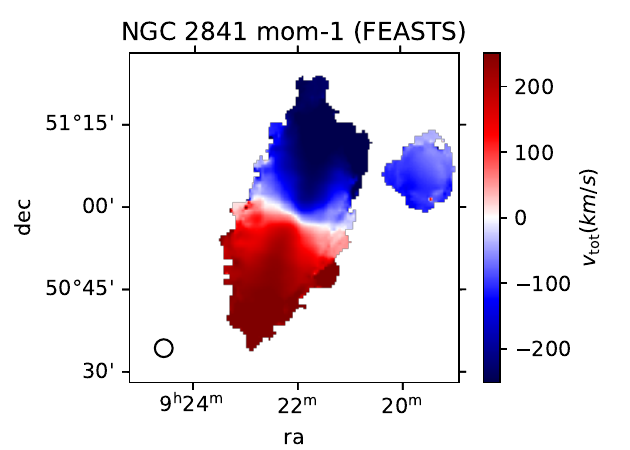}
\includegraphics[width=5.5cm]{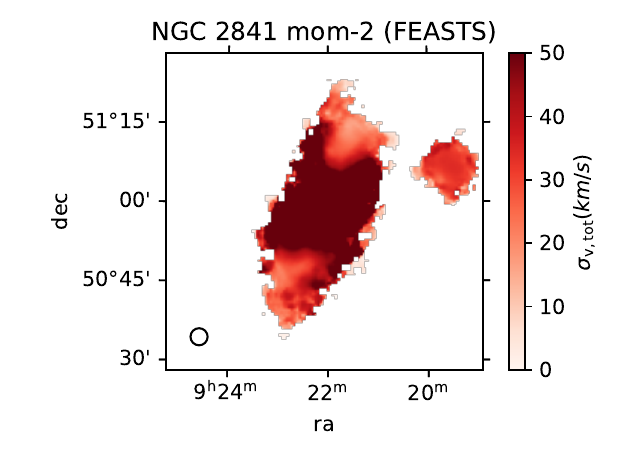}

\centering
\includegraphics[width=5.5cm]{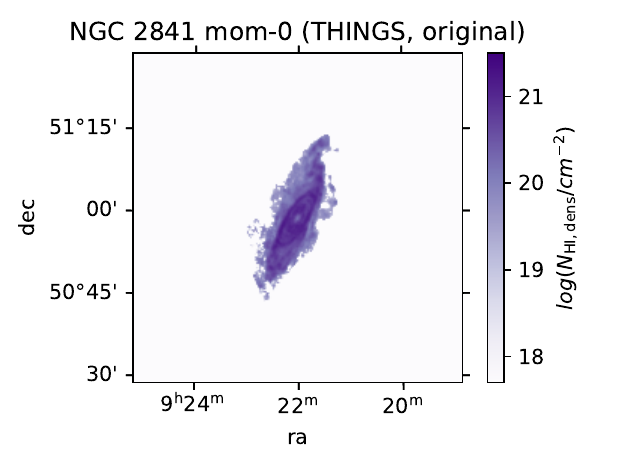}
\includegraphics[width=5.5cm]{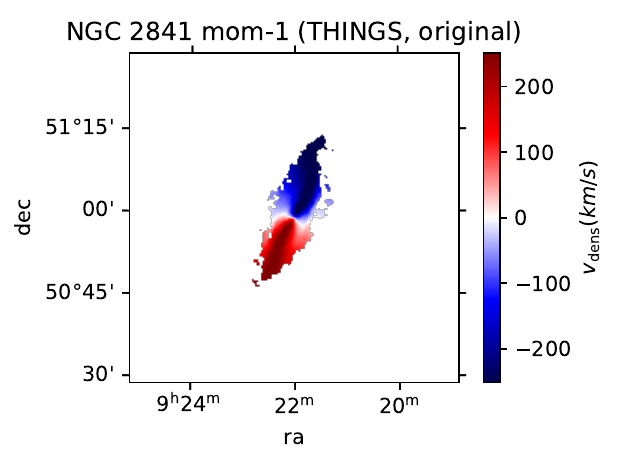}
\includegraphics[width=5.5cm]{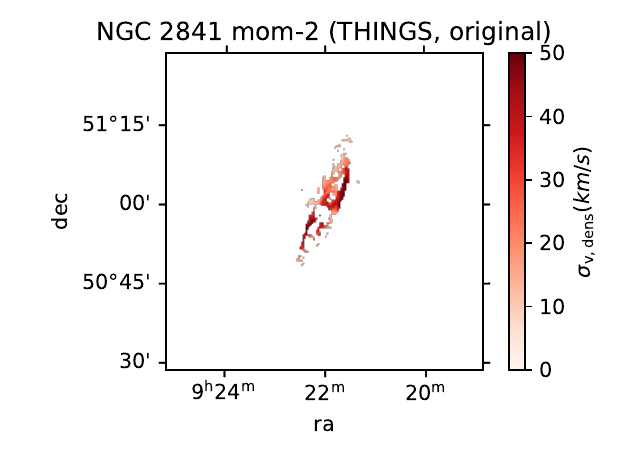}

\centering
\includegraphics[width=5.5cm]{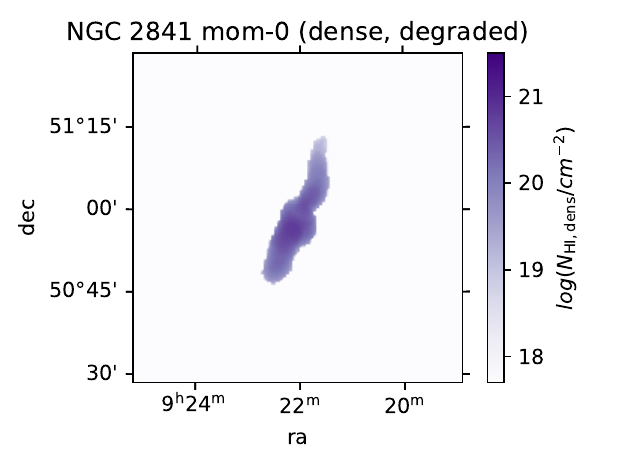}
\includegraphics[width=5.5cm]{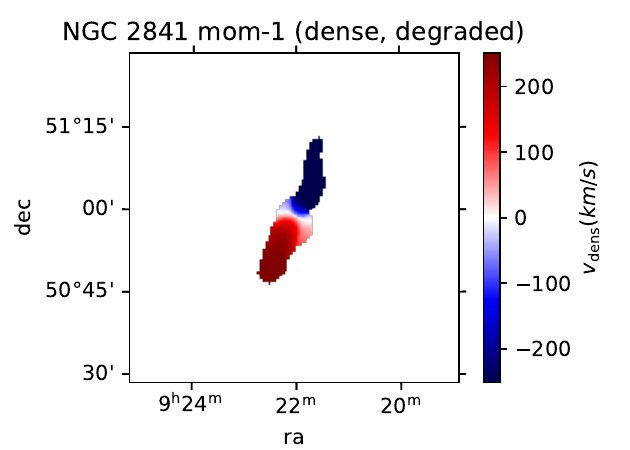}
\includegraphics[width=5.5cm]{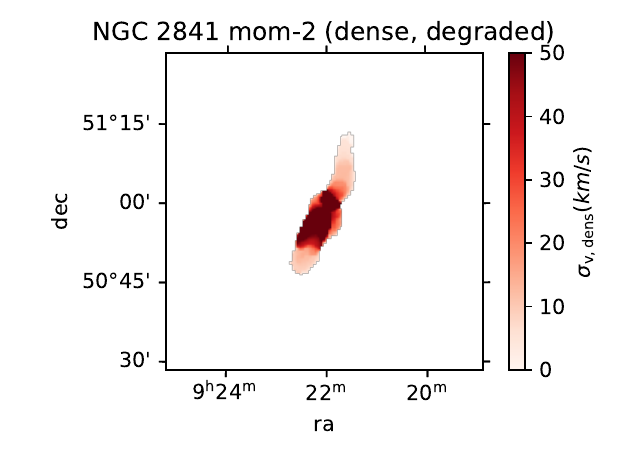}

\centering
\includegraphics[width=5.5cm]{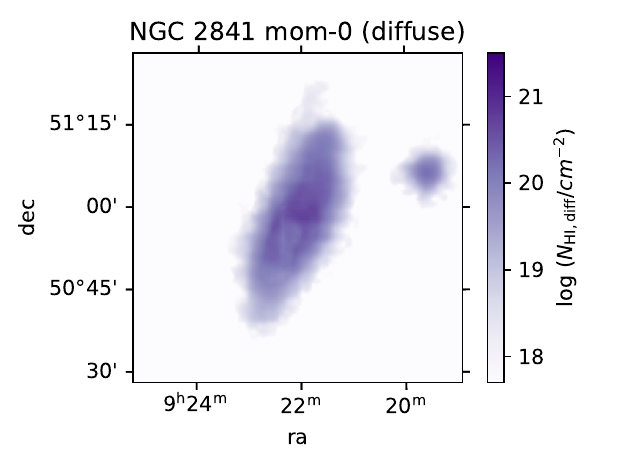}
\includegraphics[width=5.5cm]{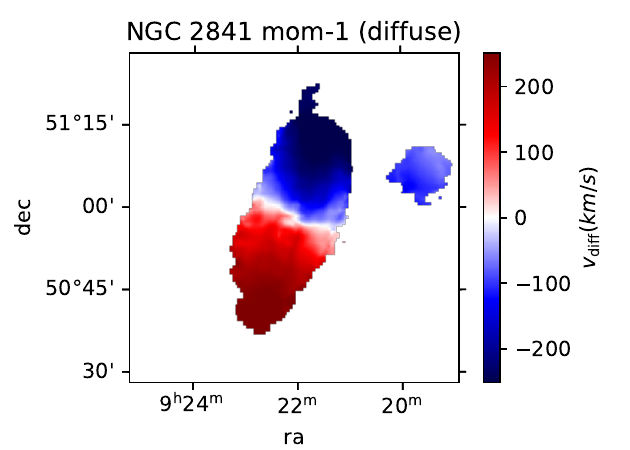}
\includegraphics[width=5.5cm]{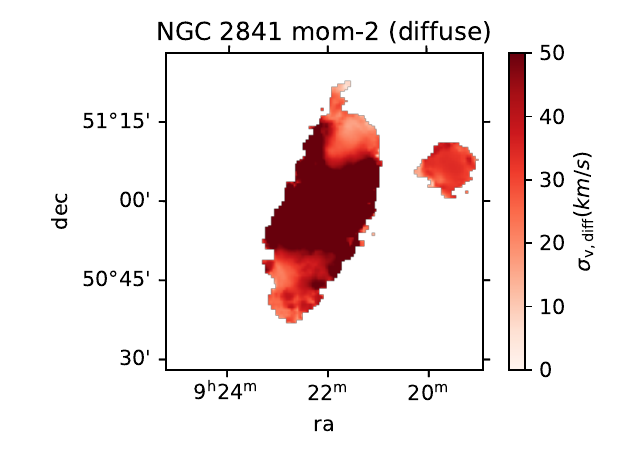}

\centering
\includegraphics[width=5.5cm]{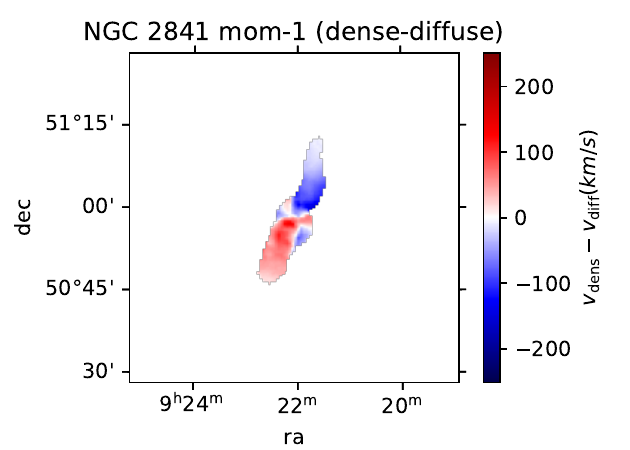}

\caption{Same as Figure~\ref{fig:mom_n628}, but for the galaxy NGC 2841. }
\label{fig:mom_n2841}
\end{figure*}

\begin{figure*} 
\centering
\includegraphics[width=5.5cm]{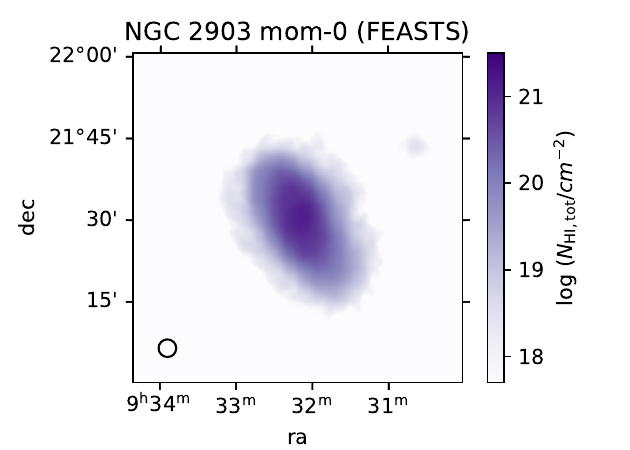}
\includegraphics[width=5.5cm]{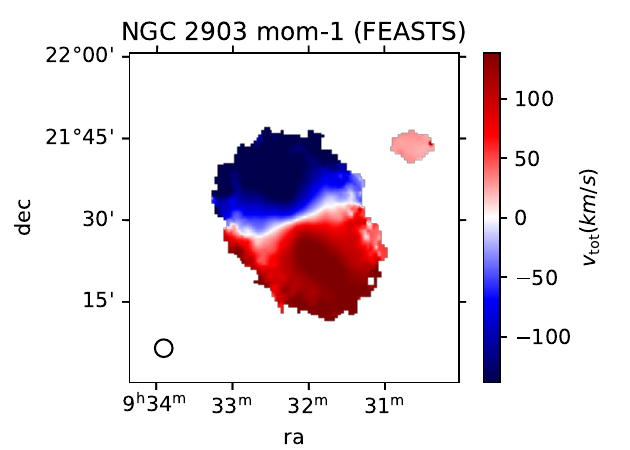}
\includegraphics[width=5.5cm]{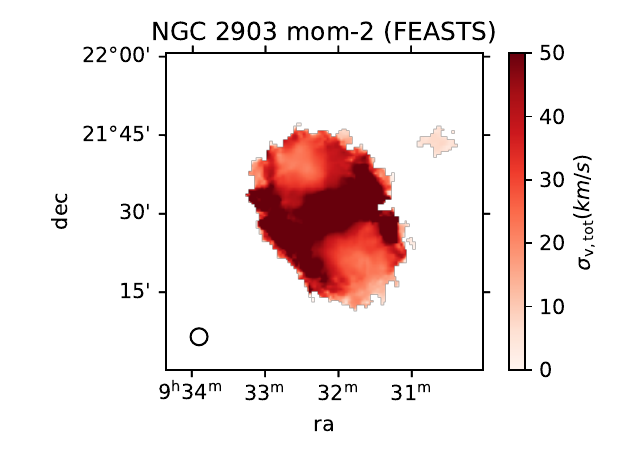}

\centering
\includegraphics[width=5.5cm]{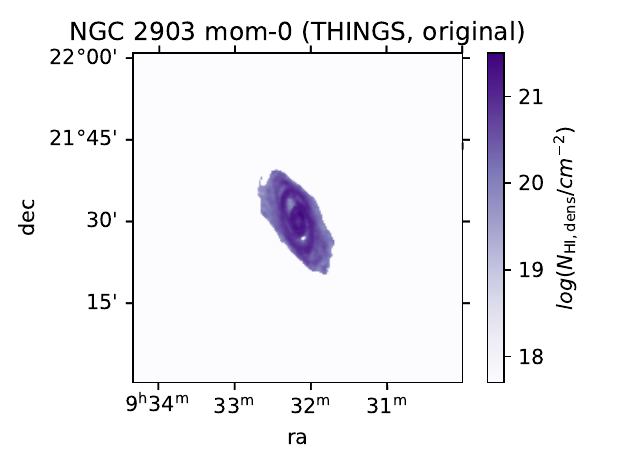}
\includegraphics[width=5.5cm]{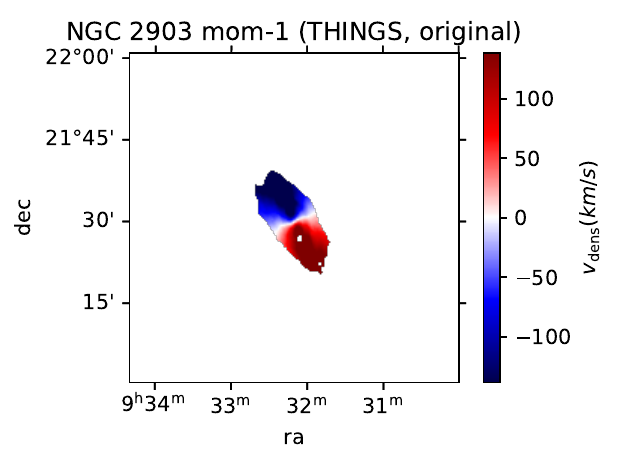}
\includegraphics[width=5.5cm]{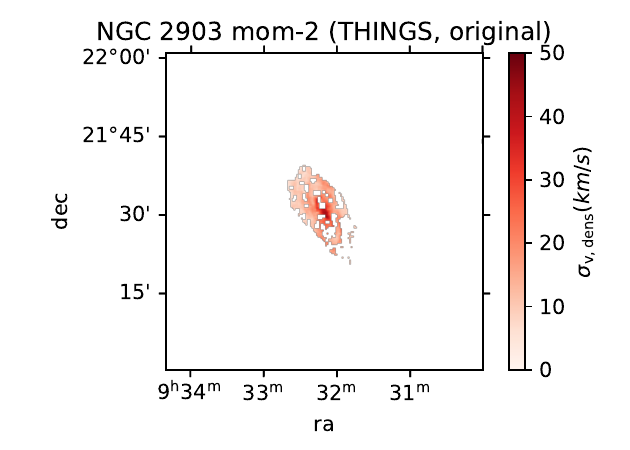}

\centering
\includegraphics[width=5.5cm]{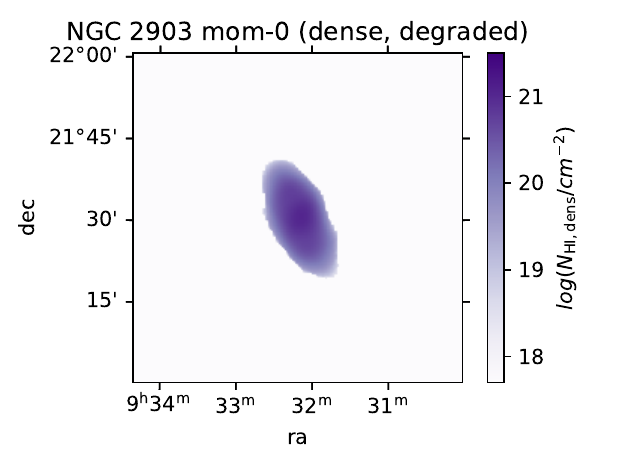}
\includegraphics[width=5.5cm]{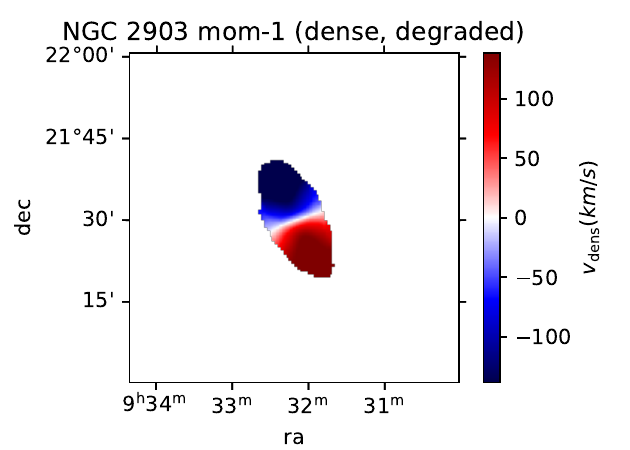}
\includegraphics[width=5.5cm]{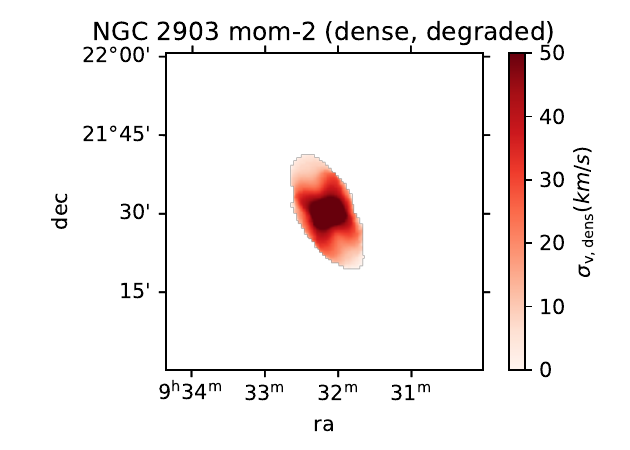}

\centering
\includegraphics[width=5.5cm]{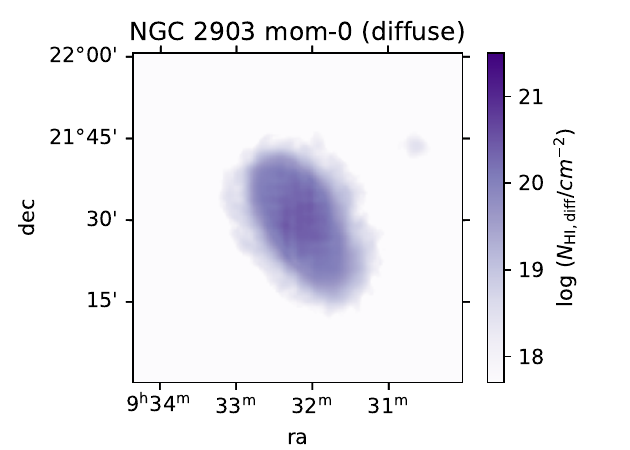}
\includegraphics[width=5.5cm]{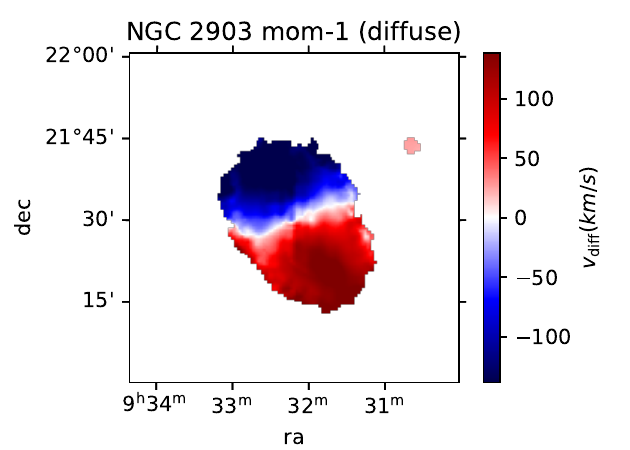}
\includegraphics[width=5.5cm]{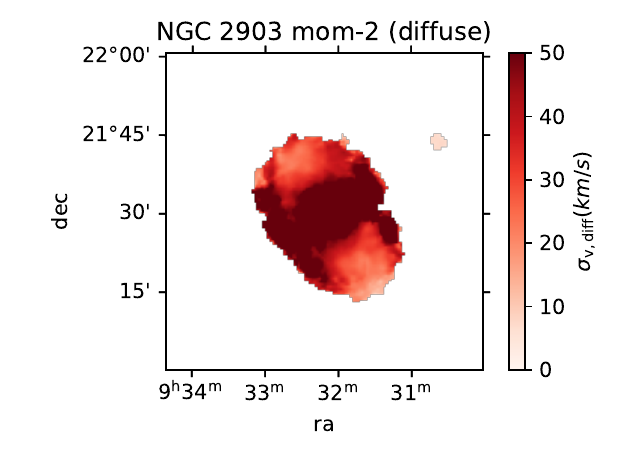}

\centering
\includegraphics[width=5.5cm]{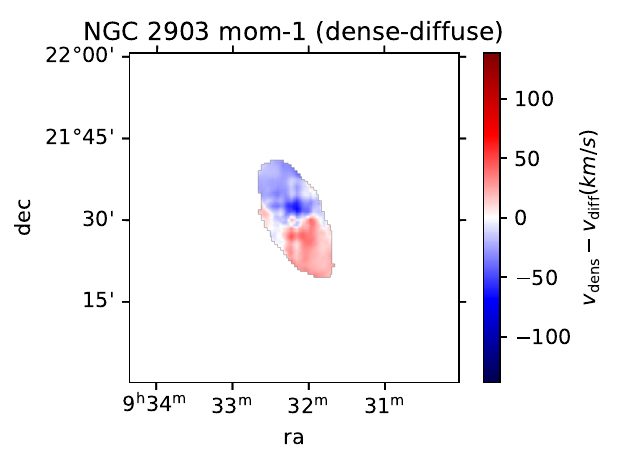}

\caption{Same as Figure~\ref{fig:mom_n628}, but for the galaxy NGC 2903. }
\label{fig:mom_n2903}
\end{figure*}

\begin{figure*} 
\centering
\includegraphics[width=5.5cm]{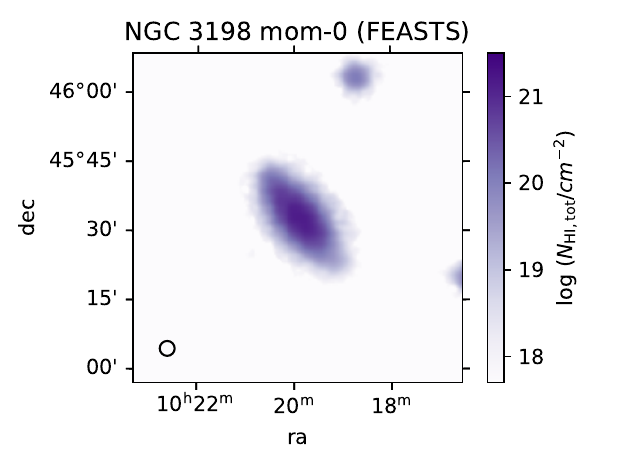}
\includegraphics[width=5.5cm]{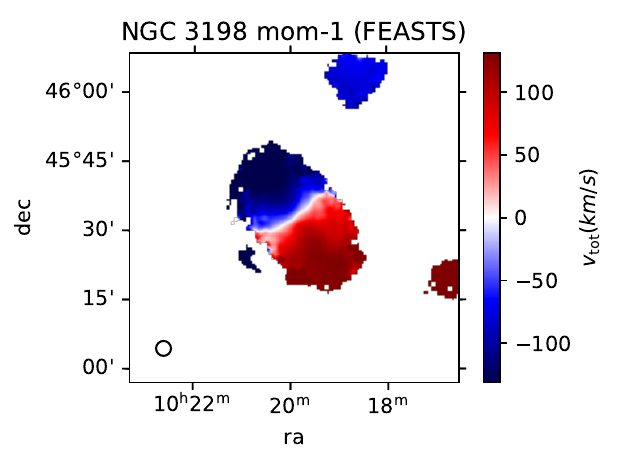}
\includegraphics[width=5.5cm]{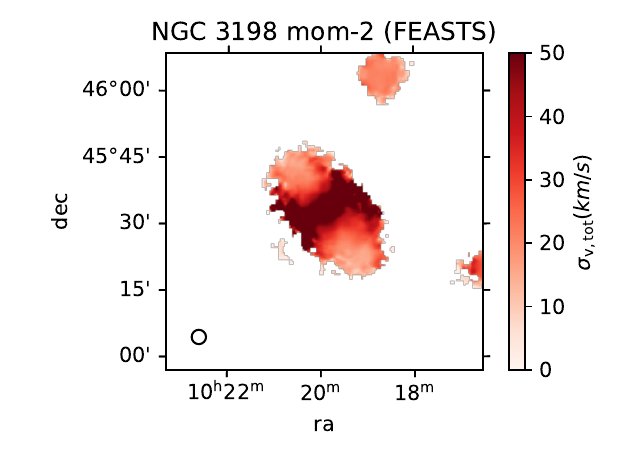}

\centering
\includegraphics[width=5.5cm]{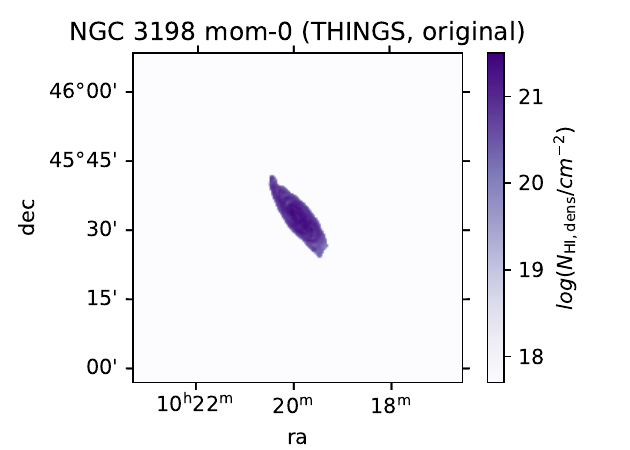}
\includegraphics[width=5.5cm]{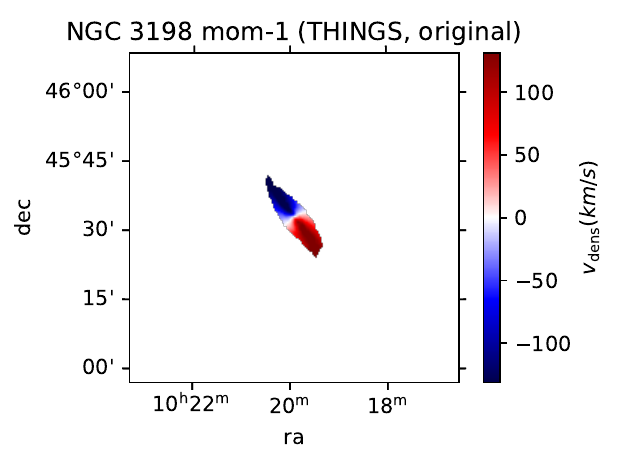}
\includegraphics[width=5.5cm]{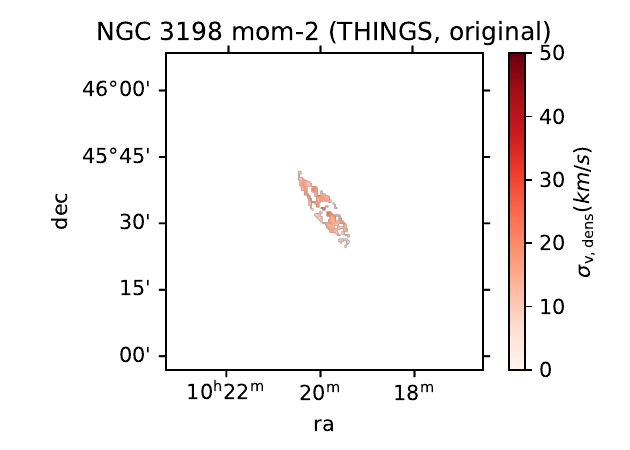}

\centering
\includegraphics[width=5.5cm]{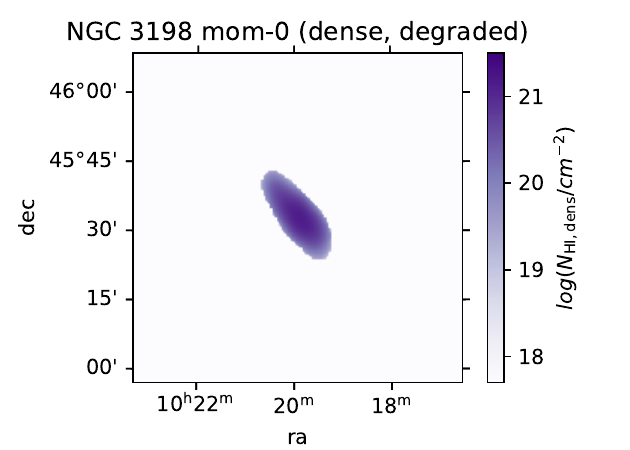}
\includegraphics[width=5.5cm]{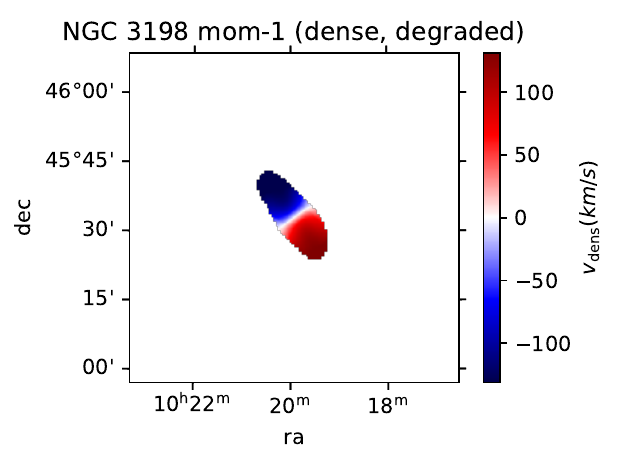}
\includegraphics[width=5.5cm]{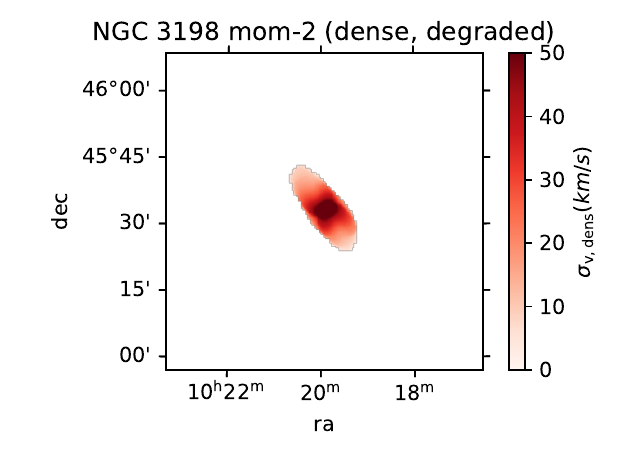}

\centering
\includegraphics[width=5.5cm]{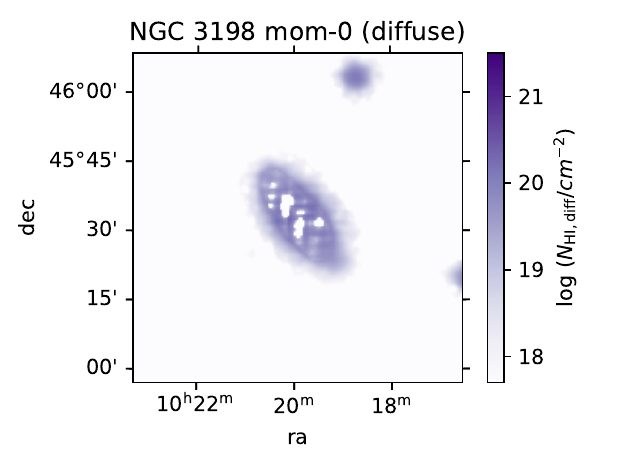}
\includegraphics[width=5.5cm]{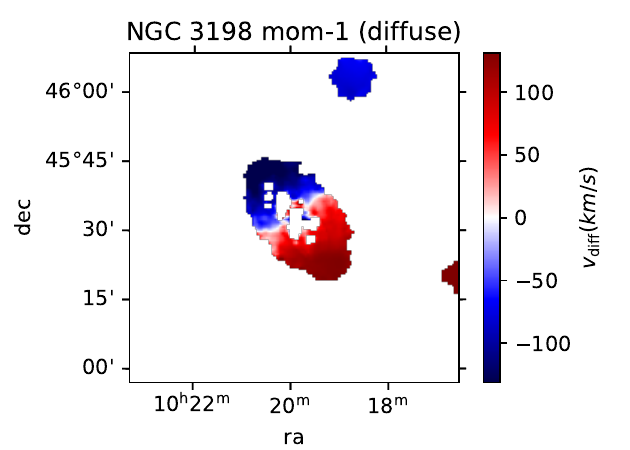}
\includegraphics[width=5.5cm]{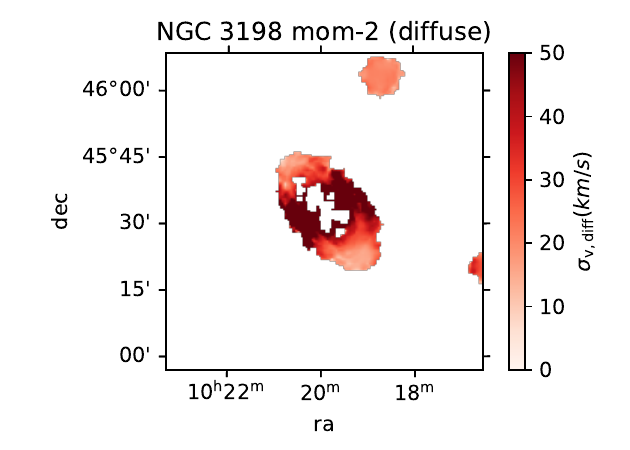}

\centering
\includegraphics[width=5.5cm]{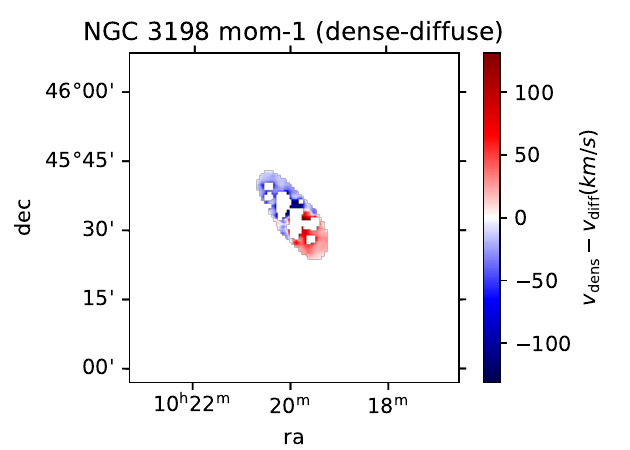}

\caption{Same as Figure~\ref{fig:mom_n628}, but for the galaxy NGC 3198. }
\label{fig:mom_n3198}
\end{figure*}

\begin{figure*} 
\centering
\includegraphics[width=5.5cm]{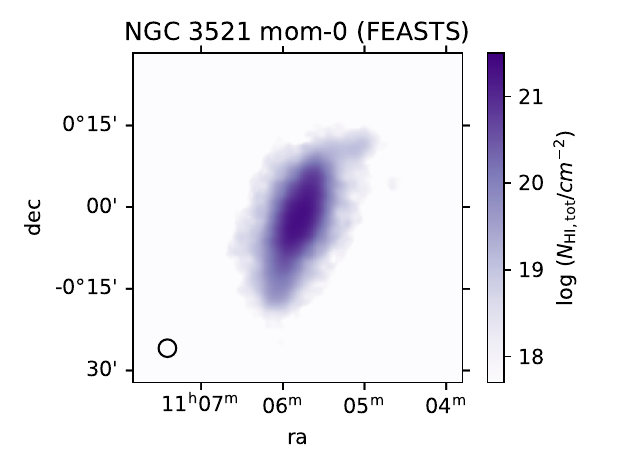}
\includegraphics[width=5.5cm]{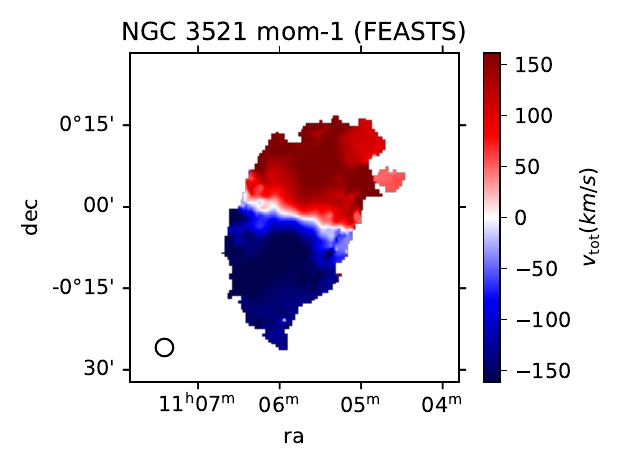}
\includegraphics[width=5.5cm]{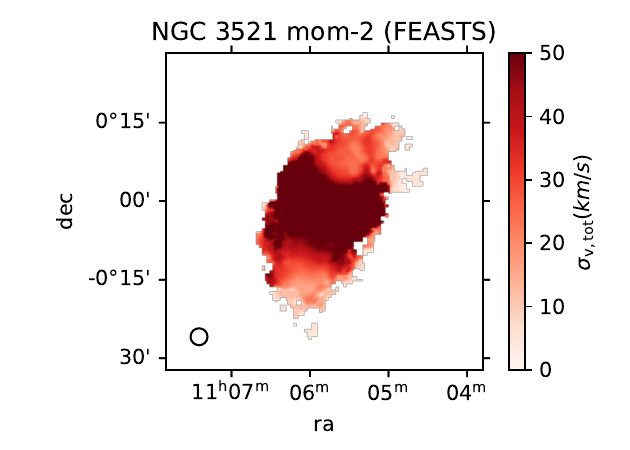}

\centering
\includegraphics[width=5.5cm]{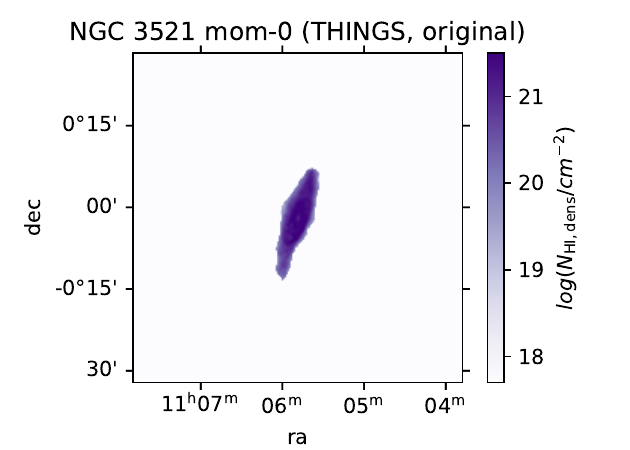}
\includegraphics[width=5.5cm]{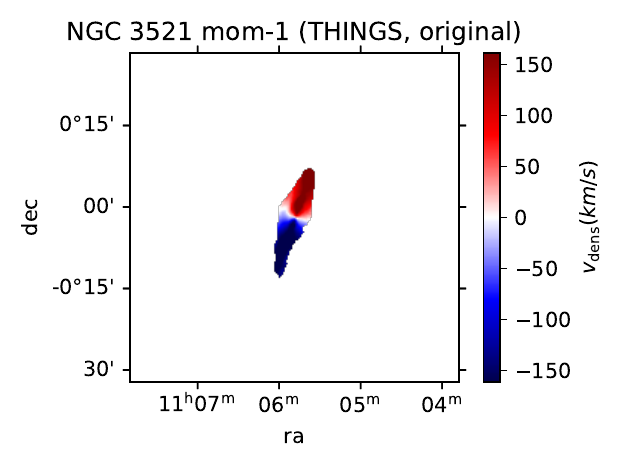}
\includegraphics[width=5.5cm]{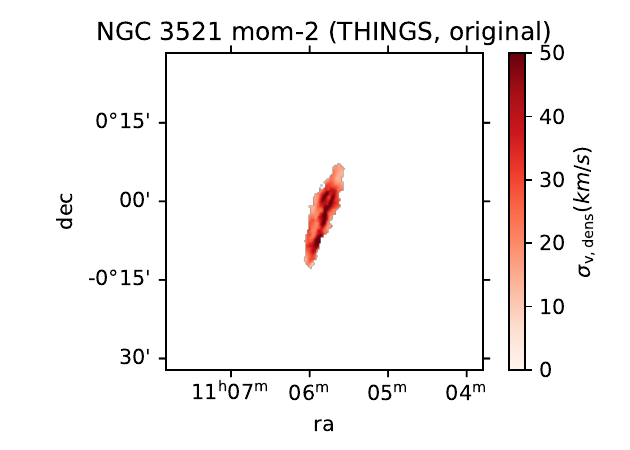}

\centering
\includegraphics[width=5.5cm]{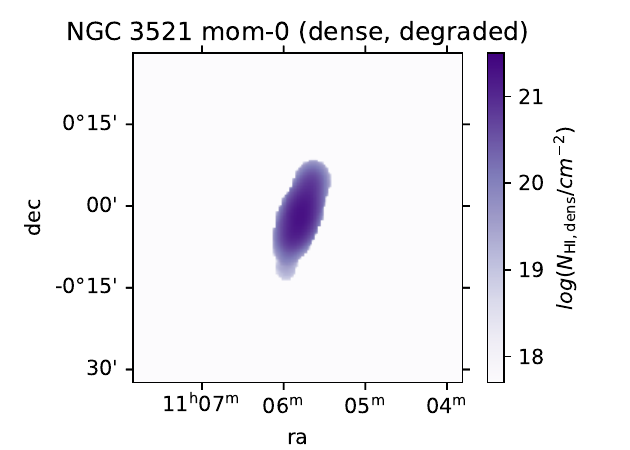}
\includegraphics[width=5.5cm]{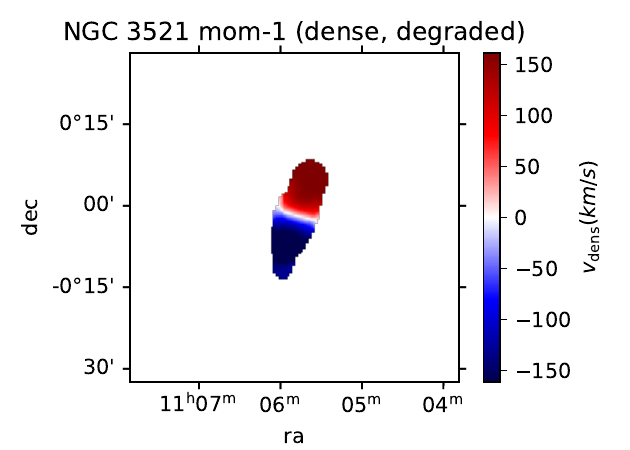}
\includegraphics[width=5.5cm]{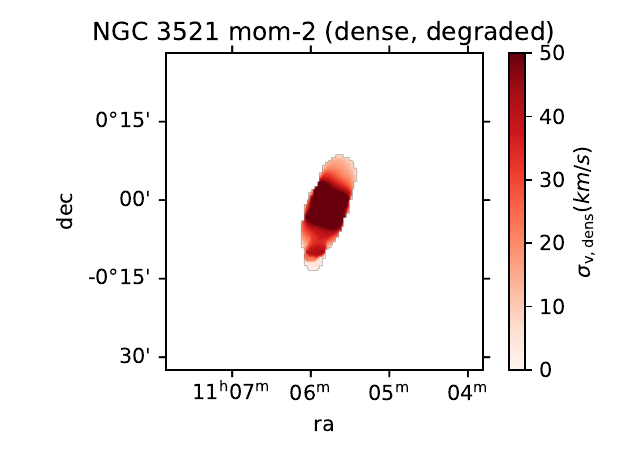}

\centering
\includegraphics[width=5.5cm]{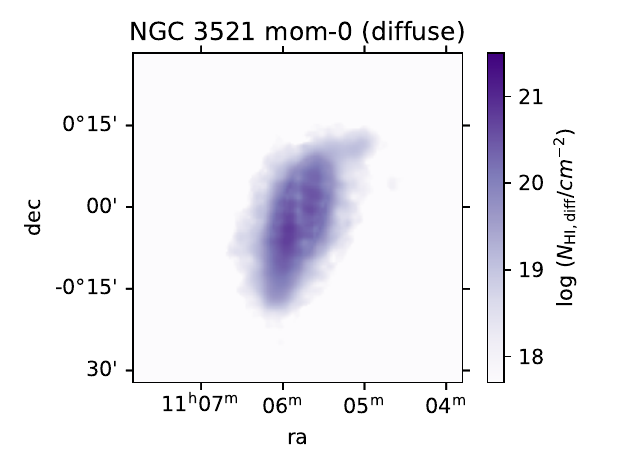}
\includegraphics[width=5.5cm]{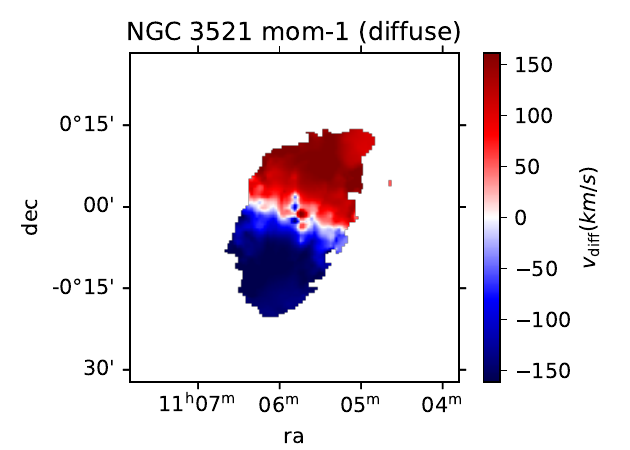}
\includegraphics[width=5.5cm]{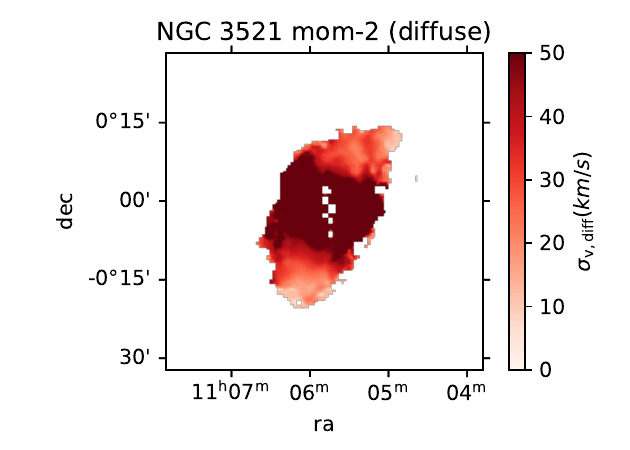}

\centering
\includegraphics[width=5.5cm]{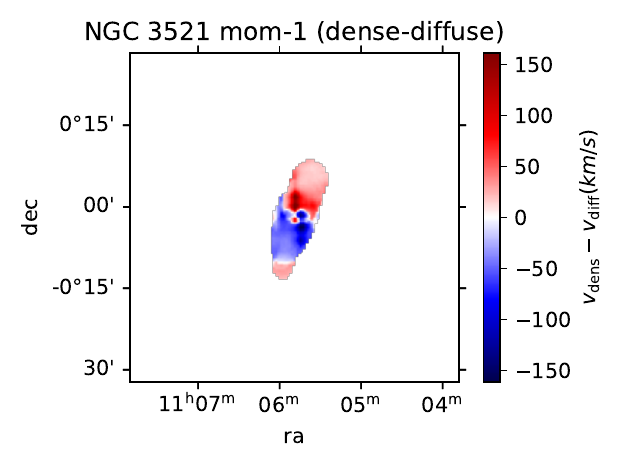}

\caption{Same as Figure~\ref{fig:mom_n628}, but for the galaxy NGC 3521. }
\label{fig:mom_n3521}
\end{figure*}

\begin{figure*} 
\centering
\includegraphics[width=5.5cm]{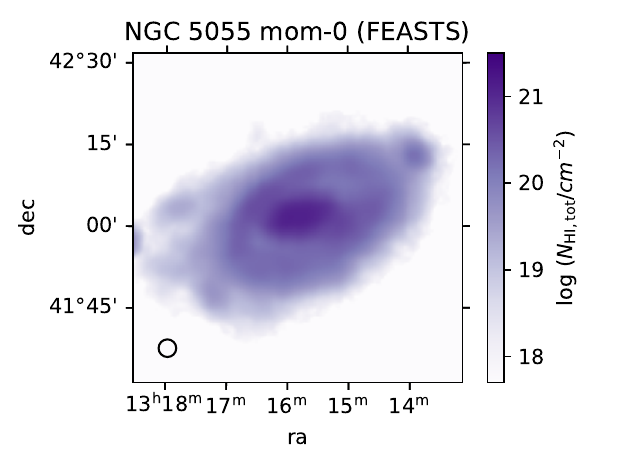}
\includegraphics[width=5.5cm]{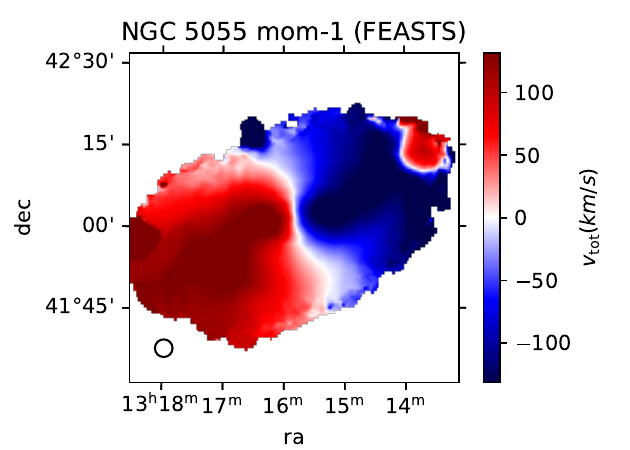}
\includegraphics[width=5.5cm]{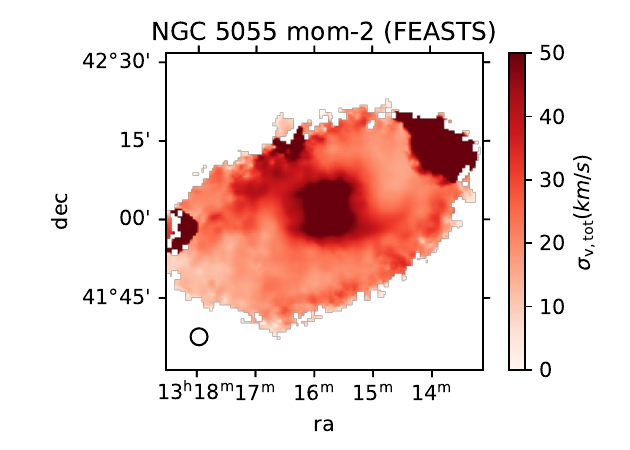}

\centering
\includegraphics[width=5.5cm]{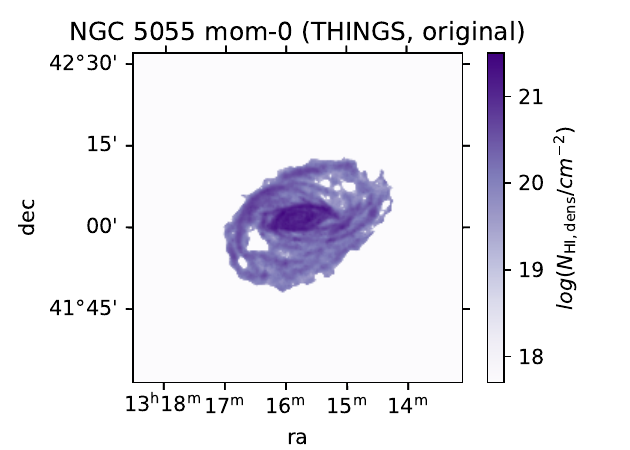}
\includegraphics[width=5.5cm]{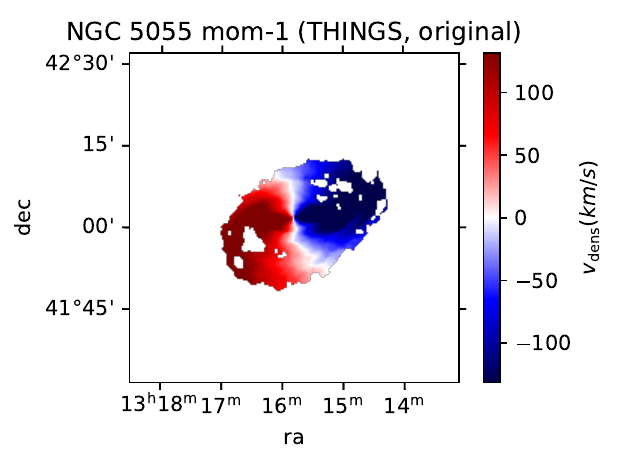}
\includegraphics[width=5.5cm]{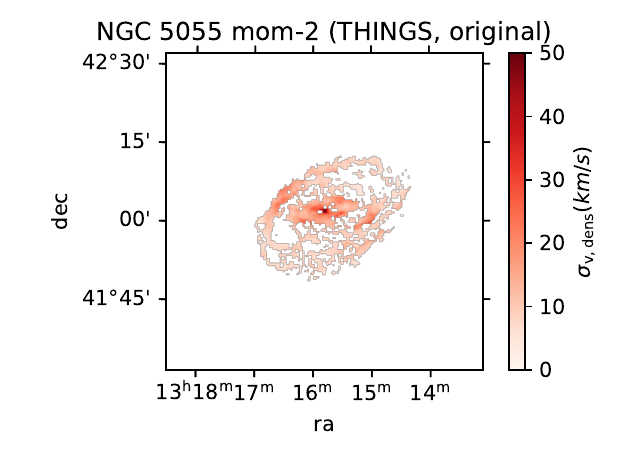}

\centering
\includegraphics[width=5.5cm]{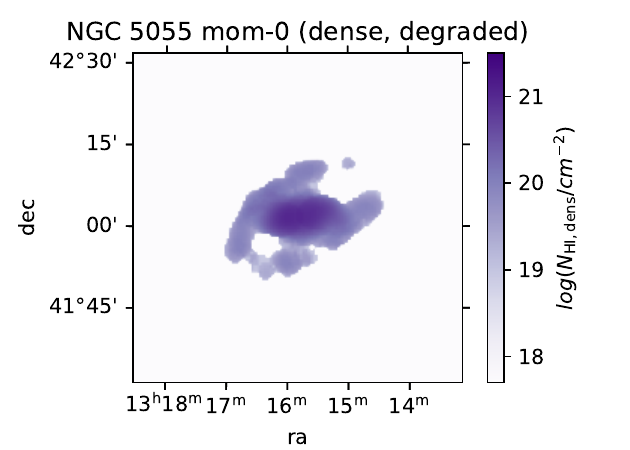}
\includegraphics[width=5.5cm]{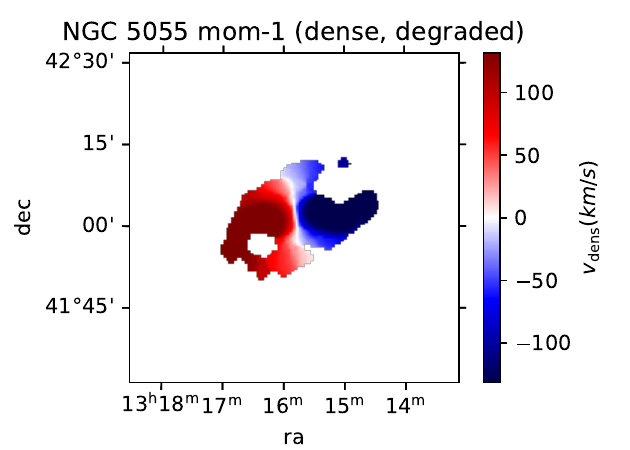}
\includegraphics[width=5.5cm]{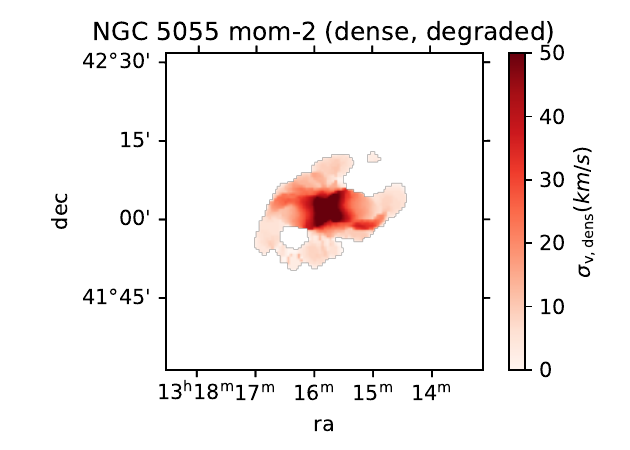}

\centering
\includegraphics[width=5.5cm]{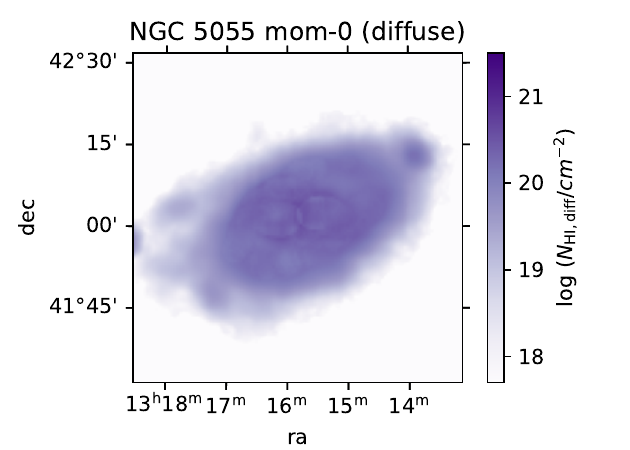}
\includegraphics[width=5.5cm]{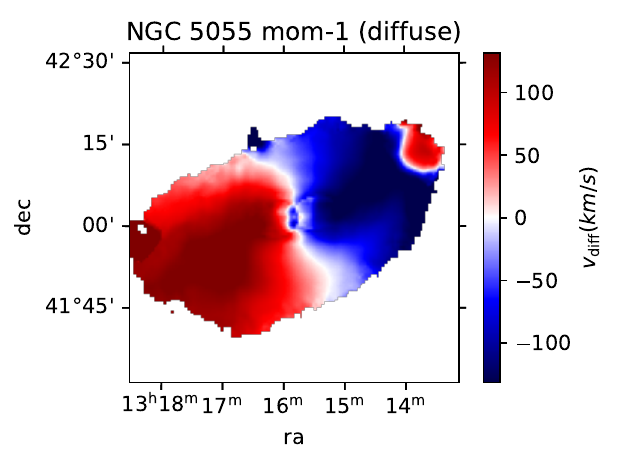}
\includegraphics[width=5.5cm]{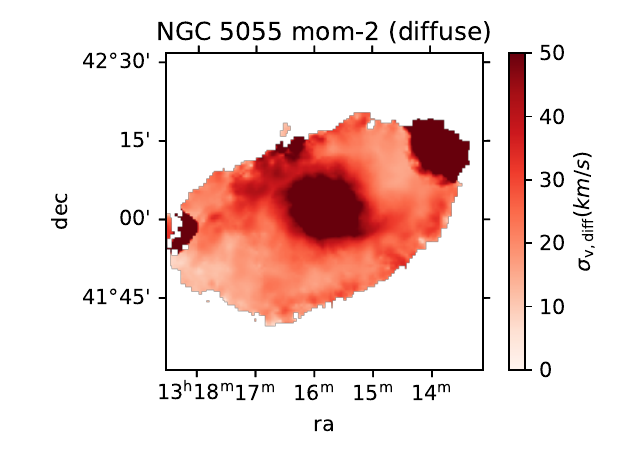}

\centering
\includegraphics[width=5.5cm]{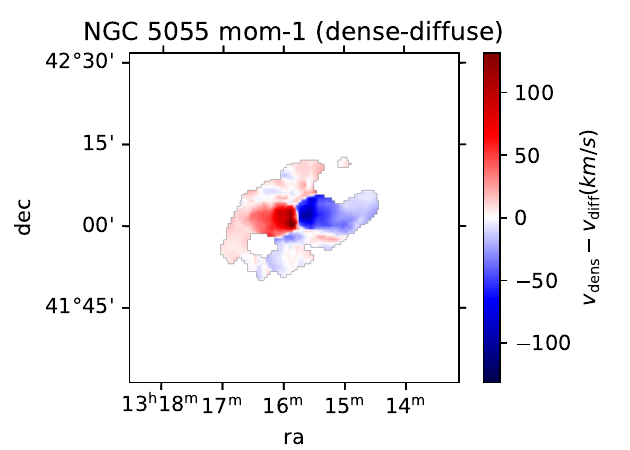}

\caption{Same as Figure~\ref{fig:mom_n628}, but for the galaxy NGC 5055. }
\label{fig:mom_n5055}
\end{figure*}

\begin{figure*} 
\centering
\includegraphics[width=5.5cm]{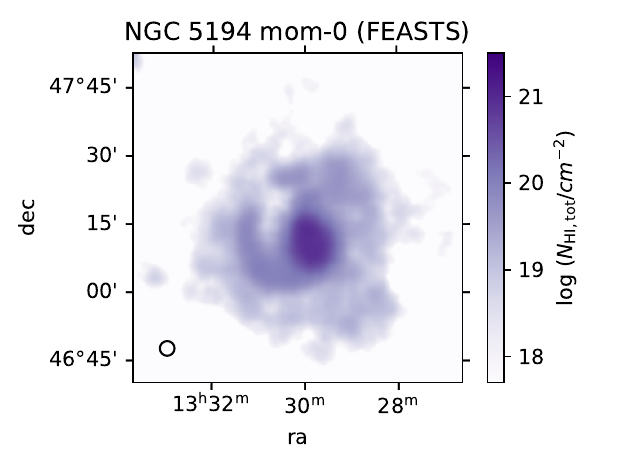}
\includegraphics[width=5.5cm]{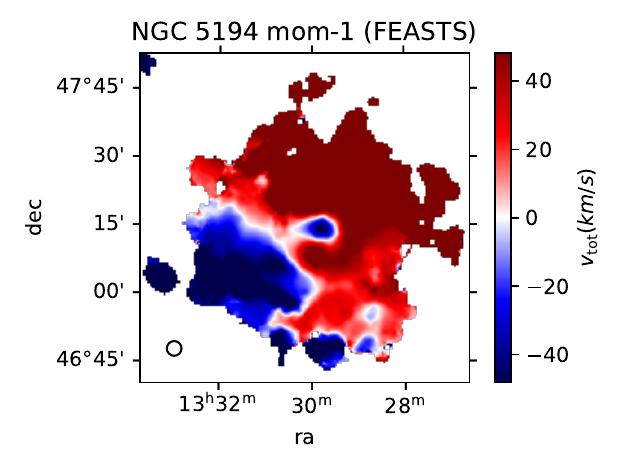}
\includegraphics[width=5.5cm]{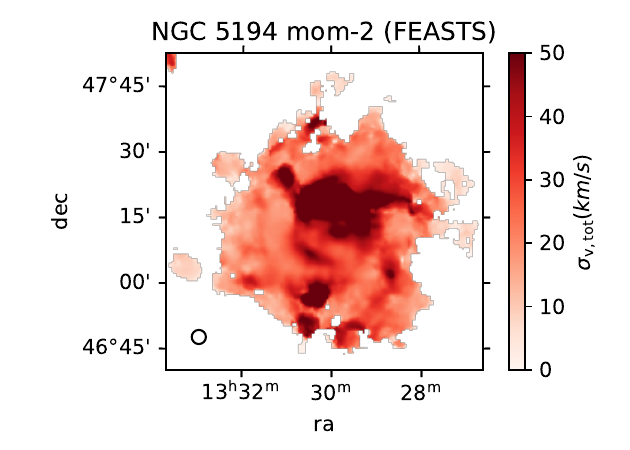}

\centering
\includegraphics[width=5.5cm]{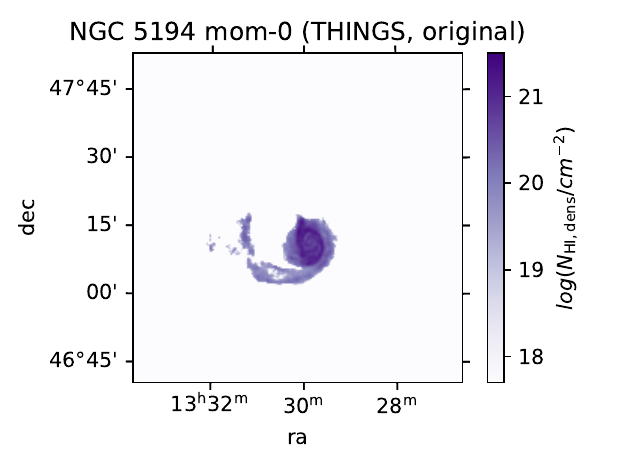}
\includegraphics[width=5.5cm]{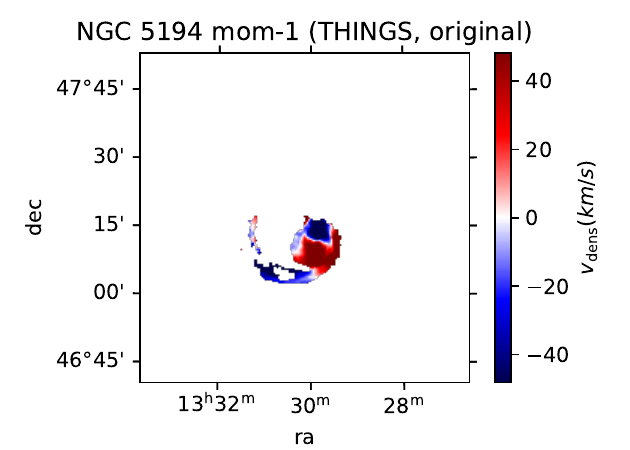}
\includegraphics[width=5.5cm]{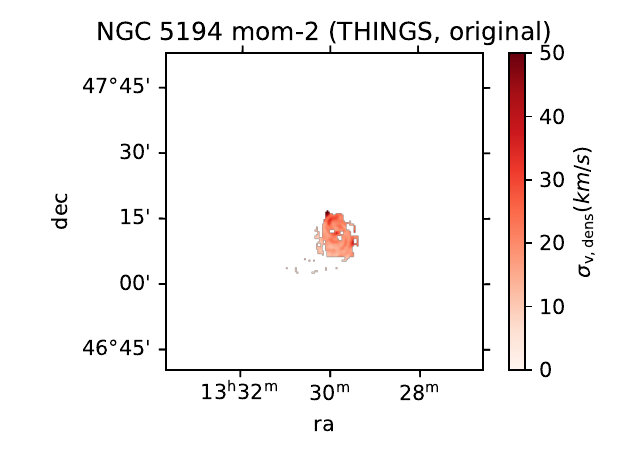}

\centering
\includegraphics[width=5.5cm]{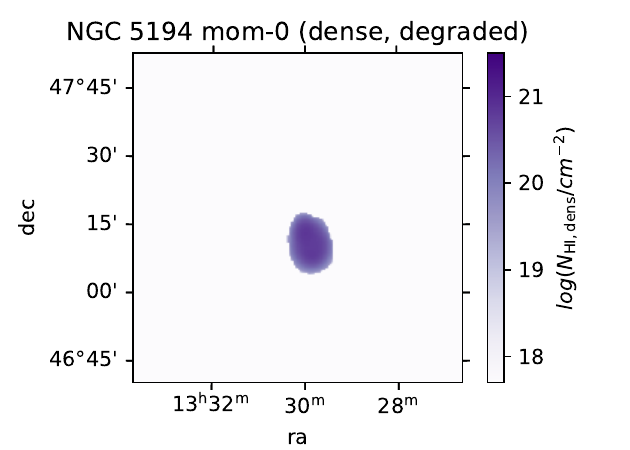}
\includegraphics[width=5.5cm]{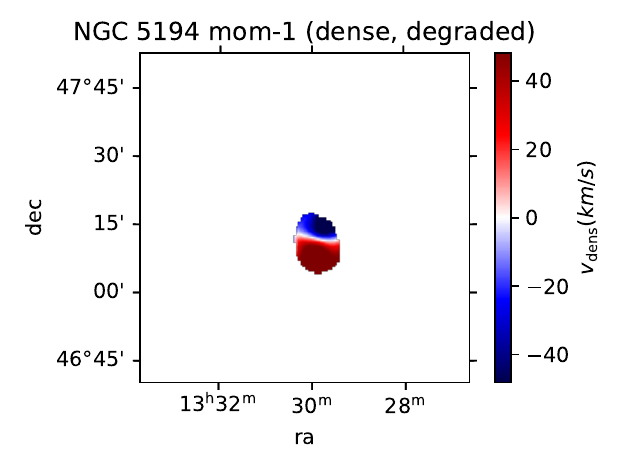}
\includegraphics[width=5.5cm]{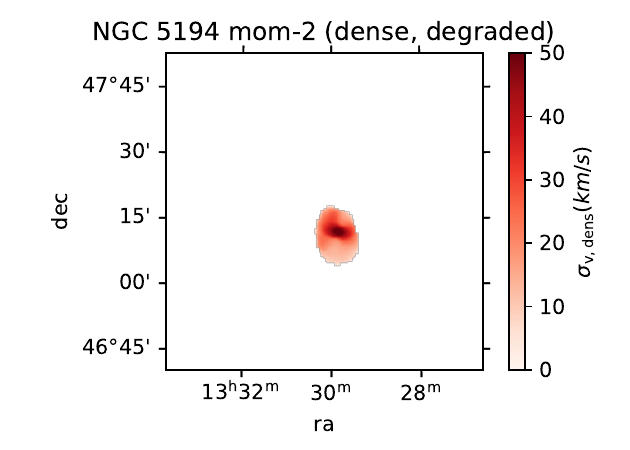}

\centering
\includegraphics[width=5.5cm]{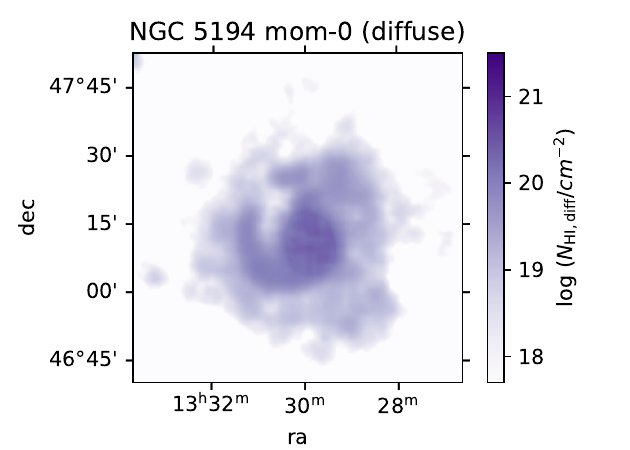}
\includegraphics[width=5.5cm]{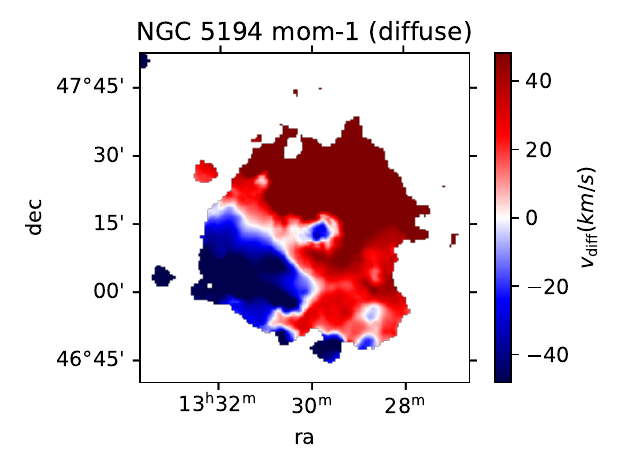}
\includegraphics[width=5.5cm]{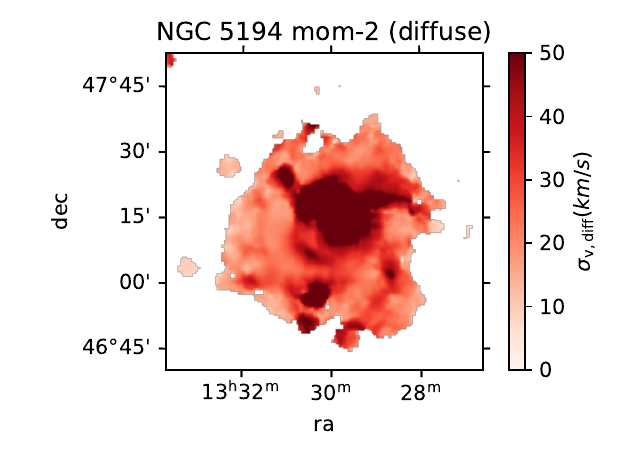}

\centering
\includegraphics[width=5.5cm]{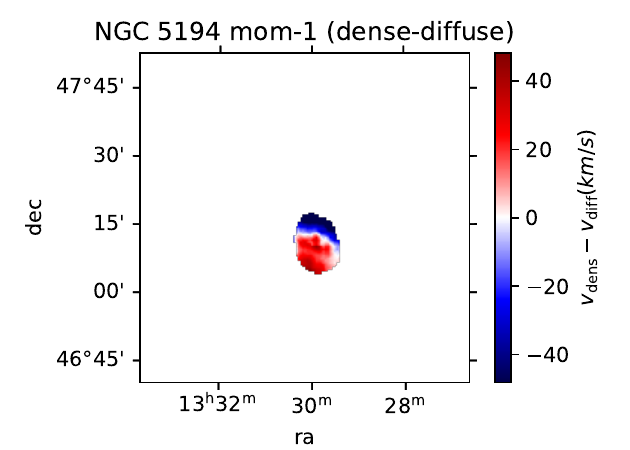}

\caption{Same as Figure~\ref{fig:mom_n628}, but for the galaxy NGC 5194. }
\label{fig:mom_n5194}
\end{figure*}

\begin{figure*} 
\centering
\includegraphics[width=5.5cm]{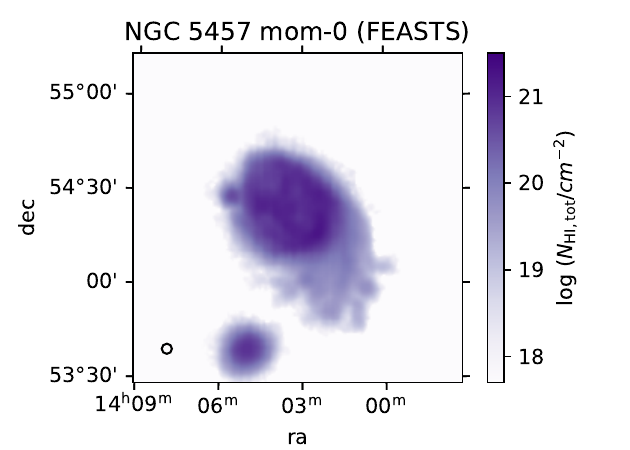}
\includegraphics[width=5.5cm]{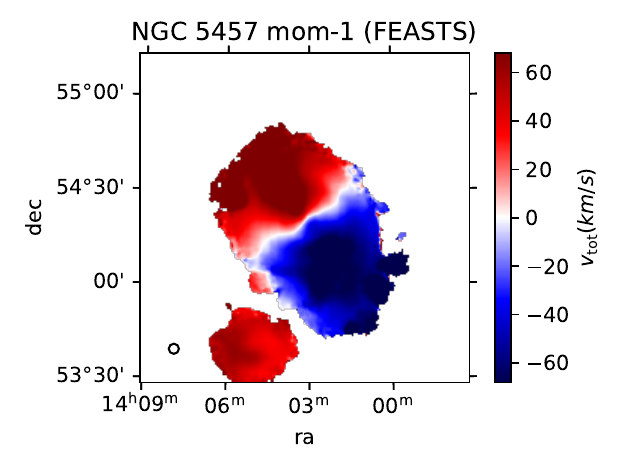}
\includegraphics[width=5.5cm]{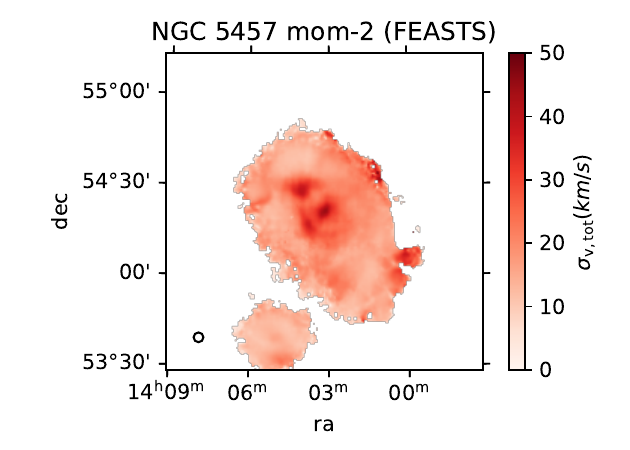}

\centering
\includegraphics[width=5.5cm]{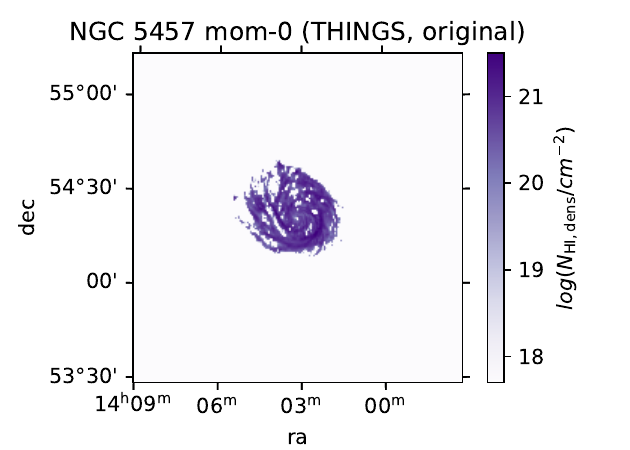}
\includegraphics[width=5.5cm]{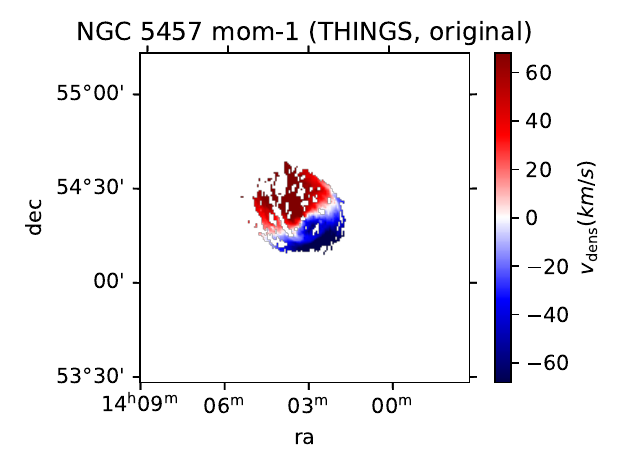}
\includegraphics[width=5.5cm]{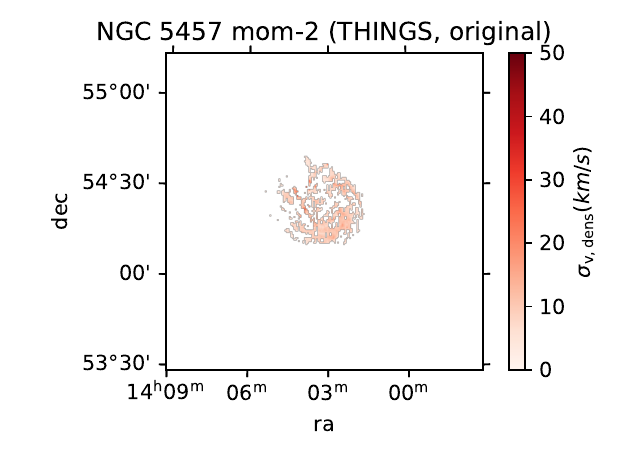}

\centering
\includegraphics[width=5.5cm]{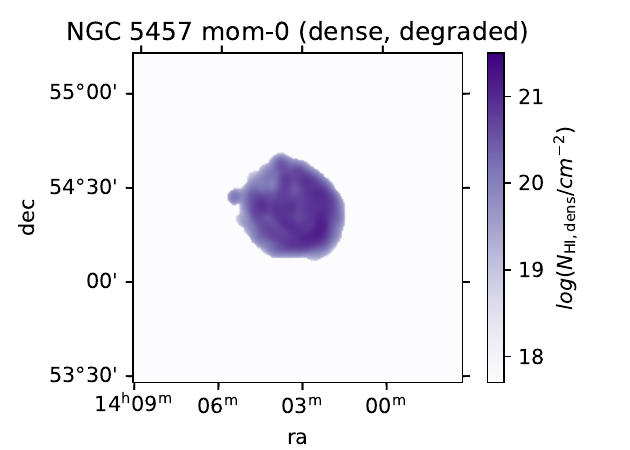}
\includegraphics[width=5.5cm]{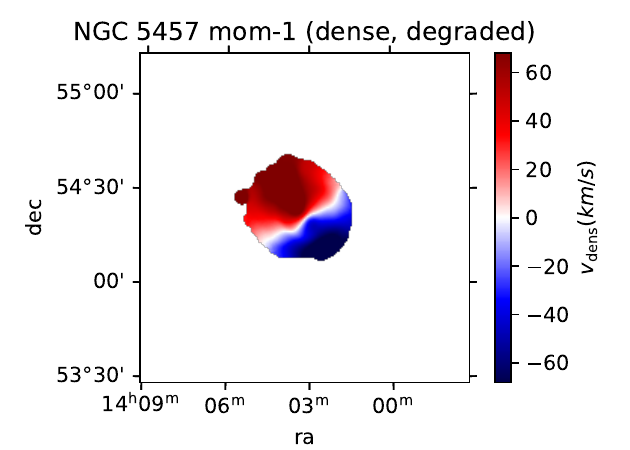}
\includegraphics[width=5.5cm]{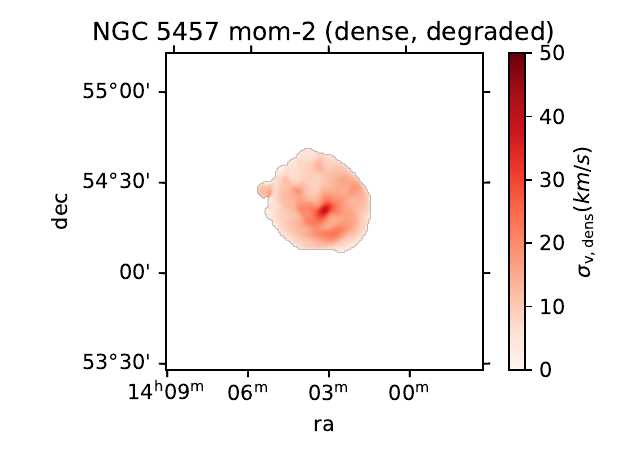}

\centering
\includegraphics[width=5.5cm]{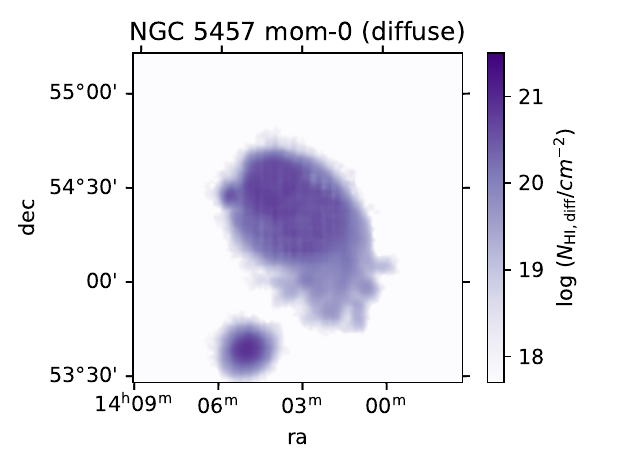}
\includegraphics[width=5.5cm]{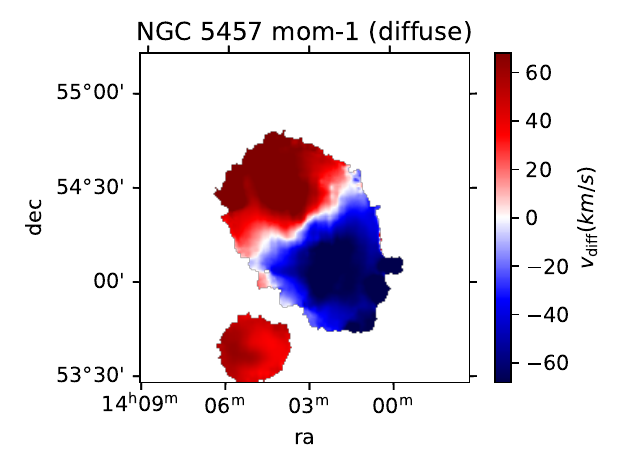}
\includegraphics[width=5.5cm]{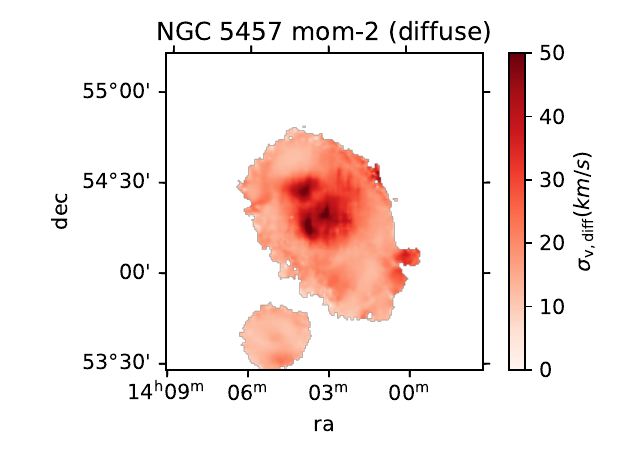}

\centering
\includegraphics[width=5.5cm]{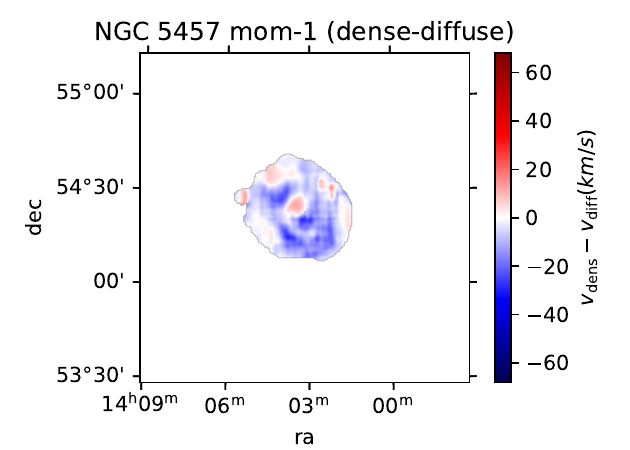}

\caption{Same as Figure~\ref{fig:mom_n628}, but for the galaxy NGC 5457. }
\label{fig:mom_n5457}
\end{figure*}

\begin{figure*} 
\centering
\includegraphics[width=5.5cm]{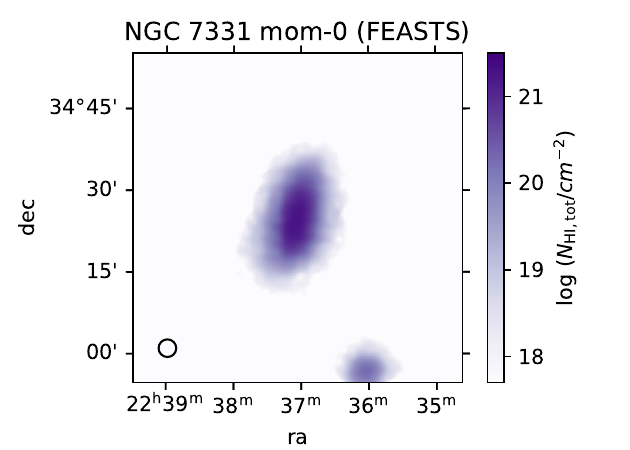}
\includegraphics[width=5.5cm]{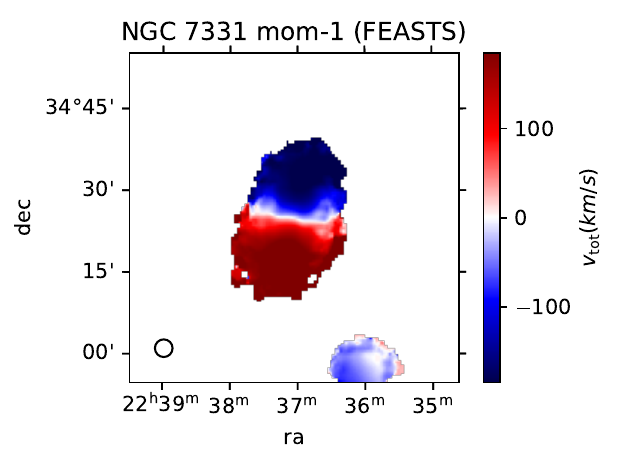}
\includegraphics[width=5.5cm]{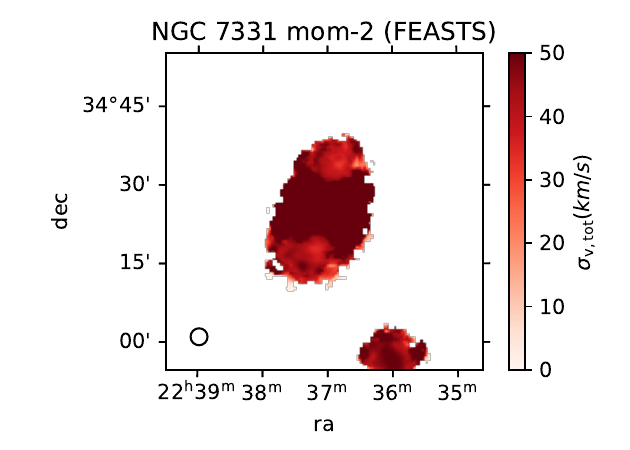}

\centering
\includegraphics[width=5.5cm]{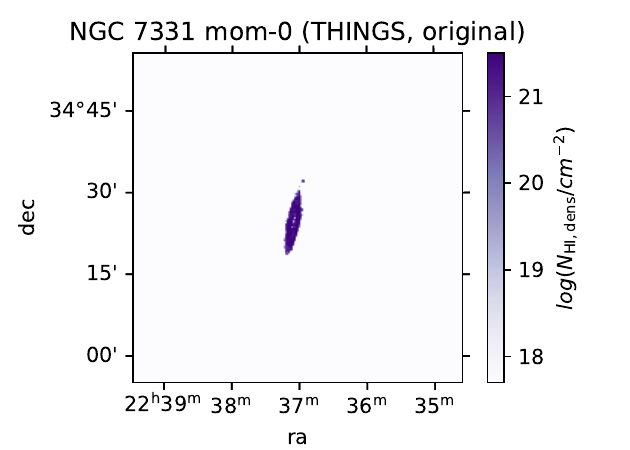}
\includegraphics[width=5.5cm]{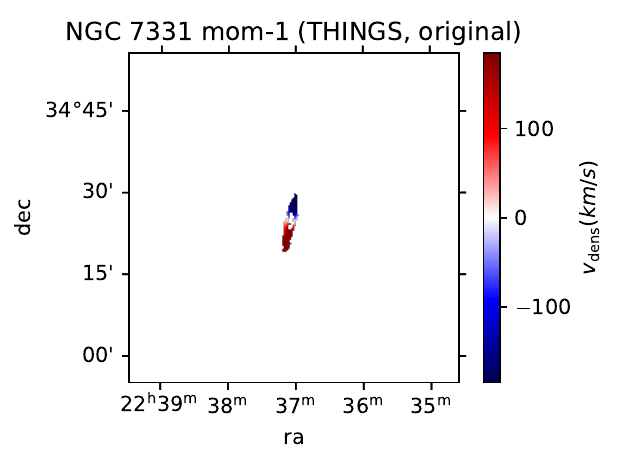}
\includegraphics[width=5.5cm]{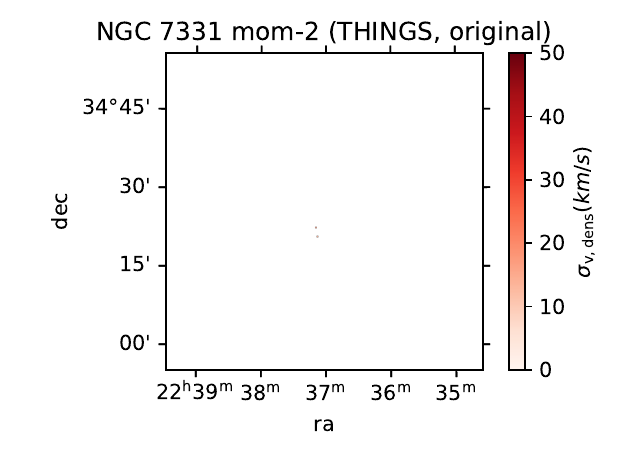}

\centering
\includegraphics[width=5.5cm]{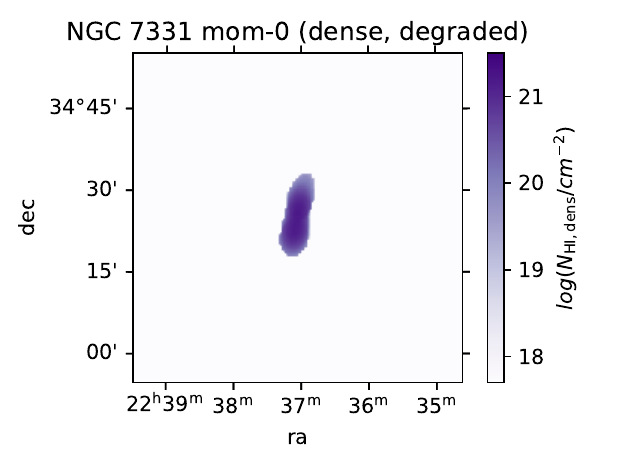}
\includegraphics[width=5.5cm]{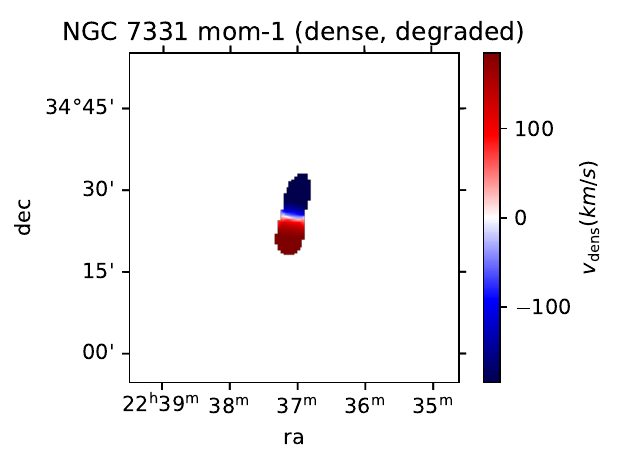}
\includegraphics[width=5.5cm]{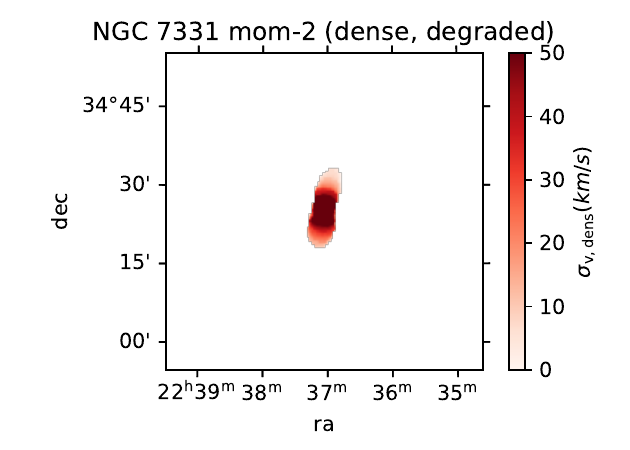}

\centering
\includegraphics[width=5.5cm]{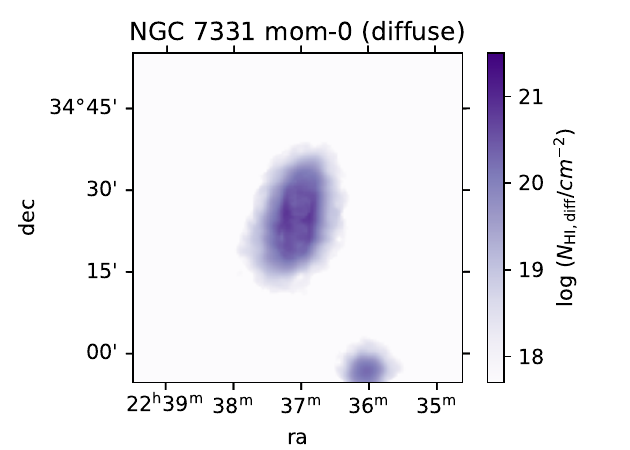}
\includegraphics[width=5.5cm]{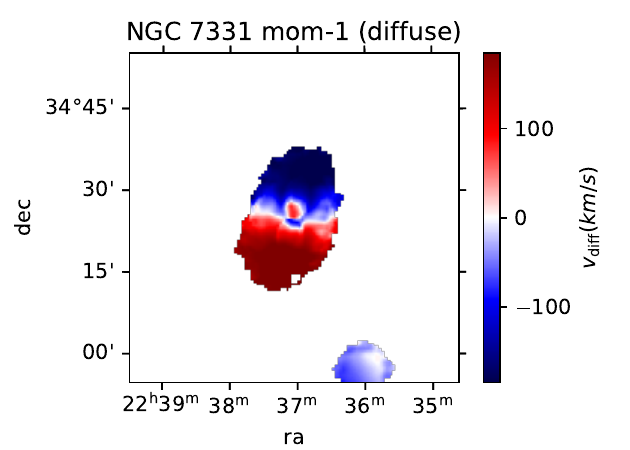}
\includegraphics[width=5.5cm]{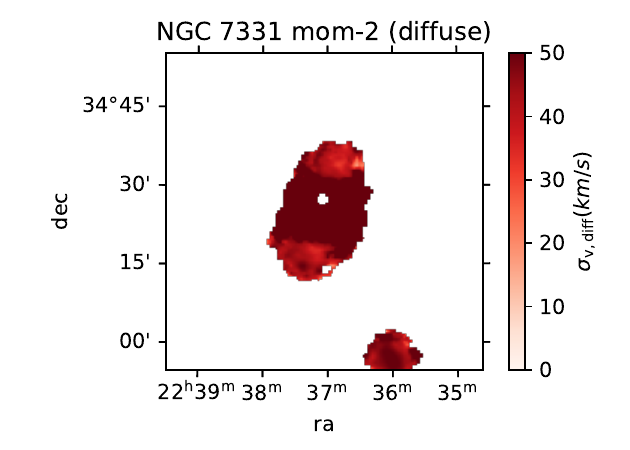}

\centering
\includegraphics[width=5.5cm]{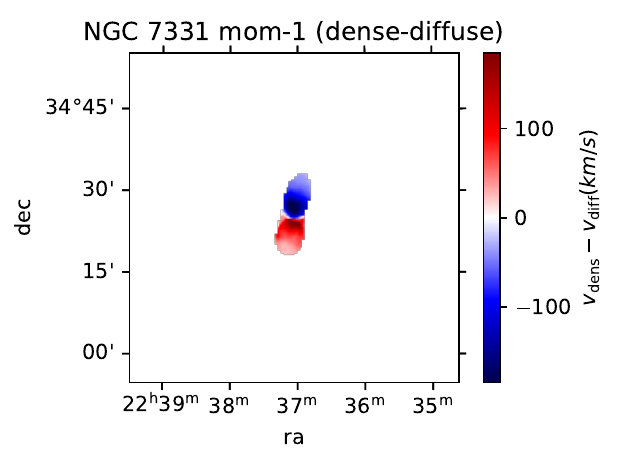}

\caption{Same as Figure~\ref{fig:mom_n628}, but for the galaxy NGC 7331. }
\label{fig:mom_n7331}
\end{figure*}

We present moment images of $\hi$ in Figure~\ref{fig:mom_n628} to \ref{fig:mom_n7331}.
The first row in each figure shows the full-extent moment 0, 1 and 2 images of the FEASTS data. 
The second to fourth row shows moment images for the original THINGS data, the degraded dense $\hi$, and the diffuse $\hi$ (see Section~\ref{sec:creat_dmom}).
The fifth row shows the difference moment-1 image of the diffuse $\hi$, which is the moment-1 image of the resolution-degraded dense $\hi$ minus the moment-1 image of the diffuse $\hi$. 

\section{Radial Profiles of Properties of the Missed HI}
\label{appendix:prof}
\begin{figure*} 
\centering
\includegraphics[width=17cm]{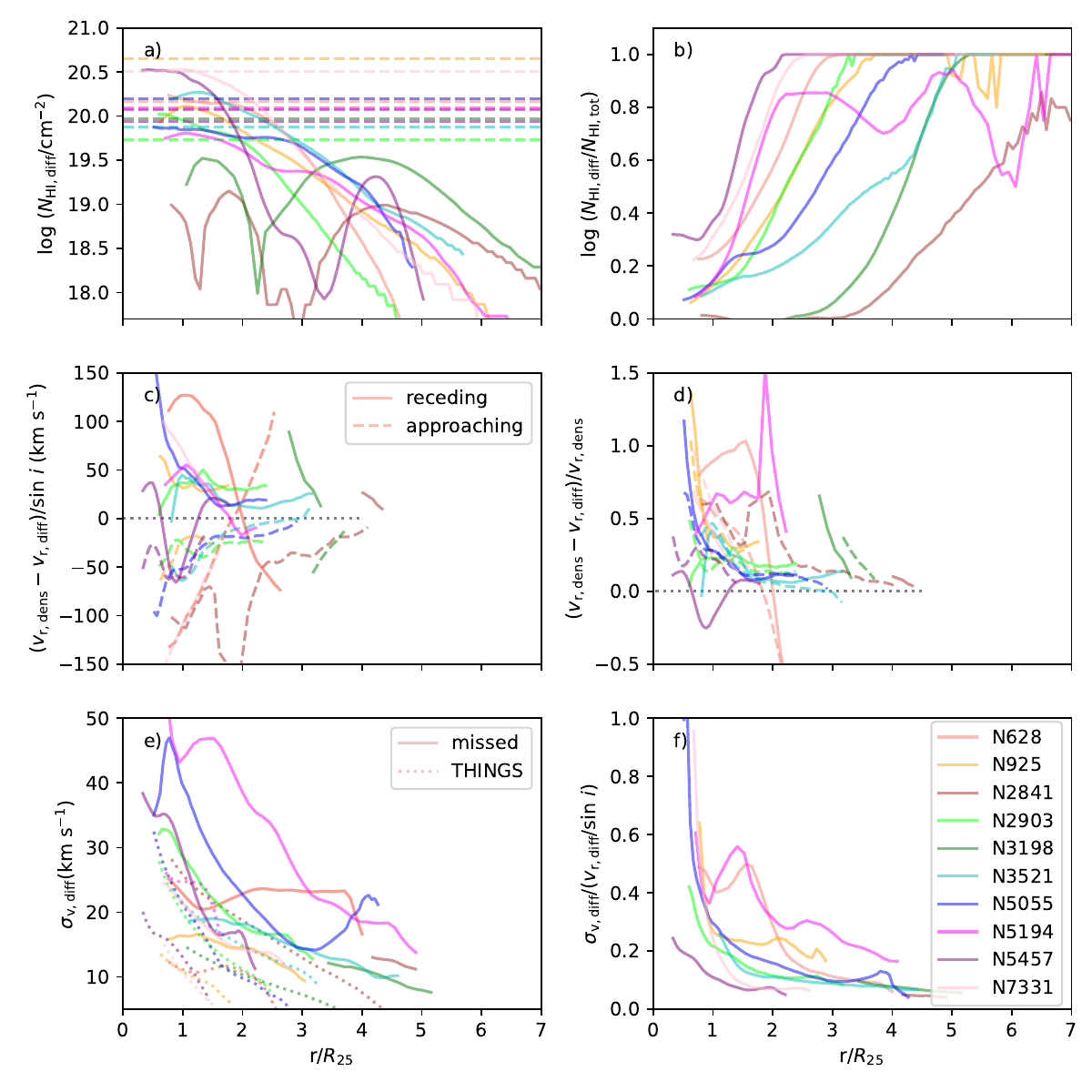}
\caption{Radial profiles of properties of the missed $\hi$. The figure is similar to Figure~\ref{fig:prof}, except that properties of the missed $\hi$ instead of the diffuse $\hi$ are plotted.
In panel a, the dashed lines show the 3-$\sigma$ column density limit of the THINGS data. 
}
\label{fig:miss_prof}
\end{figure*}

We present profiles of properties of the missed $\hi$, derived in similar ways as for the diffuse $\hi$.
The plots in Figure~\ref{fig:miss_prof} are also produced in similar ways as in Figure~\ref{fig:prof} for the diffuse $\hi$.
In panel a, most of the missed $\hi$ has column densities below the detection limit of THINGS, with the only exception NGC 5457, which has the largest $\hi$ disk among the sample.
In panel c, the relative velocity differences of NGC 925 and NGC 5055 rise above unity near the center, seemingly implying counter rotation, but the beam smearing and projection effects need be better accounted for in the future before a solid conclusion is drawn.
Otherwise, all panels show that, the distribution and kinematical properties of the missed $\hi$ are qualitatively similar to those of the diffuse $\hi$.
   
\bibliographystyle{apj}

\end{CJK*}
\end{document}